\documentclass[11pt]{article}
\usepackage{amsmath,amsfonts,amssymb,graphicx,epsfig,pdflscape,multirow}
\usepackage{cite}
\usepackage[usenames]{color}
\usepackage{pstricks}

\usepackage{subfig}

\usepackage[nomessages]{fp}

\usepackage[backref=false]{hyperref}
\hypersetup{
colorlinks=true,
citecolor=red,
linkcolor=darkblue
}
\definecolor{darkblue}{rgb}{0,0,.8}
\definecolor{red}{rgb}{1,0,0}
\definecolor{myboxcolor}{rgb}{.95,.94,.99}

\numberwithin{equation}{section}

\graphicspath{{./eps/}}
\long\def\ignore#1{}
\definecolor{purple}{rgb}{1,0,1}
\definecolor{darkpurple}{rgb}{1,.2,1}
\definecolor{pink}{rgb}{1,.7,.7}

\definecolor{AliceBlue}{RGB}{240,248,255}
\definecolor{CornflowerBlue}{RGB}{100,149,237}
\definecolor{RoyalBlue}{RGB}{60,105,225}
\definecolor{RedViolet}{RGB}{199,21,133}

\def\TLfoFrb{
\psbezier[showpoints=false,linewidth=2pt,linecolor=RoyalBlue](0,0.45)(.25,.45)(.45,.25)(0.45,0)
\psbezier[showpoints=false,linewidth=2pt,linecolor=RoyalBlue](0,0.55)(.35,.55)(.55,.35)(0.55,0)
\psbezier[showpoints=false,linewidth=2pt,linecolor=RoyalBlue](0.45,1)(0.45,0.65)(0.65,0.45)(1,.45)
\psbezier[showpoints=false,linewidth=2pt,linecolor=RoyalBlue](0.55,1)(0.55,0.75)(0.75,0.55)(1,.55)
}

\def\IfoFrb{
\rput{-90}(0,1){\TLfoFrb}
}

\def\BfoFrb{
\psline[linewidth=2pt,linecolor=RoyalBlue](0,.45)(1,.45)
\psline[linewidth=2pt,linecolor=RoyalBlue](0,.55)(1,.55)
\psline[linewidth=2pt,linecolor=RoyalBlue](.45,0)(.45,.4)
\psline[linewidth=2pt,linecolor=RoyalBlue](.55,0)(.55,.4)
\psline[linewidth=2pt,linecolor=RoyalBlue](.45,.6)(.45,1)
\psline[linewidth=2pt,linecolor=RoyalBlue](.55,.6)(.55,1)
}

\def\BinvfoFrb{
\rput{-90}(0,1){\BfoFrb}
}

\def\TLfoFp{
\psbezier[showpoints=false,linewidth=2pt,linecolor=RedViolet](0,0.45)(.25,.45)(.45,.25)(0.45,0)
\psbezier[showpoints=false,linewidth=2pt,linecolor=RedViolet](0,0.55)(.35,.55)(.55,.35)(0.55,0)
\psbezier[showpoints=false,linewidth=2pt,linecolor=RedViolet](0.45,1)(0.45,0.65)(0.65,0.45)(1,.45)
\psbezier[showpoints=false,linewidth=2pt,linecolor=RedViolet](0.55,1)(0.55,0.75)(0.75,0.55)(1,.55)
}

\def\IfoFp{
\rput{-90}(0,1){\TLfoFp}
}

\def\XfoFp{
\psbezier[showpoints=false,linewidth=2pt,linecolor=RedViolet](0,0.45)(.25,.45)(.45,.25)(0.45,0)%
\psbezier[showpoints=false,linewidth=2pt,linecolor=RedViolet](0,0.55)(.25,.55)(.45,.75)(0.45,1)
\psbezier[showpoints=false,linewidth=2pt,linecolor=RedViolet](0.55,0)(0.55,0.25)(0.75,0.45)(1,.45)
\psbezier[showpoints=false,linewidth=2pt,linecolor=RedViolet](0.55,1)(0.55,0.75)(0.75,0.55)(1,.55)%
}

\newcommand{\nc}{\newcommand}
\nc\disp{\displaystyle}
\nc{\fh}{\hat{f}}
\nc{\muh}{\hat{\mu}}
\nc{\nuh}{\hat{\nu}}
\nc{\spos}[2]{\makebox(0,0)[#1]{$\sm{#2}$}}
\nc{\sm}[1]{{\scriptstyle #1}}
\nc{\qbar}{\overline{q}}
\nc{\bib}{\bibitem}
\nc{\al}{\alpha}
\nc{\g}{\gamma}
\nc{\G}{\Gamma}
\nc{\D}{\Delta}
\nc{\eps}{\epsilon}
\nc{\la}{\lambda}
\nc{\La}{\Lambda}
\nc{\var}{\varphi}
\nc{\pa}{\partial}
\nc{\nn}{\nonumber \\ }
\nc{\hf}{\frac{1}{2}}
\nc{\dz}{\frac{dz}{2\pi i}}
\nc{\bin}[2]{\left(\!\!\!\begin{array}{c} {#1}\\ {#2} \end{array}\!\!\!\right)}
\nc{\be}{\begin{equation}}
\nc{\ee}{\end{equation}}
\nc{\bea}{\begin{eqnarray}}
\nc{\eea}{\end{eqnarray}}
\nc{\bra}[1]{\langle {#1}|}
\nc{\ket}[1]{|{#1}\rangle}
\nc{\ketw}[1]{({#1})^{\phantom{a}}_{{\cal W}}}
\nc{\chit}{\raisebox{0.25ex}{$\chi$}}
\nc{\chih}{\raisebox{0.25ex}{$\hat\chi$}}
\nc{\Db}{\mbox{\boldmath $D$}}
\nc{\Hb}{\mbox{\boldmath $H$}}
\nc{\calH}{{\cal H}}
\nc{\calR}{{\cal R}}
\nc{\calL}{{\cal L}}
\nc{\calV}{{\cal V}}
\nc{\Hc}{{\cal H}}
\nc{\Rc}{{\cal R}}
\nc{\Lc}{{\cal L}}
\nc{\Vc}{{\cal V}}
\nc{\Ib}{\mbox{\boldmath $I$}}
\nc{\qb}{\bar{q}}
\nc{\Ac}{\mathcal{A}}
\nc{\Bc}{\mathcal{B}}
\nc{\Cc}{\mathcal{C}}
\nc{\Dc}{\mathcal{D}}
\nc{\Ec}{\mathcal{E}}
\nc{\Gc}{\mathcal{G}}
\nc{\Ic}{\mathcal{I}}
\nc{\Jc}{\mathcal{J}}
\nc{\Oc}{\mathcal{O}}
\nc{\Pc}{\mathcal{P}}
\nc{\Sc}{\mathcal{S}}
\nc{\Tc}{\mathcal{T}}
\nc{\Wc}{\mathcal{W}}
\nc{\Xc}{\mathcal{X}}
\nc{\Yc}{\mathcal{Y}}
\nc{\Zc}{\mathcal{Z}}
\nc{\fus}{\mbox{}\,\hat\otimes\,\mbox{}}
\nc{\Pch}{\hat{\Pc}}
\nc{\Rch}{\hat{\Rc}}
\nc{\Dh}{\hat{\Delta}}
\nc{\rh}{\hat{r}}
\nc{\sh}{\hat{s}}
\nc{\kb}{\bar{k}}
\nc{\taub}{\bar{\tau}}
\nc{\Jcb}{\Jc_{\mathrm{b}}}
\nc{\rtt}{\mathtt{r}}
\nc{\stt}{\mathtt{s}}
\nc{\cosR}{\cos\frac{\pi p'rr'}{p}}
\nc{\cosS}{\cos\frac{\pi pss'}{p'}}
\nc{\sinR}{\sin\frac{\pi p'rr'}{p}}
\nc{\sinS}{\sin\frac{\pi pss'}{p'}}
\def\vvdots{\mathinner{\mkern1mu\raise1pt\vbox{\kern7pt\hbox{.}}\mkern2mu
  \raise4pt\hbox{.}\mkern2mu\raise7pt\hbox{.}\mkern1mu}}
\nc{\gauss}[2]{\left[\!\!\begin{array}{c} {#1}\\ {#2} \end{array}\!\!\right]}
\nc{\sbin}[2]{\left\{\!\!\!\begin{array}{c} {#1}\\ {#2} 
\end{array}\!\!\!\right\}}
\nc{\sbinlr}[2]{\Big\langle\!\!\begin{array}{c} {#1}\\ {#2} 
\end{array}\!\!\Big\rangle}
\nc{\bino}[2]{\left(\!\!\begin{array}{c} {#1}\\ {#2} \end{array}\!\!\right)}
\def\half {\mbox{$\textstyle \frac{1}{2}$} }
\def\vec#1{\mbox {\boldmath $#1$}}

\definecolor{lightblue}{rgb}{.61,.61,1}
\definecolor{midblue}{rgb}{.7,.7,1}
\definecolor{lightlightblue}{rgb}{.9,.9,1}
\definecolor{lightestblue}{rgb}{.96,.96,1}
\definecolor{lightpurple}{rgb}{1,.65,1}

\def\loopa{
\psframe[linewidth=.25pt](0,0)(1,1)
\psarc[linewidth=1.5pt,linecolor=blue](1,0){.5}{90}{180}
\psarc[linewidth=1.5pt,linecolor=blue](0,1){.5}{-90}{0}
}
\def\loopb{
\psframe[linewidth=.25pt](0,0)(1,1)
\psarc[linewidth=1.5pt,linecolor=blue](0,0){.5}{0}{90}
\psarc[linewidth=1.5pt,linecolor=blue](1,1){.5}{180}{270}
}
\def\braida{
\psframe[linewidth=.25pt](0,0)(1,1)
\psline[linewidth=1.5pt,linecolor=blue](.5,0)(.5,.4)
\psline[linewidth=1.5pt,linecolor=blue](.5,.6)(.5,1)
\psline[linewidth=1.5pt,linecolor=blue](0,.5)(1,.5)
}
\def\braidb{
\psframe[linewidth=.25pt](0,0)(1,1)
\psline[linewidth=1.5pt,linecolor=blue](.5,0)(.5,1)
\psline[linewidth=1.5pt,linecolor=blue](0,.5)(.4,.5)
\psline[linewidth=1.5pt,linecolor=blue](.6,.5)(1,.5)
}

\def\facegrid#1#2{
\psframe[fillstyle=solid,fillcolor=lightlightblue,linewidth=0pt]#1#2
\psgrid[gridlabels=0pt,subgriddiv=1]#1#2
}

\nc{\ch}{{\rm ch}}
\nc{\R}{{\cal R}}
\nc{\dkk}{\delta_{j,\{k,k'\}}^{(2)}}
\nc{\drr}{\delta_{j,\{r,r'\}}^{(2)}}
\nc{\ddkk}{\delta_{j,\{k,k'\}}^{(4)}}
\nc{\dddkk}{\delta_{j,\{k,k'\}}^{(8)}}
\nc{\dnn}{\delta_{j,\{n,n'\}}^{(2)}}
\nc{\ddnn}{\delta_{j,\{n,n'\}}^{(4)}}
\nc{\dddnn}{\delta_{j,\{n,n'\}}^{(8)}}

\nc{\chh}{\widehat{\mathrm{ch}}}
\nc{\ph}{\hat{p}}
\def\equiveq{=}

\definecolor{pink}{rgb}{1,.65,.65}

\thicklines

\thicklines

\def \fusedX{
\psellipse[linewidth=1pt](0,0)(.4,.15)
\psellipse[linewidth=1pt](2,0)(.4,.15)
\psellipse[linewidth=1pt](0,2)(.4,.15)
\psellipse[linewidth=1pt](2,2)(.4,.15)
\psbezier[linewidth=1pt](.4,0)(.9,.75)(1.1,.75)(1.6,0)
\psbezier[linewidth=1pt](-.4,0)(.6,.7)(.6,1.3)(-.4,2)
\psbezier[linewidth=1pt](.4,2)(.9,1.25)(1.1,1.25)(1.6,2)
\psbezier[linewidth=1pt](2.4,0)(1.4,.7)(1.4,1.3)(2.4,2)
 }

\def \fusedI{
\psellipse[linewidth=1pt](0,0)(.4,.15)
\psellipse[linewidth=1pt](2,0)(.4,.15)
\psellipse[linewidth=1pt](0,2)(.4,.15)
\psellipse[linewidth=1pt](2,2)(.4,.15)
\multirput(0,0)(2,0){2}{\multirput(0,0)(.8,0){2}{\psline[linewidth=1pt](-.4,0)(-.4,2)}}
}

\def \fusedE{
\psellipse[linewidth=1pt](0,0)(.4,.15)
\psellipse[linewidth=1pt](2,0)(.4,.15)
\psellipse[linewidth=1pt](0,2)(.4,.15)
\psellipse[linewidth=1pt](2,2)(.4,.15)
\psbezier[linewidth=1pt,showpoints=false](-.4,0)(.5,1.2)(1.5,1.2)(2.4,0)
\psbezier[linewidth=1pt,showpoints=false](.4,0)(.7,.5)(1.3,.5)(1.6,0)
\psbezier[linewidth=1pt,showpoints=false](-.4,2)(.5,.8)(1.5,.8)(2.4,2)
\psbezier[linewidth=1pt,showpoints=false](.4,2)(.7,1.5)(1.3,1.5)(1.6,2)
}

\def \fusedB{
\psellipse[linewidth=1pt](0,0)(.4,.15)
\psellipse[linewidth=1pt](2,0)(.4,.15)
\psellipse[linewidth=1pt](0,2)(.4,.15)
\psellipse[linewidth=1pt](2,2)(.4,.15)
\psline[linewidth=1pt](-.4,2)(1.6,0)
\psline[linewidth=1pt](.4,2)(2.4,0)
\pspolygon[linewidth=0pt,fillstyle=solid,fillcolor=white,linecolor=white](.35,1.2)(.85,1.6)(1.65,.8)(1.15,.4)
\psline[linewidth=1pt](-.4,0)(1.6,2)
\psline[linewidth=1pt](.4,0)(2.4,2)
}

\def \fusedBinv{
\psellipse[linewidth=1pt](0,0)(.4,.15)
\psellipse[linewidth=1pt](2,0)(.4,.15)
\psellipse[linewidth=1pt](0,2)(.4,.15)
\psellipse[linewidth=1pt](2,2)(.4,.15)
\psline[linewidth=1pt](-.4,0)(1.6,2)
\psline[linewidth=1pt](.4,0)(2.4,2)
\pspolygon[linewidth=2pt,fillstyle=solid,fillcolor=white,linecolor=white](1.6,1.2)(1.2,1.6)(.4,.8)(.8,.4)
\psline[linewidth=1pt](-.4,2)(1.6,0)
\psline[linewidth=1pt](.4,2)(2.4,0)
}

\def \Monoid{
\rput(0,0){\MonoidCap}
\rput(0,1){\MonoidCup}
}

\def \MonoidCap{
\psbezier[linewidth=1pt](0,0)(0,.55)(1,.55)(1,0)
}

\def \MonoidCup{
\psbezier[linewidth=1pt](0,0)(0,-.55)(1,-.55)(1,0)
}

\def \Braid{
\psline[linewidth=1pt](0,1)(1,0)
\pspolygon[linewidth=1pt,linecolor=white,fillstyle=solid,fillcolor=white](.3,.5)(.5,.7)(.7,.5)(.5,.3)
\psline[linewidth=1pt](0,0)(1,1)
}

\def \invBraid{
\psline[linewidth=1pt](0,0)(1,1)
\pspolygon[linewidth=1pt,linecolor=white,fillstyle=solid,fillcolor=white](.3,.5)(.5,.7)(.7,.5)(.5,.3)
\psline[linewidth=1pt](0,1)(1,0)
}

\def \cabledTwo{
\psbezier[linewidth=1pt,showpoints=false](.4,0)(.7,.75)(1.3,.75)(1.6,0)
\psbezier[linewidth=1pt,showpoints=false](-.4,0)(.2,1.6)(1.8,1.6)(2.4,0)
}

\def \cabledThree{
\multirput(0,0)(2,0){2}{\psbezier[linewidth=1pt,showpoints=false](.4,0)(.7,.75)(1.3,.75)(1.6,0)}
\psbezier[linewidth=1pt,showpoints=false](-.4,0)(.2,2.3)(3.8,2.3)(4.4,0)
}

\def \cabledFour{
\multirput(0,0)(2,0){3}{\psbezier[linewidth=1pt,showpoints=false](.4,0)(.7,.75)(1.3,.75)(1.6,0)}
\psbezier[linewidth=1pt,showpoints=false](-.4,0)(.2,2.7)(5.8,2.7)(6.4,0)
}

\def \cabledDefOne[#1]{
\psline[linewidth=1pt](-.4,0)(-.4,#1)
\psline[linewidth=1pt](.4,0)(.4,#1)
}

\def \link{
\psbezier[linewidth=1pt,showpoints=false](.4,0)(.7,.75)(1.3,.75)(1.6,0)
}

\def \linkLow{
\psbezier[linewidth=1pt,showpoints=false](.4,0)(.6,.6)(1.4,.6)(1.6,0)
}

\def \linkMed{
\psbezier[linewidth=1pt,showpoints=false](.4,0)(1,1.5)(3,1.5)(3.6,0)
}

\def \linkMeder{
\psbezier[linewidth=1pt,showpoints=false](-.4,0)(.2,2.3)(3.8,2.3)(4.4,0)
}

\def \linkLonger{
\psbezier[linewidth=1pt,showpoints=false](-.4,0)(.2,2.7)(5.8,2.7)(6.4,0)
}

\def \linkLongx{
\psbezier[linewidth=1pt,showpoints=false](.4,0)(1,3)(7,3)(7.6,0)
}

\def \linkLongX{
\psbezier[linewidth=1pt,showpoints=false](-.4,0)(.2,3.8)(7.8,3.8)(8.4,0)
}

\def \loopProj{
\psellipse[linewidth=1pt](0,0)(.4,.15)
}

\def \loopProjMed{
\psellipse[linewidth=1pt](.5,0)(.9,.15)
}

\def \loopProjLong{
\psellipse[linewidth=1pt](1,0)(1,.15)
}

\def \loopProjLonger{
\psellipse[linewidth=1pt](1,0)(1.4,.15)
}

\def \squareProj[#1,#2,#3,#4,#5]{
\psline[linewidth=1pt](.4,0)(.4,#2)
\psline[linewidth=1pt](1.6,0)(1.6,#3)
\psline[linewidth=1pt](1.6,0)(1.6,-#4)
\psline[linewidth=1pt](.4,0)(.4,-#5)
\pspolygon[linewidth=1pt,fillstyle=solid,fillcolor=white](0,0)(1,1)(2,0)(1,-1)
\rput(1,0){$#1\lambda$}
}

\def \singleDef[#1]{
\psline(-.4,0)(-.4,#1)
}

\def \singleDefR[#1]{
\psline(.4,0)(.4,#1)
}

\def \proj[#1]{
\psellipse[linewidth=1pt](#1,0)(#1,.2)
}

\def \Xtile{
\psarc[linecolor=blue,linewidth=1.5pt](0,0){.4}{0}{90}
\psarc[linecolor=blue,linewidth=1.5pt](1,0){.4}{90}{180}
\psarc[linecolor=blue,linewidth=1.5pt](1,1){.4}{180}{270}
\psarc[linecolor=blue,linewidth=1.5pt](0,1){.4}{270}{360}
}

\def \TLtile{
\psarc[linecolor=blue,linewidth=1.5pt](1,0){.4}{90}{180}
\psarc[linecolor=blue,linewidth=1.5pt](1,0){.6}{90}{180}
\psarc[linecolor=blue,linewidth=1.5pt](0,1){.6}{270}{360}
\psarc[linecolor=blue,linewidth=1.5pt](0,1){.4}{270}{360}
}

\nc\drtm{{\vec D}}      		
\nc\face{\mathbb{X}} 		
\nc\faceK{\mathbb{K}} 		
\nc\genface[1]{\mathbb{X}^{(#1)}} 
\nc \ham{{\mathcal H}}			

\nc \nface{\mathbb{\hat X}}	

\nc\TLw{\omega} 			
\nc\TLb{\beta} 				

\nc\BMWw{\omega_2}		
\nc\BMWb{\beta_2}			

\nc\fTLw{\omega_2} 			
\nc\fTLb{\beta_2} 			

\nc{\genw}[1]{\omega_{#1}}  	
\nc{\Genw}[2]{\omega_{#1}^{(#2)}}
\nc{\genb}[1]{\beta_{#1}}       	

\def \trinomial[#1][#2][#3][#4]{\left[{#1\atop #2,#3,#4}\right]}

\def \superTrinomial[#1][#2]{
\left({#1 \atop #2} \right)_{\!2}
}

\def \qTrinomial[#1][#2][#3][#4]{\left[{#1\atop #2,#3,#4}\right]_{\!q}}

\begin{document}

\topmargin -5mm
\oddsidemargin 5mm

\setcounter{page}{1}

\vspace{8mm}
\begin{center}
{\huge {\bf Logarithmic Superconformal Minimal Models}}

\vspace{10mm}
{\Large Paul A. Pearce$^\ast$, J{\o}rgen Rasmussen$^\dagger$, Elena Tartaglia$^\ast$}
\\[.4cm]
{\em {}$^\ast$Department of Mathematics and Statistics, University of Melbourne}\\
{\em Parkville, Victoria 3010, Australia}
\\[.4cm]
{\em {}$^\dagger$School of Mathematics and Physics, University of Queensland}\\
{\em St Lucia, Brisbane, Queensland 4072, Australia}
\\[.4cm]
{\tt p.pearce\,@\,ms.unimelb.edu.au}
\qquad
{\tt j.rasmussen\,@\,uq.edu.au}
\qquad
{\tt elena.tartaglia\,@\,unimelb.edu.au}
\end{center}

\vspace{10mm}
\centerline{{\bf{Abstract}}}
\vskip.4cm
\noindent
The higher fusion level logarithmic minimal models ${\cal LM}(P,P';n)$ have recently been constructed as the diagonal GKO cosets 
${(A_1^{(1)})_k\oplus(A_1^{(1)})_n}/{(A_1^{(1)})_{k+n}}$ where $n\ge 1$ is an integer fusion level 
and \mbox{$k=\frac{nP}{P'-P}-2$} is a fractional level. For $n=1$,  these are the well-studied logarithmic minimal \mbox{models} ${\cal LM}(P,P')\equiv {\cal LM}(P,P';1)$. For $n\ge 2$, we argue that these critical theories are realized on the lattice by $n\times n$ fusion of the $n=1$ models. 
We study the critical fused lattice models ${\cal LM}(p,p')_{n\times n}$ within a lattice approach and focus our study on the $n=2$ models. We call these logarithmic superconformal minimal models 
${\cal LSM}(p,p')\equiv {\cal LM}(P,P';2)$ where \mbox{$P=|2p-p'|$}, \mbox{$P'=p'$} and $p,p'$ are coprime. These models share the 
central charges $c=c^{P,P';2}=\frac{3}{2}\big(1-\frac{2(P'-P)^2}{P P'}\big)$ of the \mbox{rational} superconformal minimal models ${\cal SM}(P,P')$. 
Lattice realizations of these theories are constructed by fusing $2\times 2$ blocks of the elementary face operators of the $n=1$ logarithmic minimal models ${\cal LM}(p,p')$. 
Algebraically, this entails the fused planar Temperley-Lieb algebra which is a \mbox{spin-1} Birman-Murakami-Wenzl tangle algebra with loop fugacity $\beta_2=[x]_3=x^2+1+x^{-2}$ and twist $\omega=x^4$ where $x=e^{i\lambda}$ and $\lambda=\frac{(p'-p)\pi}{p'}$. 
The first two members of this $n=2$ series are superconformal dense polymers ${\cal LSM}(2,3)$ with $c=-\frac{5}{2}$, $\beta_2=0$ and superconformal percolation ${\cal LSM}(3,4)$ with $c=0$, $\beta_2=1$. 
We calculate the bulk and \mbox{boundary} free energies analytically. 
By numerically studying finite-size conformal spectra on the strip with \mbox{appropriate} boundary conditions we argue that, in the continuum scaling limit, these lattice models are associated with the logarithmic superconformal models ${\cal LM}(P,P';2)$. 
For system size $N$, we propose finitized Kac character formulas of the form 
$q^{-\frac{c^{P,P';2}}{24}+\Delta^{P,P';2}_{r,s,\ell}}\hat{\chi}^{(N)}_{r,s;\ell}(q)$ for $s$-type boundary conditions with $r=1$, $s=1,2,3,\ldots$, $\ell=0,1,2$. The $P,P'$ dependence enters only in the fractional power of $q$ in the prefactor and $\ell=0,2$ labels the Neveu-Schwarz sectors ($r+s$ even) and $\ell=1$ labels the Ramond sectors ($r+s$ odd). 
Combinatorially, the finitized characters involve Motzkin and Riordan polynomials defined in terms of $q$-trinomial coefficients. 
Using the Hamiltonian limit and the finitized characters we argue, from examples of finite lattice calculations, that there exist reducible yet indecomposable representations for which the Virasoro dilatation operator $L_0$ exhibits rank-$2$ Jordan cells confirming that these theories are indeed logarithmic. 
We relate these results to the $N=1$ superconformal representation theory.

\newpage
\tableofcontents

\newpage
\hyphenpenalty=30000

\section{Introduction}

The simplest logarithmic Conformal Field Theories (CFTs)~\cite{DFMS,Gurarie93} are by now well studied from an algebraic and a lattice perspective. 
The current status can be seen in the special issue~\cite{SpecialIssue}. Some of the articles more relevant to this work include \cite{CardyLimits,CreutzigRidout,GJRSV}.
In the Virasoro picture, that is assuming that the conformal algebra is the Virasoro algebra and not an extended symmetry algebra, it has recently been argued~\cite{PRcoset13} that general logarithmic minimal models ${\cal LM}(P,P';n)$ at arbitrary integer fusion level $n$ are constructed as diagonal $su(2)$ GKO cosets~\cite{GKO85,GKO86,ACT91}. For $n=1$, these logarithmic CFTs are realized on the square lattice by the logarithmic minimal models ${\cal LM}(p,p')$~\cite{PRZ} where $p,p'$ are coprime integers. As loop models, these coincide with special ``rational" points on the critical line of $O(n)$ models~\cite{Nienhuis1990}. 
The first members of this series include critical dense polymers ${\cal LM}(1,2)$~\cite{PR2007,PRV0910,PRV1210,MDPRtorus} and critical (bond) percolation ${\cal LM}(2,3)$~\cite{BroadHamm57}. For $n=2$, the coset theories are logarithmic superconformal field theories. Such theories have been considered, from within the algebraic approach, by several authors~\cite{KhorramiEtAl,KheirandishKh,MavromatosEtAl01,MavromatosEtAl03,RasmussenLog04,Nagi}. 

In this paper, we adopt a lattice approach and study exactly solvable two-dimensional lattice models~\cite{BaxBook} associated with the $n>1$ logarithmic minimal models ${\cal LM}(P,P';n)$. On the square lattice, the Boltzmann face weights of the fused models ${\cal LM}(p,p')_{n\times n}$~\cite{Tartaglia} are obtained by using a process of fusion~\cite{KulishReshetikhinSklyanin1981} to form $n\times n$ fused blocks of elementary face operators of the ${\cal LM}(p,p')$ models. For integrable boundary conditions on a strip, commuting double row transfer matrices for the fused models can be constructed analogously to the associated rational theories~\cite{BPO96}. 
In this way, we obtain families of Yang-Baxter integrable loop models which realize the complete family of higher fusion level logarithmic minimal models ${\cal LM}(P,P';n)$ in the continuum scaling limit.

We develop a general framework but focus our study here on the logarithmic superconformal minimal models 
as $2\times 2$ fusions of the logarithmic minimal models ${\cal LM}(p,p')$
\bea
{\cal LSM}(p,p'):={\cal LM}(p,p')_{2\times 2}\equiv{\cal LM}(P,P';2),\qquad P=|2p-p'|,\ \; P'=p'\label{super}
\eea
The explicit identification of these theories with $n=2$ logarithmic minimal cosets is consistent with the duality relation $p\leftrightarrow p'-p$ of Section~\ref{Duality} and is confirmed by direct numerics in Section~\ref{numerical}. For $p\ge 2$, the central charges of these logarithmic theories are shared with the associated nonunitary minimal models so the identifications ${\cal SM}(p,p')\equiv{\cal M}(P,P';2)$ also apply naturally to these rational minimal models. For the unitary minimal models with  $p=p'-1$, this is in accord with the analytic results of 
Kl\"umper and Pearce~\cite{KlumperPearce1992}. 
The first members of the superconformal series (\ref{super}) include superconformal dense polymers ${\cal LSM}(2,3)$ and superconformal percolation ${\cal LSM}(3,4)$ which are lattice models that are expected to be of independent interest in statistical mechanics. 

Mathematically, the Yang-Baxter algebras underlying the general ${\cal LM}(P,P';n)$ models are planar braid-monoid algebras~\cite{JonesPlanar} in the form of $n\times n$ fused Temperley-Lieb algebras~\cite{TL,MartinBook,AkutsuWadati87_I,AkutsuDeguchiWadati87_II,DeguchiAkutsuWadati88_III,AkutsuDeguchiWadati88_IV,DeguchiWadatiAkutsu88_V}. For the $n=2$ superconformal models, this algebra is a one-parameter specialization of the two-parameter Birman-Wenzl-Murakami (BMW)~\cite{BW,Mura} tangle algebra.

The layout of the paper is as follows. In Section~\ref{Sec:LM}, we recall the coset construction~\cite{PRcoset13} of the logarithmic minimal models ${\cal LM}(P,P';n)$ and summarize their conformal data including their central charges, conformal dimensions and Kac characters. 

In Section~\ref{Sec:TL}, we describe the $n\times n$ fused Temperley-Lieb algebra, its generators in the form of braids and generalized monoids and the construction of the $n\times n$ face operators. 
The standard fusion procedure described in the existing literature~\cite{AkutsuWadati87_I,AkutsuDeguchiWadati87_II,DeguchiAkutsuWadati88_III,AkutsuDeguchiWadati88_IV,DeguchiWadatiAkutsu88_V} uses 
Wenzl-Jones projectors~\cite{Wenzl1987,Jones1985} which only exist for $n<p'$. To generalize fusion to all $n\in\mathbb{N}$, we develop a new and general diagrammatic fusion procedure and explicitly implement it in the $2\times 2$ and $n\times n$ cases. The standard procedure is reviewed in Sections~\ref{Sec:LinTL} and~\ref{Sec:FusTL}  for the purpose of establishing that the new diagrammatic fusion procedure agrees with the standard procedure when $n<p'$.
Specializing to the $n=2$ logarithmic superconformal models, we also discuss the matrix representations of the generators through their action on link states in the Neveu-Schwarz ($r+s$ even) and Ramond ($r+s$ odd) sectors labelled by the quantum numbers $(r,s,\ell)$. The form of these link states, their relation to the quantum numbers $(r,s,\ell)$ and their combinatorial counting in terms of trinomial coefficients are all new.

In Section~\ref{Sec:LSM}, we address Yang-Baxter integrability in the presence of suitable boundaries. The various subsections build on familiar constructs and concepts such as bulk and boundary Yang-Baxter equations, transfer matrices and their quantum Hamiltonian limits, duality, inversion relations and finitized characters, but we generalize them to the new context in which a number of new features and subtleties arise. 
More explicitly, restricting to the case $n=2$ with $r=1$, we construct the commuting families of double row transfer matrices and their Hamiltonian limits and observe that they exhibit a duality under the involution $p\leftrightarrow p'-p$. One new feature that arises is the appearance, in Ramond sectors, of a new boundary field $\eta$ which has not been seen previously. 
We calculate analytically the bulk and boundary free energies in the $r=1$  sectors. Also, for $n=2$, we use combinatorial arguments to conjecture general expressions for the $r=1$ finitized Kac characters. These involve polynomial generalizations of Motzkin and Riordan numbers~\cite{Motzkin,Riordan,CatMR} defined in terms of $q$-trinomial coefficients. 

In Section~\ref{numericalSection}, we confirm numerically that the logarithmic superconformal models arising from the $2\times 2$ fusion of the elementary ${\cal LM}(p,p')$ lattice models indeed are to be identified with the ${\cal LM}(P,P';2)$ CFT coset models where $P=|2p-p'|$ and $P'=p'$. The relation between the parameters $p,p'$ of the lattice model and the parameters $P,P'$ of the CFT coset is not predicted by theory and is far from obvious. Specifically, employing this identification and using estimates based on finite-size corrections, we confirm the predicted values of the central charges $c$ and the first few conformal dimensions $\Delta_{1,s;\ell}^{P,P';2}$ in the $r=1$ column of the Kac tables~(\ref{genLogConfWts}). 
The numerics also confirms that we have correctly constructed explicit boundary conditions conjugate to the Kac operators, at positions $(1,s,\ell)$, in the $r=1$ column of the infinitely extended Kac tables.

In Section~\ref{Sec:JordanRep}, using the Hamiltonian limit, we argue that there exist reducible yet indecomposable representations for which the Virasoro dilatation operator $L_0$ exhibits rank-$2$ Jordan cells confirming that these theories are indeed logarithmic. We also relate these results to the $N=1$ superconformal representation theory.
We conclude with some general comments in Section~\ref{Sec:Conclusion}.

\section{Logarithmic Minimal CFTs ${\cal LM}(P,P';n)$}
\label{Sec:LM}

\subsection{Coset construction and central charges of ${\cal LM}(P,P';n)$}

Algebraically, the logarithmic minimal models are constructed~\cite{PRcoset13} as cosets
\be
 {\cal LM}(P,P';n)\simeq \mbox{COSET}\Big(\frac{nP}{P'\!-\!P}-2,n\Big),\quad \mathrm{gcd}\Big[P,\frac{P'\!-\!P}{n}\Big]=1,\quad 
    1\leq P<P',\quad n,P,P'\in\mathbb{N}
\label{Clog}
\ee
where $n=1,2,\ldots$ is an integer fusion level and $\mathbb{N}$ denotes the set of positive integers. The diagonal GKO coset~\cite{GKO85,GKO86,ACT91} takes the form
\be
 \mbox{COSET}(k,n):\quad \frac{(A_1^{(1)})_k\oplus(A_1^{(1)})_n}{(A_1^{(1)})_{k+n}},
 \qquad k=\frac{\ph}{\ph'}-2,\qquad \mathrm{gcd}[\ph,\ph']=1,\qquad n,\ph,\ph'\in\mathbb{N}
\label{cosetAAA}
\ee
where $k$ is a fractional fusion level and the subscripts on the 
affine $su(2)$ current algebra $A_1^{(1)}$ denote the respective levels $k$, $n$ and $k+n$.
The central charge of the coset Virasoro algebra is thus given by
\be
 c=c_k+c_n-c_{k+n}
  =\frac{3kn(k+n+4)}{(k+2)(n+2)(k+n+2)},\qquad c_k=\frac{3k}{k+2}
\label{ck}
\ee
where $c_k$ is the central charge of the affine current algebra $(A_1^{(1)})_k$. 
The central charges of the logarithmic minimal models ${\cal LM}(P,P';n)$ are thus
\bea
c^{P,P';n}=\frac{3n}{n+2} \Big(1-\frac{2(n+2)(P'-P)^2}{n^2 P P'}\Big),\qquad \mbox{gcd}\Big[P,\frac{P'\!-\!P}{n}\Big]=1,\qquad 0<P<P'
\label{centralcharges}
\eea
The usual logarithmic minimal models~\cite{PRZ} are given by $n=1$. The logarithmic superconformal minimal models are given by the specialization $n=2$ 
with central charges
\bea
c^{P,P';2}=\frac{3}{2} \Big(1-\frac{2(P'-P)^2}{P P'}\Big),\qquad \mbox{gcd}\Big[P,\frac{P'\!-\!P}{2}\Big]=1,\qquad 0<P<P'
\label{supercentralcharges}
\eea

\subsection{Branching functions and logarithmic minimal Kac characters}

The Kac characters of the logarithmic minimal models ${\cal LM}(P,P';n)$ are given by the branching functions~\cite{ACT91,BMSW9702,PRcoset13} 
of the logarithmic coset (\ref{Clog}).
These are expressible in terms of the
string functions~\cite{KP84,JM84,GQ87,DQ90,HNY90} of $\mathbb{Z}_n$ 
parafermions with central charge $c=\frac{2n-2}{n+2}$. For the fundamental domain
\be
 0\leq m\leq\ell\leq n,\qquad \ell-m\in2\mathbb{Z},\qquad m,\ell=0,1,\ldots,n,\qquad 
n\in {\mathbb N}
\label{stringdomain}
\ee 
the string functions are given by
\bea
 c_m^\ell(q)&=&\frac{q^{-\frac{1}{24}\frac{2n-2}{n+2}+\frac{\ell(\ell+2)}{4(n+2)}-\frac{m^2}{4n}}}{(q)_\infty^3}
  \sum_{i,j=0}^\infty(-1)^{i+j}q^{ij(n+1)+\frac{1}{2}i(i+1)+\frac{1}{2}j(j+1)}\nn
&&\qquad\quad\mbox{}\times\big[q^{\frac{i}{2}(\ell+m)+\frac{j}{2}(\ell-m)}-q^{n-\ell+1+\frac{i}{2}(2n+2-\ell-m)+\frac{j}{2}(2n+2-\ell+m)}\big]\label{string}
\eea
where the dependence on $n$ has been suppressed and we use the $q$-factorials
\bea
(q)_n=\prod_{k=1}^n (1-q^k),\qquad (q)_\infty=\prod_{k=1}^\infty (1-q^k)\label{qfactorials}
\eea
The notation for the string functions should not be confused with the notation for the central charges.
The fundamental domain of definition (\ref{stringdomain}) of the string functions
is extended to the domain
\be
 \ell=0,1,\ldots,n,
 \qquad m\in\mathbb{Z},\qquad n\in\mathbb N
\label{elldomain}
\ee
by setting $c_m^\ell(q)=0$ for $\ell-m\notin 2\mathbb{Z}$ and using the symmetries
\bea
 c_m^\ell(q)=c_{-m}^\ell(q)=c_{n-m}^{n-\ell}(q)=c_{m+2n}^\ell(q)\label{stringsym}
\eea
so that $c_m^\ell(q)$ is even and periodic in $m$ with period $2n$. 

Explicitly, the Kac characters of the logarithmic minimal models are given~\cite{PRcoset13} by the branching functions of the GKO coset (\ref{Clog}) 
\begin{align}
\mbox{}\hspace{-6pt}\chit_{r,s;\ell}^{\,P,P'\!;n}(q)
 &\!=\!q^{\Delta_{r,s}^{P,P'\!;n}\!-\!\frac{c^{P,P'\!;n}}{24}+\frac{n-1}{12(n+2)}}
  \big[c_{r-s}^{\ell}(q)-q^{\frac{rs}{n}}c_{r+s}^{\ell}(q)\big],\ \ 
   r\!+\!s\equiveq \ell\ \mathrm{mod}\ 2,\ \; r,s\in\mathbb{N},\ \; 0\leq\ell\leq n
\end{align}
where the conformal weights $\Delta_{r,s}^{P,P'\!;n}$ are defined in the next section. These branching functions satisfy the logarithmic branching rules (see \cite{PRcoset13} for notations)
\be
 \chih_{r,s}^{\,\ph,\ph'}(q,z)\,\chh_{\rho,0}^{n+2,1}(q,z)
  =\!\!\sum_{\mbox{\scriptsize$\sigma\in\mathbb{N}$}\atop \mbox{\scriptsize $\sigma\equiveq r\!+\!\ell$ mod 2}}
    \!\!\!\!\!\chit_{r,\sigma;\ell}^{\,\ph,\ph+n\ph'\!;n}(q)\,\chih_{\sigma,s}^{\,\ph+n\ph'\!,\ph'}\!(q,z),\qquad 
\ell=\begin{cases}
n\!+\!1\!-\!\rho,&\mbox{$s$ odd}\\[-3pt]
\rho\!-\!1,&\mbox{$s$ even}
\end{cases}
\label{branchL2}
\ee

For $n=1$, we recover the quasi-rational Kac characters of the logarithmic minimal models
\bea
\chi^{P,P'\!;1}_{r,s;0}(q)=q^{-\frac{c^{P,P';1}}{24}+\Delta_{r,s}^{P,P'\!;1}}\,\frac{(1-q^{rs})}{(q)_\infty}
\eea
For $n=2$, the branching functions simplify to
\bea
\begin{array}{rl}
\hspace{-6pt}NS\!:\,\ell=0,2;\  r\pm s\;\mbox{even:}&\chi^{P,P'\!,2}_{r,s,\ell}(q)=\!
\begin{cases}
q^{-\frac{c^{P,P'\!;2}}{24}+\frac{1}{48}+\Delta_{r,s}^{P,P'\!;2}}\,(1-q^{\frac{rs}{2}})\,c_{m_-}^\ell,&\!\!m_-=m_+
\\[2pt]
q^{-\frac{c^{P,P'\!;2}}{24}+\frac{1}{48}+\Delta_{r,s}^{P,P'\!;2}}[c_{m_-}^\ell(q)\!-\!q^{\frac{rs}{2}}c_{m_+}^\ell(q)],&\!\!m_+=2\!-\!m_-
\end{cases}\\[18pt]
\hspace{-6pt}R\!:\;\ \ \ell=1;\ \; r\pm s\;\mbox{odd:}&\chi^{P,P'\!;2}_{r,s,1}(q)
=q^{-\frac{c^{P,P'\!;2}}{24}+\frac{1}{48}+\Delta_{r,s}^{P,P'\!,2}}\,(1-q^{\frac{rs}{2}})\,c_1^1(q)
\end{array}
\label{superKac}
\eea
where $m_-=0,2=r-s\mbox{\;mod\;$4$}$ and $m_+=0,2=r+s\mbox{\;mod\;$4$}$. 
In this case, there are three independent string functions which are related to the three irreducible Virasoro characters $\ch_\D(q)$, $\D\in\{0,\frac{1}{16},\frac{1}{2}\}$, of the rational Ising model with central charge $c=c^{3,4;1}=\frac{1}{2}$ by
\begin{align}
\begin{array}{rl}
&c^0_0(q)\disp= c^2_2(q)=\frac{\ch_0(q)}{(q)_\infty}=\frac{q^{-\frac{1}{48}}}{2(q)_\infty}\Big[ \prod_{k=1}^\infty (1+q^{k-\frac{1}{2}}) + \prod_{k=1}^\infty (1-q^{k-\frac{1}{2}}) \Big] = \frac{q^{-\frac{1}{48}}}{(q)_\infty}\sum^\infty_{j=0\atop j\text{\tiny\;even}} \frac{q^{\frac{j^2}{2}}}{(q)_j}
\\[12pt]
&c^2_0(q)\disp= c^0_2(q)=\frac{\ch_{\frac{1}{2}}(q)}{(q)_\infty}=\frac{q^{-\frac{1}{48}}}{2(q)_\infty}\Big[ \prod_{k=1}^\infty (1+q^{k-\frac{1}{2}}) - \prod_{k=1}^\infty (1-q^{k-\frac{1}{2}}) \Big] = \frac{q^{-\frac{1}{48}}}{(q)_\infty}\sum^\infty_{j=1\atop j\text{\tiny\;odd}} \frac{q^{\frac{j^2}{2}}}{(q)_j}
\\[12pt]
&c^1_1(q)\disp=\frac{\ch_{\frac{1}{16}}(q)}{(q)_\infty}= \frac{q^{\frac{1}{24}}}{\:(q)_\infty}\prod_{k=1}^\infty (1+q^k) 
=\frac{q^{\frac{1}{24}}}{(q)_\infty}\sum_{j=0\atop j \text{ even}}^\infty \frac{q^{\frac{j^2-j}{2}}}{(q)_{j}}=
\frac{q^{\frac{1}{24}}}{(q)_\infty}\sum_{j=1\atop j \text{ odd}}^\infty \frac{q^{\frac{j^2-j}{2}}}{(q)_{j}}
\end{array}
\label{Isingfermionic}
\end{align}
For later use, we have recalled the fermionic forms (positive coefficient $q$-series) of these string functions.

It is to be stressed that, in this paper, we work throughout with the Virasoro algebra as the chiral conformal algebra and not the superconformal or ${\cal W}$-extended chiral algebra. In the superconformal picture, the sectors $\ell$ and $n-\ell$ are combined into symmetric and anti-symmetric super-characters. For $n=2$ in the $NS$ sector, this gives the superconformal characters
\bea
\hspace{-6pt}NS\!:&\chi^{P,P'\!;2}_{r,s,0}(q)\pm \chi^{P,P'\!;2}_{r,s;2}(q)=\!
\begin{cases}
q^{-\frac{c^{P,P'\!;2}}{24}+\frac{1}{48}+\Delta_{r,s}^{P,P'\!;2}}(\pm1)^{\frac{m_-}{2}}(1-q^{\frac{rs}{2}})[c_0^0(q)\pm c_2^0(q)],&\!\!m_-=m_+\\[2pt]
q^{-\frac{c^{P,P'\!;2}}{24}+\frac{1}{48}+\Delta_{r,s}^{P,P'\!;2}}(\pm1)^{\frac{m_-}{2}}(1\mp q^{\frac{rs}{2}})[c_0^0(q)\pm c_2^0(q)],&\!\!m_+=2-m_-
\end{cases}
\label{superCharspm}
\eea
where the combinations of string functions are related through the irreducible Virasoro Ising characters to simple infinite products
\bea
c_0^0(q)\pm c_2^0(q)=\frac{\ch_0(q)\pm \ch_{\frac{1}{2}}(q)}{(q)_\infty}
=\frac{q^{-\frac{1}{48}}}{(q)_\infty}\prod_{n=1}^\infty (1\pm q^{n-\frac{1}{2}}),
\eea
We call our models {\em  superconformal minimal models} because the underlying lattice models are the same with or without an extended symmetry. The only difference is that, with a superconformal or ${\cal W}$-extended symmetry, different boundary conditions must be constructed on the lattice to respect the enlarged symmetry~\cite{RichardP,PRR,RP2008,Ras2009}.

\subsection{Conformal weights of ${\cal LM}(P,P';n)$}

\begin{table}[tb]
{\vspace{0in}\psset{unit=.91cm}
{\small
\begin{center}
\begin{pspicture}(0,0)(7,11)
\psframe[linewidth=0pt,fillstyle=solid,fillcolor=lightlightblue](0,0)(7,11)
\multirput(0,0)(0,2){6}{\multirput(0,0)(2,0){4}{\psframe[linewidth=0pt,fillstyle=solid,fillcolor=lightestblue](0,0)(1,1)}}
\multirput(0,0)(0,2){5}{\multirput(1,1)(2,0){3}{\psframe[linewidth=0pt,fillstyle=solid,fillcolor=lightestblue](0,0)(1,1)}}
\psgrid[gridlabels=0pt,subgriddiv=1]
\rput(.5,10.65){$\vdots$}\rput(1.5,10.65){$\vdots$}\rput(2.5,10.65){$\vdots$}\rput(3.5,10.65){$\vdots$}\rput(4.5,10.65){$\vdots$}\rput(5.5,10.65){$\vdots$}\rput(6.5,10.5){$\vvdots$}
\rput(.5,9.5){$\frac{31}{16}$}\rput(1.5,9.5){$\frac 12,1$}\rput(2.5,9.5){$-\frac 1{16}$}\rput(3.5,9.5){$\frac 12,0$}\rput(4.5,9.5){$\frac{15}{16}$}\rput(5.5,9.5){$\frac 52,3$}\rput(6.5,9.5){$\cdots$}
\rput(.5,8.5){$\frac{4}3,\!\frac{11}6$}\rput(1.5,8.5){$\frac{13}{48}$}\rput(2.5,8.5){$\frac{1}3,\!-\!\frac 16$}\rput(3.5,8.5){$\frac{13}{48}$}\rput(4.5,8.5){$\frac 43,\!\frac{11}{6}$}\rput(5.5,8.5){$\frac {157}{48}$}\rput(6.5,8.5){$\cdots$}
\rput(.5,7.5){$\frac{15}{16}$}\rput(1.5,7.5){$\frac 12,0$}\rput(2.5,7.5){$-\frac 1{16}$}\rput(3.5,7.5){$\frac 12,1$}\rput(4.5,7.5){$\frac{31}{16}$}\rput(5.5,7.5){$\frac 92,4$}\rput(6.5,7.5){$\cdots$}
\rput(.5,6.5){$1,\frac 12$}\rput(1.5,6.5){$-\frac 1{16}$}\rput(2.5,6.5){$0,\frac 12$}\rput(3.5,6.5){$\frac{15}{16}$}\rput(4.5,6.5){$3,\frac 52$}\rput(5.5,6.5){$\frac{79}{16}$}\rput(6.5,6.5){$\cdots$}
\rput(.5,5.5){$\frac{13}{48}$}\rput(1.5,5.5){$-\!\frac 16,\!\frac 13$}\rput(2.5,5.5){$\frac {13}{48}$}\rput(3.5,5.5){$\frac{11}{6},\!\frac 43$}\rput(4.5,5.5){$\frac{157}{48}$}\rput(5.5,5.5){$\frac{35}{6},\!\frac{19}3$}\rput(6.5,5.5){$\cdots$}
\rput(.5,4.5){$0,\frac 12$}\rput(1.5,4.5){$-\frac 1{16}$}\rput(2.5,4.5){$1,\frac 12$}\rput(3.5,4.5){$\frac{31}{16}$}\rput(4.5,4.5){$4,\frac 92$}\rput(5.5,4.5){$\frac{111}{16}$}\rput(6.5,4.5){$\cdots$}
\rput(.5,3.5){$-\frac{1}{16}$}\rput(1.5,3.5){$\frac 12,0$}\rput(2.5,3.5){$\frac{15}{16}$}\rput(3.5,3.5){$\frac{5}2,3$}\rput(4.5,3.5){$\frac{79}{16}$}\rput(5.5,3.5){$\frac{17}2,8$}\rput(6.5,3.5){$\cdots$}
\rput(.5,2.5){$\frac 13,\!-\!\frac 16$}\rput(1.5,2.5){$\frac{13}{48}$}\rput(2.5,2.5){$\frac 43,\!\frac{11}6$}\rput(3.5,2.5){$\frac{157}{48}$}\rput(4.5,2.5){$\frac{19}3,\!\frac{35}6$}\rput(5.5,2.5){$\frac{445}{48}$}\rput(6.5,2.5){$\cdots$}
\rput(.5,1.5){$-\frac{1}{16}$}\rput(1.5,1.5){$\frac 12,1$}\rput(2.5,1.5){$\frac{31}{16}$}\rput(3.5,1.5){$\frac{9}2,4$}\rput(4.5,1.5){$\frac{111}{16}$}\rput(5.5,1.5){$\frac{21}2,\!11$}\rput(6.5,1.5){$\cdots$}
\rput(.5,.5){$0,\frac 32$}\rput(1.5,.5){$\frac{15}{16}$}\rput(2.5,.5){$3,\frac 52$}\rput(3.5,.5){$\frac{79}{16}$}\rput(4.5,.5){$8,\frac{17}2$}\rput(5.5,.5){$\frac{191}{16}$}\rput(6.5,.5){$\cdots$}
{\color{blue}
\rput(.5,-.5){$1$}
\rput(1.5,-.5){$2$}
\rput(2.5,-.5){$3$}
\rput(3.5,-.5){$4$}
\rput(4.5,-.5){$5$}
\rput(5.5,-.5){$6$}
\rput(6.5,-.5){$r$}
\rput(-.5,.5){$1$}
\rput(-.5,1.5){$2$}
\rput(-.5,2.5){$3$}
\rput(-.5,3.5){$4$}
\rput(-.5,4.5){$5$}
\rput(-.5,5.5){$6$}
\rput(-.5,6.5){$7$}
\rput(-.5,7.5){$8$}
\rput(-.5,8.5){$9$}
\rput(-.5,9.5){$10$}
\rput(-.5,10.5){$s$}}
\end{pspicture}
\hspace{1in}
\begin{pspicture}(0,0)(7,11)
\psframe[linewidth=0pt,fillstyle=solid,fillcolor=lightlightblue](0,0)(7,11)
\multirput(0,0)(0,2){6}{\multirput(0,0)(2,0){4}{\psframe[linewidth=0pt,fillstyle=solid,fillcolor=lightestblue](0,0)(1,1)}}
\multirput(0,0)(0,2){5}{\multirput(1,1)(2,0){3}{\psframe[linewidth=0pt,fillstyle=solid,fillcolor=lightestblue](0,0)(1,1)}}
\psgrid[gridlabels=0pt,subgriddiv=1]
\rput(.5,10.65){$\vdots$}\rput(1.5,10.65){$\vdots$}\rput(2.5,10.65){$\vdots$}\rput(3.5,10.65){$\vdots$}\rput(4.5,10.65){$\vdots$}\rput(5.5,10.65){$\vdots$}\rput(6.5,10.5){$\vvdots$}
\rput(.5,9.5){$4$}\rput(1.5,9.5){$\frac{35}{16},\!\frac{43}{16}$}\rput(2.5,9.5){$1$}\rput(3.5,9.5){$\frac{11}{16},\!\frac{3}{16}$}\rput(4.5,9.5){$0$}\rput(5.5,9.5){$\frac{3}{16},\!\frac{11}{16}$}\rput(6.5,9.5){$\cdots$}
\rput(.5,8.5){$3,\frac{7}2$}\rput(1.5,8.5){$\frac{25}{16}$}\rput(2.5,8.5){$1,\frac 12$}\rput(3.5,8.5){$\frac{1}{16}$}\rput(4.5,8.5){$0,\frac 12$}\rput(5.5,8.5){$\frac{9}{16}$}\rput(6.5,8.5){$\cdots$}
\rput(.5,7.5){$\frac 94$}\rput(1.5,7.5){$\frac {23}{16},\!\frac{15}{16}$}\rput(2.5,7.5){$\frac 14$}\rput(3.5,7.5){$\!-\!\frac{1}{16}\!,\!\frac{7}{16}$}\rput(4.5,7.5){$\frac 14$}\rput(5.5,7.5){$\frac{23}{16},\!\frac{15}{16}$}\rput(6.5,7.5){$\cdots$}
\rput(.5,6.5){$2,\frac 32$}\rput(1.5,6.5){$\frac {9}{16}$}\rput(2.5,6.5){$0,\frac 12$}\rput(3.5,6.5){$\frac{1}{16}$}\rput(4.5,6.5){$1,\frac 12$}\rput(5.5,6.5){$\frac{25}{16}$}\rput(6.5,6.5){$\cdots$}
\rput(.5,5.5){$1$}\rput(1.5,5.5){$\frac {3}{16},\!\frac{11}{16}$}\rput(2.5,5.5){$0$}\rput(3.5,5.5){$\frac{11}{16},\!\frac{3}{16}$}\rput(4.5,5.5){$1$}\rput(5.5,5.5){$\frac{35}{16},\!\frac{43}{16}$}\rput(6.5,5.5){$\cdots$}
\rput(.5,4.5){$\frac 12,1$}\rput(1.5,4.5){$\frac{1}{16}$}\rput(2.5,4.5){$\frac 12,0$}\rput(3.5,4.5){$\frac{9}{16}$}\rput(4.5,4.5){$\frac 32,2$}\rput(5.5,4.5){$\frac{49}{16}$}\rput(6.5,4.5){$\cdots$}
\rput(.5,3.5){$\frac 14$}\rput(1.5,3.5){$\frac{7}{16}\!,\!\!-\!\frac{1}{16}$}\rput(2.5,3.5){$\frac 14$}\rput(3.5,3.5){$\frac{15}{16},\!\frac{23}{16}$}\rput(4.5,3.5){$\frac 94$}\rput(5.5,3.5){$\frac{71}{16},\!\frac{63}{16}$}\rput(6.5,3.5){$\cdots$}
\rput(.5,2.5){$\frac 12,0$}\rput(1.5,2.5){$\frac 1{16}$}\rput(2.5,2.5){$\frac 12,1$}\rput(3.5,2.5){$\frac{25}{16}$}\rput(4.5,2.5){$\frac{7}2,3$}\rput(5.5,2.5){$\frac{81}{16}$}\rput(6.5,2.5){$\cdots$}
\rput(.5,1.5){$0$}\rput(1.5,1.5){$\frac{3}{16},\!\frac{11}{16}$}\rput(2.5,1.5){$1$}\rput(3.5,1.5){$\frac{43}{16},\!\frac{35}{16}$}\rput(4.5,1.5){$4$}\rput(5.5,1.5){$\frac{99}{16},\!\!\frac{107}{16}$}\rput(6.5,1.5){$\cdots$}
\rput(.5,.5){$0,\!\frac{3}{2}$}\rput(1.5,.5){$\frac {9}{16}$}\rput(2.5,.5){$2,\!\frac{3}{2}$}\rput(3.5,.5){$\frac{49}{16}$}\rput(4.5,.5){$5,\!\frac{11}{2}$}\rput(5.5,.5){$\frac{121}{16}$}\rput(6.5,.5){$\cdots$}
{\color{blue}
\rput(.5,-.5){$1$}
\rput(1.5,-.5){$2$}
\rput(2.5,-.5){$3$}
\rput(3.5,-.5){$4$}
\rput(4.5,-.5){$5$}
\rput(5.5,-.5){$6$}
\rput(6.5,-.5){$r$}
\rput(-.5,.5){$1$}
\rput(-.5,1.5){$2$}
\rput(-.5,2.5){$3$}
\rput(-.5,3.5){$4$}
\rput(-.5,4.5){$5$}
\rput(-.5,5.5){$6$}
\rput(-.5,6.5){$7$}
\rput(-.5,7.5){$8$}
\rput(-.5,8.5){$9$}
\rput(-.5,9.5){$10$}
\rput(-.5,10.5){$s$}}
\end{pspicture}
\end{center}}}
\caption{The infinitely extended Kac tables of conformal weights (\ref{genLogConfWts}) for superconformal dense polymers ${\cal LSM}(2,3)\equiv{\cal LM}(1,3;2)$ with $c=-\frac{5}{2}$ and $\beta_2=0$ and superconformal percolation ${\cal LSM}(3,4)\equiv{\cal LM}(2,4;2)$ with $c=0$ and $\beta_2=1$. The Neveu-Schwarz (NS) sectors with $r+s$ even correspond to $\ell=0,2$ and are shown as $\Delta_{r,s;0}^{P,P'\!;2},\Delta_{r,s;2}^{P,P'\!;2}$. The conformal weights in these two sectors differ by half-odd integers reflecting the presence of supersymmetry. The Ramond (R) sectors with $r+s$ odd correspond to $\ell=1$. }
\end{table}

The conformal weights of the logarithmic minimal models ${\cal LM}(P,P';n)$ are given explicitly~\cite{PRcoset13} by
\be
 \D_{r,s;\ell}^{P,P'\!;n}= \D_{r,s}^{P,P'\!;n}+\D_{r-s}^{\ell;n}+\mbox{Max}[\half(\ell\!+\!2\!-\!r\!-\!s),0],\qquad
 r+s\equiveq \ell\ \mathrm{mod}\ 2,\qquad r,s=1,2,\ldots
 \label{genLogConfWts}
\ee
The first term on the right side is 
\be
 \D_{r,s}^{P,P'\!;n}=\frac{(rP'-sP)^2-(P'-P)^2}{4nP P'},\qquad r,s= 1,2,\ldots
\label{DrsMMn}
\ee
Setting $m'=m$ mod $2n$, the second term is the conformal weight of the string function $c_m^\ell(q)$ 
\be
\hspace{-6pt}\Delta_m^{\ell;n}=\mbox{Max}[\Delta(m',\ell,n),\Delta(2n\!-\!m',\ell,n),\Delta(n\!-\!m',n\!-\!\ell,n)],\qquad
\Delta(m,\ell,n)=\frac{\ell(\ell+2)}{4(n+2)}-\frac{m^2}{4n}
\label{stringConfWts}
\ee
folded into the fundamental domain (\ref{stringdomain}).
The third term  
only gives a nonzero contribution for $r+s\le\ell\le n$. 
These conformal weights are thus organized into $n+1$ layered and infinitely extended Kac tables each displaying the 
checkerboard pattern $ r+s\equiveq \ell\ \mathrm{mod}\ 2$. In accord with the fact that these theories are nonunitary, the minimal conformal weight is
\be
\Delta^{P,P';n}_{\mathrm{min}}=\begin{cases}\Delta^{P,P';n}_{nP,nP';0},&\mbox{$P'-P$ even}\\[8pt]
\Delta^{P,P';n}_{nP,nP';n},&\mbox{$P'-P$ odd}
\end{cases}=\Delta^{P,P';n}_{nP,nP'}=-\frac{(P'-P)^2}{4n PP'}<0
\ee
It follows that the effective central charge is independent of $P,P'$ and given by the central charge of the affine current algebra $(A_1^{(1)})_n$
\be
 c_{\mathrm{eff}}^{\,P,P'\!;n}=c^{\,P,P'\!;n}-24\,\Delta^{P,P';n}_{\mathrm{min}}=c_n=\frac{3n}{n+2}
\ee
The logarithmic superconformal minimal models with $n=2$ therefore all have the effective central charge $c_{\text{eff}}=c_2=\frac{3}{2}$.

The infinitely extended Kac tables of conformal weights are shown in Table~1 for superconformal dense polymers ${\cal LM}(1,3;2)$ with $c=-\frac{5}{2}$ and $\beta_2=0$ and superconformal percolation ${\cal LM}(2,4;2)$ with $c=0$ and $\beta_2=1$.

\section{Fused Temperley-Lieb Algebra}
\label{Sec:TL}

The planar Temperley-Lieb (TL) algebra~\cite{JonesPlanar}, is a diagrammatic algebra generated by the two tiles or 2-tangles
\bea
\psset{unit=.9cm}
\begin{pspicture}[shift=-.45](1,1)
\facegrid{(0,0)}{(1,1)}
\rput[bl](0,0){\loopa}
\end{pspicture}\qquad\qquad
\begin{pspicture}[shift=-.42](1,1)
\facegrid{(0,0)}{(1,1)}
\rput[bl](0,0){\loopb}
\end{pspicture}
\eea
Within the planar algebra, these tiles are multiplied together (in arbitrary directions) by connecting the nodes at the midpoints of the edges of the faces by a planar web of connectivities. Fixing the direction for multiplication in the planar algebra leads to the loop representation of the correspondng linear TL algebra~\cite{TL}.

\subsection{Linear Temperley-Lieb algebra}
\label{Sec:LinTL}

The linear Temperley-Lieb (TL) algebra~\cite{TL}  ${\cal T\!L}(x;N)$ is a one-parameter algebra generated by the  identity $I$ and the monoids $e_j$, $j=1,\ldots,N$  subject to the relations
\bea
e_j^2&=&\TLb e_j \label{eq:TL1}\\
e_je_{j\pm 1}e_j&=&e_j \label{eq:TL2}\\
e_ie_j&=&e_je_i\qquad\quad |i-j|\ge 2\label{eq:TL3}
\eea
The parameter $x=e^{i\lambda}$ is a (complex) phase. For generic loop models $\lambda\in\mathbb{R}$ whereas, for the logarithmic minimal models ${\cal LM}(p,p')$, the crossing parameter $\lambda=\frac{(p'-p)\pi}{p'}$ is restricted to rational fractions of $\pi$. A faithful representation of the (linear) TL algebra is given by the loop representation with generators on a set of $N$ parallel strings
\be
I = \;
\psset{unit=.7cm}
\begin{pspicture}[shift=-.75](0,-.45)(3.5,1)
\rput(0,-.4){1}\rput(1,-.4){2}\rput(2.5,-.4){$N\!-\!1$}\rput(3.5,-.4){$N$}
\rput(1.75,.5){\ldots}
\multirput(0,0)(2.5,0){2}{\multirput(0,0)(1,0){2}{\psline(0,0)(0,1)}}
\end{pspicture}\: , \qquad\qquad
e_j=\;
\begin{pspicture}[shift=-.75](0,-.45)(8,1)
\rput(0,-.4){1}\rput(1,-.4){2}\rput(3.5,-.4){$j$}\rput(4.5,-.4){$j\!+\!1$}\rput(7,-.4){$N\!-\!1$}\rput(8,-.4){$N$}
\multirput(1.75,.5)(4.5,0){2}{\ldots}
\multirput(0,0)(7,0){2}{\multirput(0,0)(1,0){2}{\psline(0,0)(0,1)}}
\multirput(2.5,0)(3,0){2}{\psline(0,0)(0,1)}
\rput(3.5,0){\Monoid}
\end{pspicture}
\ee
that act by vertical concatenation. In diagrams, closed loops are removed and replaced with the scalar loop fugacity $\TLb$. For example, diagrammatically, relation \eqref{eq:TL1} becomes
\be
\psset{unit=.7cm}
\begin{pspicture}[shift=-1.25](0,-.45)(8,2)
\rput(0,-.4){1}\rput(1,-.4){2}\rput(3.5,-.4){$j$}\rput(4.5,-.4){$j\!+\!1$}\rput(7,-.4){$N\!-\!1$}\rput(8,-.4){$N$}
\multirput(1.75,1)(4.5,0){2}{\ldots}
\multirput(0,0)(0,1){2}{
\multirput(0,0)(7,0){2}{\multirput(0,0)(1,0){2}{\psline(0,0)(0,1)}}
\multirput(2.5,0)(3,0){2}{\psline(0,0)(0,1)}
\rput(3.5,0){\Monoid}
}
\end{pspicture}
\ =\  \TLb\ \;
\begin{pspicture}[shift=-.75](0,-.45)(8,1)
\rput(0,-.4){1}\rput(1,-.4){2}\rput(3.5,-.4){$j$}\rput(4.5,-.4){$j\!+\!1$}\rput(7,-.4){$N\!-\!1$}\rput(8,-.4){$N$}
\multirput(1.75,.5)(4.5,0){2}{\ldots}
\multirput(0,0)(7,0){2}{\multirput(0,0)(1,0){2}{\psline(0,0)(0,1)}}
\multirput(2.5,0)(3,0){2}{\psline(0,0)(0,1)}
\rput(3.5,0){\Monoid}
\end{pspicture}
\ee
The \textit{loop fugacity} $\beta$ is a scalar weight assigned to closed loops
\bea
\psset{unit=.75cm}
\TLb=\beta_1=\begin{pspicture}[shift=-.4](0,0)(1,1)
\pscircle(.5,.5){.5}
\end{pspicture}
= \begin{pspicture}[shift=-.4](0,0)(1,1)
\psarc[arrowsize=6pt,arrowlength=1]{->}(.5,.5){.5}{10}{380}
\end{pspicture}
\;+\begin{pspicture}[shift=-.4](0,0)(1,1)
\psarc[arrowsize=6pt,arrowlength=1]{<-}(.5,.5){.5}{-25}{375}
\end{pspicture}\;
=x+x^{-1}=2\cos\lambda,\qquad \beta_{n-1}=[x]_n=\frac{x^n-x^{-n}}{x-x^{-1}}=\frac{\sin n\lambda}{\sin\lambda}
\label{fugacities}
\eea
The TL algebra is associated with $(A^{(1)}_1)_1$ or $su(2)$ at level $n=1$, so the loop segments carry a spin-$\half$ charge.

The TL algebra ${\cal T\!L}(x;N)$ encompasses a braid-monoid algebra where the braids $b_j$ and inverse braids $b_j^{-1}$ are defined by
\be
b_j=i(x^{1/2}e_j-x^{-1/2} I),\qquad  b_j^{-1}=i(x^{1/2}I-x^{-1/2} e_j)  \label{eq:TLbbinv}
\ee
with diagrammatic representations
\be
b_j=\ 
\psset{unit=.7cm}
\begin{pspicture}[shift=-.75](0,-.45)(8,1)
\rput(0,-.4){1}\rput(1,-.4){2}\rput(3.5,-.4){$j$}\rput(4.5,-.4){$j\!+\!1$}\rput(7,-.4){$N\!-\!1$}\rput(8,-.4){$N$}
\multirput(1.75,.5)(4.5,0){2}{\ldots}
\multirput(0,0)(7,0){2}{\multirput(0,0)(1,0){2}{\psline(0,0)(0,1)}}
\multirput(2.5,0)(3,0){2}{\psline(0,0)(0,1)}
\rput(3.5,0){\Braid}
\end{pspicture}\qquad\ \ 
b_j^{-1}=\ 
\begin{pspicture}[shift=-.75](0,-.45)(8,1)
\rput(0,-.4){1}\rput(1,-.4){2}\rput(3.5,-.4){$j$}\rput(4.5,-.4){$j\!+\!1$}\rput(7,-.4){$N\!-\!1$}\rput(8,-.4){$N$}
\multirput(1.75,.5)(4.5,0){2}{\ldots}
\multirput(0,0)(7,0){2}{\multirput(0,0)(1,0){2}{\psline(0,0)(0,1)}}
\multirput(2.5,0)(3,0){2}{\psline(0,0)(0,1)}
\rput(3.5,0){\invBraid}
\end{pspicture}
\ee
Although braid operators are not strictly planar objects, in the planar algebra setting~\cite{JonesPlanar} they are viewed as additional rigid tiles (2-tangles) that are connected by planar webs of connectivities. 
Assuming the $e_j$ are real in a given matrix representation so that $\overline{e_j}=e_j$, we see the inverse braids are given by complex conjugation $b_j^{-1}=\overline{b_j}$. 
The braids satisfy the quadratic relation
\be
(b_j+i x^{-1/2} I)(b_j-i x^{3/2} I)=0 \label{eq:TLquad}
\ee
The additional relations for the TL braid-monoid algebra are 
\begin{align}
b_jb_j^{-1}&=I\nn
b_jb_{j+1}b_j&=b_{j+1}b_jb_{j+1}\nn
b_ib_j&=b_jb_i,\qquad\quad  |i-j|\ge 2\nn
b_ie_j&=e_jb_i, \qquad\quad|i-j|\ge 2\\
e_jb_{j\pm 1}b_j&=b_{j\pm 1}b_je_{j\pm 1}=e_je_{j\pm 1}\nn
b_je_j=e_jb_j&=\TLw e_j\nonumber
\end{align}
where the scalar phase $\TLw=ix^{3/2}$ is called the \emph{twist}, as it enters diagrammatically by undoing a twist
\be
\psset{unit=.6cm}
\begin{pspicture}[shift=-.5](0,0)(1,1.4)
\rput(0,0){\Braid}
\rput(0,1){\MonoidCap}
\end{pspicture}
\;=\; \TLw\;
\begin{pspicture}[shift=0](0,0)(1,1)
\rput(0,0){\MonoidCap}
\end{pspicture}
\ee
The twist relation follows using the definition of the braid operator \eqref{eq:TLbbinv} and the property that closed loops can be removed 
\be
b_je_j=(ix^{1/2}e_j-ix^{-1/2} I)e_j=[ix^{1/2}(x+x^{-1})-ix^{-1/2}]e_j=ix^{3/2}e_j=\TLw e_j
\ee
A similar (complex conjugate) relation holds for undoing a twist created by acting with the inverse braid on the monoid giving
\bea
b_j^{-1}e_j=e_jb_j^{-1}=\TLw^{-1} e_j, \qquad
\psset{unit=.6cm}
\begin{pspicture}[shift=-.5](0,0)(1,1.4)
\rput(0,0){\invBraid}
\rput(0,1){\MonoidCap}
\end{pspicture}
\;=\; \TLw^{-1}\;
\begin{pspicture}[shift=0](0,0)(1,1)
\rput(0,0){\MonoidCap}
\end{pspicture}
\eea

\subsection{Fused Temperley-Lieb algebra}
\label{Sec:FusTL}

\subsubsection{$2\times 2$ fused linear Temperley-Lieb algebra}

Assume that $\lambda\ne\frac{\pi}{2}$ so that $x=e^{i\lambda}\ne i$ and $\beta\ne 0$. To define the $2\times 2$ fused TL algebra ${\cal T\!L}(x;2,N)$, we use the first non-trivial Wenzl-Jones~\cite{Wenzl1987,Jones1985} projector
\be
p_j=I-\frac{1}{\TLb}\,e_j,\qquad  p_j^2=p_j,\qquad \beta\ne 0\label{eq:projTL}
\ee
This projector annihilates the monoid
\be
p_je_j=e_jp_j=0
\ee
It is represented diagrammatically by two fused or cabled strings as
\be
p_j\ = 
\psset{unit=.4cm}
\begin{pspicture}[shift=-.85](0,0)(2,2)
\psline[linewidth=1pt](.4,0)(.4,2)
\psline[linewidth=1pt](1.6,0)(1.6,2)
\psellipse[linewidth=1pt](1,1)(.6,.2)
\end{pspicture}=
\begin{pspicture}[shift=-.85](0,0)(2,2)
\psline[linewidth=1pt](.4,0)(.4,2)
\psline[linewidth=1pt](1.6,0)(1.6,2)
\end{pspicture}-{\TLb}^{-1}
\begin{pspicture}[shift=-.85](0,0)(2,2)
\psbezier[linewidth=1pt,showpoints=false](.4,0)(.5,1)(1.5,1)(1.6,0)
\psbezier[linewidth=1pt,showpoints=false](.4,2)(.5,1)(1.5,1)(1.6,2)
\end{pspicture}
\ee
and satisfies
\be
\psset{unit=.4cm}
p_j^2=\begin{pspicture}[shift=-.85](0,0)(2,2)
\psline[linewidth=1pt](.4,0)(.4,2)
\psline[linewidth=1pt](1.6,0)(1.6,2)
\multirput(0,0)(0,.8){2}{\psellipse[linewidth=1pt](1,.6)(.6,.2)}
\end{pspicture}=
\begin{pspicture}[shift=-.85](0,0)(2,2)
\psline[linewidth=1pt](.4,0)(.4,2)
\psline[linewidth=1pt](1.6,0)(1.6,2)
\psellipse[linewidth=1pt](1,1)(.6,.2)
\end{pspicture}=p_j, \qquad
\psset{unit=.7cm}
\begin{pspicture}[shift=-.4](-.5,-.5)(.5,.5)
\psellipse[linewidth=1pt](0,0)(.4,.1)
\rput(0,0){\psbezier(-.4,0)(-.4,.6)(.4,.6)(.4,0)}
\end{pspicture}
= \begin{pspicture}[shift=-.4](-.5,-.5)(.5,.5)
\psellipse[linewidth=1pt](0,0)(.4,.1)
\rput(0,0){\psbezier(-.4,0)(-.4,-.6)(.4,-.6)(.4,0)}
\end{pspicture}
=0=p_j e_j=e_j p_j
\label{eq:projProp}
\ee

In the $2\times 2$ fused TL algebra ${\cal T\!L}(x;2,N)$, the consecutive pairs of $2N$ strings are fused together to form $N$ compound strings each carrying spin-1 charge. The identity in the $2\times 2$ fused TL algebra is
\bea
I=\prod_{j=1}^N p_{2j-1}
\eea
The fused loop fugacity is
\be
\psset{unit=.4cm}
\begin{pspicture}[shift=-.5cm](0,0)(3,3)
\pscircle(1.5,1.5){1.5}
\pscircle(1.5,1.5){.75}
\psellipse[linewidth=1pt](.37,1.5)(.38,.2)
\psellipse[linewidth=1pt](2.6,1.5)(.38,.2)
\end{pspicture}=
\begin{pspicture}[shift=-.5cm](0,0)(3,3)
\pscircle(1.5,1.5){1.5}
\pscircle(1.5,1.5){.75}
\psellipse[linewidth=1pt](2.6,1.5)(.38,.2)
\end{pspicture} =
\begin{pspicture}[shift=-.5cm](0,0)(3,3)
\pscircle(1.5,1.5){1.5}
\pscircle(1.5,1.5){.75}
\end{pspicture} - \frac{1}{\TLb}\;
\begin{pspicture}[shift=-.5cm](0,0)(3,3)
\pscircle(1.5,1.5){1.5}
\pscircle(1.5,1.5){.75}
\pspolygon[fillstyle=solid,fillcolor=white,linecolor=white](2.15,1.2)(2.2,1.8)(3.1,1.8)(3.1,1.2)
\psbezier[showpoints=false](2.15,1.8)(2.25,1.5)(2.85,1.5)(2.93,1.85)
\psbezier[showpoints=false](2.14,1.15)(2.27,1.5)(2.87,1.5)(2.93,1.15)
\end{pspicture}
 =\TLb^2-\TLb^{-1}(\TLb)=\TLb^2-1=x^2+1+x^{-2} \label{eq:fTLbDeriv}
\ee
This spin-1 property reflects the fact that this algebra is associated with $su(2)$ at level $n=2$
\bea
\psset{unit=.4cm}
\begin{pspicture}[shift=-.5cm](0,0)(3,3)
\pscircle(1.5,1.5){1.5}
\pscircle(1.5,1.5){.75}
\psellipse[linewidth=1pt](.37,1.5)(.38,.2)
\psellipse[linewidth=1pt](2.6,1.5)(.38,.2)
\end{pspicture}\;=\;
\psset{unit=.85cm}
\begin{pspicture}[shift=-.4](0,0)(1,1)
\psarc[arrowsize=6pt,arrowlength=.8]{->>}(.5,.55){.5}{10}{380}
\end{pspicture}
\;+\;\begin{pspicture}[shift=-.4](0,0)(1,1)
\psarc(.5,.55){.5}{-18}{375}
\end{pspicture}
\;+\;\begin{pspicture}[shift=-.4](0,0)(1,1)
\psarc[arrowsize=6pt,arrowlength=.8]{<<-}(.5,.55){.5}{-18}{375}
\end{pspicture}\;
=x^2+1+x^{-2}=[x]_3=\frac{\sin3\lambda}{\sin\lambda}=\TLb_2
\eea
The fused $2\times 2$ braids and monoids are constructed by using the projector $p_j$ to fuse~\cite{KulishReshetikhinSklyanin1981} the elementary TL generators
\begin{align}
B_j&=p_{2j-1}p_{2j+1}b_{2j}b_{2j-1}b_{2j+1}b_{2j}p_{2j-1}p_{2j+1}\nn
B_j^{-1}&=p_{2j-1}p_{2j+1}b_{2j}^{-1}b_{2j-1}^{-1}b_{2j+1}^{-1}b_{2j}^{-1}p_{2j-1}p_{2j+1}\\
E_j&=p_{2j-1}p_{2j+1}e_{2j}e_{2j-1}e_{2j+1}e_{2j}p_{2j-1}p_{2j+1}\nonumber
\end{align}
Acting on 2 cabled strings, the operators are represented diagrammatically as
\be
I=
\psset{unit=.45cm} 
\begin{pspicture}[shift=-.85](-.4,0)(2.4,2)\fusedI\end{pspicture}\:, \quad
B_j=
\psset{unit=.35cm}
\begin{pspicture}[shift=-1.4](0,0)(3,3)
\psellipse[linewidth=1pt](.5,0)(.5,.175)
\psellipse[linewidth=1pt](2.5,0)(.5,.175)
\psellipse[linewidth=1pt](.5,3)(.5,.175)
\psellipse[linewidth=1pt](2.5,3)(.5,.175)
\multirput(1,0)(0,2){2}{\Braid}
\multirput(0,1)(2,0){2}{\Braid}
\multiput(0,0)(0,2){2}{\psline[linewidth=1pt](0,0)(0,1)}
\multiput(3,0)(0,2){2}{\psline[linewidth=1pt](0,0)(0,1)}
\end{pspicture}=
\psset{unit=.45cm}
\begin{pspicture}[shift=-.85](-.4,0)(2.4,2)\fusedB \end{pspicture}, \quad
B_j^{-1}=
\psset{unit=.35cm}
\begin{pspicture}[shift=-1.4](0,0)(3,3)
\psellipse[linewidth=1pt](.5,0)(.5,.175)
\psellipse[linewidth=1pt](2.5,0)(.5,.175)
\psellipse[linewidth=1pt](.5,3)(.5,.175)
\psellipse[linewidth=1pt](2.5,3)(.5,.175)
\multirput(1,0)(0,2){2}{\invBraid}
\multirput(0,1)(2,0){2}{\invBraid}
\multiput(0,0)(0,2){2}{\psline[linewidth=1pt](0,0)(0,1)}
\multiput(3,0)(0,2){2}{\psline[linewidth=1pt](0,0)(0,1)}
\end{pspicture}=
\psset{unit=.45cm}
\begin{pspicture}[shift=-.85](-.4,0)(2.4,2)\fusedBinv \end{pspicture},\quad
E_j=
\psset{unit=.35cm}
\begin{pspicture}[shift=-1.4](0,0)(3,3)
\psellipse[linewidth=1pt](.5,0)(.5,.175)
\psellipse[linewidth=1pt](2.5,0)(.5,.175)
\psellipse[linewidth=1pt](.5,3)(.5,.175)
\psellipse[linewidth=1pt](2.5,3)(.5,.175)
\multirput(0,1)(2,0){2}{\psbezier[linewidth=1pt](0,0)(.3,.5)(.7,.5)(1,0)
\psbezier[linewidth=1pt](0,1)(.3,.5)(.7,.5)(1,1)}
\multirput(0,0)(0,2){2}{\psbezier[linewidth=1pt](1,0)(1.3,.5)(1.7,.5)(2,0)
\psbezier[linewidth=1pt](1,1)(1.3,.5)(1.7,.5)(2,1)}
\multiput(0,0)(0,2){2}{\psline[linewidth=1pt](0,0)(0,1)}
\multiput(3,0)(0,2){2}{\psline[linewidth=1pt](0,0)(0,1)}
\end{pspicture}=
\psset{unit=.45cm}
\begin{pspicture}[shift=-.85](-.4,0)(2.4,2) \fusedE\end{pspicture}
\ee
Not surprisingly for spin-1, the fused braids satisfy a cubic relation 
\be
(B_j-x^4I)(B_j+x^2I)(B_j-x^{-2}I)=0
\ee
and the monoids are given as quadratics in the braids by the skein relation\\
\be
 (x^2-x^{-2})(E_j-I)=B_j-B_j^{-1}  
\label{eq:srf}
\ee

The two-parameter Birman-Wenzl-Murakami (BMW) algebra ${\cal BMW}(x,y;N)$~\cite{Mura,BW} is a more general braid-monoid algebra with cubic and skein relations
\be
(B_j+y I)(B_j-x^2yI)(B_j-y^{-1}I)=0,\qquad  (y-y^{-1})(E_j-I)=B_j-B_j^{-1}
\ee
and braid-monoid relations 
\begin{align}
B_jB_j^{-1}&=I\nn
B_jB_{j+1}B_j&=B_{j+1}B_jB_{j+1}\nn
B_iB_j&=B_jB_i\qquad\quad  |i-j|\ge 2\nn
E_j^2&=\BMWb E_j\nn
E_jE_{j\pm 1}E_j&=E_j\\
E_iE_j&=E_jE_i\qquad\quad |i-j|\ge 2\nn
E_iB_j&=B_jE_i\qquad\quad |i-j|\ge 2\nn
E_jB_{j\pm 1}B_j&=B_{j\pm 1}B_jE_{j\pm 1}=E_jE_{j\pm 1}\nn
B_jE_j=E_jB_j&=\BMWw E_j\nonumber
\end{align}
where
\be
\BMWb=1+\frac{x^2y+x^{-2}y^{-1}}{y-y^{-1}}, \qquad \BMWw = x^2y \label{eq:BMWparams}
\ee
The $2\times 2$ fused TL algebra ${\cal T\!L}(x;2,N)\equiv {\cal BMW}(x,x^2;N)$ is thus a one-parameter specialization of  the BMW algebra with $y=x^2$.

It is often convenient  to use an alternative presentation of the fused TL algebra ${\cal T\!L}(x;2,N)$ by defining~\cite{FendleyRead2002,Fendley2006,FendleyJacobsen2008,FendleyKruskal2010} a generalized monoid operator
\be
X_j=p_{2j-1}p_{2j+1}e_{2j}p_{2j-1}p_{2j+1}
\ee
represented diagrammatically as 
\bea
\psset{unit=.45cm}
X_j\;=\;\begin{pspicture}[shift=-.85](-.4,0)(2.4,2) \fusedX \end{pspicture}
\label {Xoperator}
\eea
In this presentation, as opposed to the algebraically equivalent braid-monoid presentation, the generators are all strictly planar objects. 
The braids and inverse braids are expressed in terms of the operators $I, E_j, X_j$ by
\be
B_j=x^2E_j-(x+x^{-1})X_j+x^{-2}I, \qquad
B_j^{-1}=x^{-2}E_j-(x+x^{-1})X_j+x^2I \label{eq:braidEIX}
\ee
These combine to give the form
\be
X_j=\frac{(x^2+x^{-2})}{2(x+x^{-1})}(I+E_j)-\frac{1}{2(x+x^{-1})}(B_j+B_j^{-1}) \label{eq:X_jdefn}
\ee
which is manifestly invariant under crossing symmetry (rotation through 90 degrees)
\be
E_j\leftrightarrow I, \quad B_j\leftrightarrow B_j^{-1}, \quad X_j\leftrightarrow X_j
\label{crossingsymm}
\ee
Eliminating the braids from the braid-monoid relations gives the relations~\cite{FendleyRead2002}
\begin{align}
E_j^2&=\fTLb E_j\nn
E_jE_{j\pm1}E_j&=E_j\nn
\TLb^2X_j^2&=\genb{3}X_j+E_j \nn
\TLb X_jE_j&=\TLb E_jX_j=\fTLb E_j \nn
E_iE_j&=E_jE_i\qquad\quad  |i-j|\ge 2\nn
X_iX_j&=X_jX_i\qquad\quad  |i-j|\ge 2\nn
X_iE_j&=E_jX_i\qquad\quad  |i-j|\ge 2\\
X_jE_{j\pm1}X_j&=X_{j\pm1}E_jX_{j\pm1}\nn
\TLb E_jX_{j\pm1}E_j&=\fTLb E_j\nn
X_jE_{j\pm1}E_j&=X_{j\pm1}E_j\nn
\TLb^2X_jX_{j\pm1}E_j&=\genb{3}X_{j\pm1}E_j+E_j\nn
\TLb^3(X_jX_{j+1}X_j-X_{j+1}X_jX_{j+1})
&=\TLb\big(E_jX_{j+1}-E_{j+1}X_j+X_{j+1}E_j\nn
&\quad\  -X_jE_{j+1}+X_j-X_{j+1}\big)-E_j+E_{j+1}\nonumber
\end{align}
where $\beta_3$ is defined in (\ref{fugacities}). All of these relations hold under time-reversal, that is, with the order of the products of $E_j$ and $X_j$ reversed on both sides. We note that replacing $X_j$ with $\tilde X_j=\beta X_j$ and using the identity $\beta^{-1}\beta_3=\beta_2-1$, gives a planar generalized monoid algebra with the fused loop fugacity $\beta_2$ as the only parameter.

\subsubsection{$n\times n$ fused Temperley-Lieb algebra}

The general Wenzl-Jones projectors\cite{Jones1985,Wenzl1985,Wenzl1987,Wenzl1988} $p^{(n)}_j$ are defined recursively by
\be
p^{(1)}_j=I, \quad p^{(2)}_j=p_j,\quad p^{(n+1)}_j=p^{(n)}_j-\frac{[x]_n}{[x]_{n+1}}\,p^{(n)}_j e_{n+j-1}\,p^{(n)}_j,
\qquad \beta_n=[x]_{n+1}\ne 0
\ee
These projectors act on $n$ strings with $1\leq n\leq N$ so, diagrammatically, the recursion is
\be
\psset{unit=.3cm}
\begin{pspicture}[shift=-2.75](0,-2)(4,2.5)
\multirput(0,0)(1,0){5}{\psline(0,0)(0,2.5)}
\rput(0,1.25){\proj[2]}
\rput(2,-.6){$\underbrace{\hspace{1.2cm}}$}
\rput(2,-1.5){\small$n+1$}
\end{pspicture}=
\begin{pspicture}[shift=-2.75](0,-2)(4,2.5)
\multirput(0,0)(1,0){5}{\psline(0,0)(0,2.5)}
\rput(0,1.25){\proj[1.5]}
\rput(1.5,-.6){$\underbrace{\hspace{.9cm}}$}
\rput(1.5,-1.5){\small$n$}
\end{pspicture}
-\frac{[x]_n}{[x]_{n+1}}\:
\begin{pspicture}[shift=-1](0,0)(4,2.5)
\multirput(0,0)(1,0){3}{\psline(0,0)(0,2.5)}
\multirput(0,0)(0,2.5){2}{\proj[1.5]}
\rput(3,0){\MonoidCap}\rput(3,2.5){\MonoidCup}
\rput(1,-.6){$\underbrace{\hspace{.5cm}}$}
\rput(1,-1.5){\small$n-1$}
\end{pspicture} \label{eq:WJprojrec}
\ee
which is an expression of the spin-\half fusion rule $n\otimes2=(n+1)\oplus (n-1)$. For general $n$, the generalized loop fugacities defined to be the closure of the Wenzl-Jones projector on $n$ strands, are given by the Chebyshev polynomials of the second kind
\be
\genb{n}:=\psset{unit=.3cm}
\begin{pspicture}[shift=-1.5](0,0)(4,4)
\pscircle(2.,2.){.5}\pscircle(2.,2.){1}\pscircle(2.,2.){1.5}\pscircle(2.,2.){2}\rput(0,2.){\proj[.75]}
\end{pspicture} 
= U_n\Big(\frac{\TLb}{2}\Big) = \frac{x^{n+1}-x^{-n-1}}{x-x^{-1}}=[x]_{n+1} \label{eq:genb}
\ee
This holds trivially for $n=1$ since $\genb{1}=\TLb=x+x^{-1}=[x]_2$. It holds for general $n$ by induction
\begin{align}
\genb{n+1}&=
\psset{unit=.3cm}
\begin{pspicture}[shift=-2](0,0)(5,5)
\pscircle(2.5,2.5){.5}\pscircle(2.5,2.5){1}\pscircle(2.5,2.5){1.5}\pscircle(2.5,2.5){2}\pscircle(2.5,2.5){2.5}
\rput(0,2.5){\proj[1]}
\end{pspicture}
=\psset{unit=.3cm}
\begin{pspicture}[shift=-2](0,0)(5,5)
\pscircle(2.5,2.5){.5}\pscircle(2.5,2.5){1}\pscircle(2.5,2.5){1.5}\pscircle(2.5,2.5){2}\pscircle(2.5,2.5){2.5}
\rput(0,2.5){\proj[.75]}
\end{pspicture}
-\frac{[x]_n}{[x]_{n+1}}\:
\psset{unit=.3cm}
\begin{pspicture}[shift=-2](0,.5)(6.5,6)
\multirput(0,2)(1,0){3}{\psline[linecolor=black](0,0)(0,2.5)}
\multirput(0,2)(0,2.5){2}{\proj[1.5]}
\rput(3,2){\MonoidCap}\rput(3,4.5){\MonoidCup}
\psbezier[showpoints=false,linecolor=black](4,2)(4.5,2.5)(4.5,4)(4,4.5)
\pscurve[showpoints=false,linecolor=black](3,2)(3.8,1.7)(4.5,2)(4.5,4.5)(3.8,4.8)(3,4.5)
\pscurve[showpoints=false,linecolor=black](2,2)(3.9,1.4)(4.8,1.8)(4.8,4.7)(3.9,5.1)(2,4.5)
\pscurve[showpoints=false,linecolor=black](1,2)(3,1.1)(5.3,1.8)(5.3,4.7)(3,5.4)(1,4.5)
\pscurve[showpoints=false,linecolor=black](0,2)(3.2,0.7)(5.8,1.8)(5.8,4.7)(3.2,5.8)(0,4.5)
\end{pspicture}
=\psset{unit=.3cm}
\begin{pspicture}[shift=-2](0,0)(5,5)
\pscircle(2.5,2.5){.5}\pscircle(2.5,2.5){1}\pscircle(2.5,2.5){1.5}\pscircle(2.5,2.5){2}\pscircle(2.5,2.5){2.5}
\rput(0,2.5){\proj[.75]}
\end{pspicture}
-\frac{[x]_n}{[x]_{n+1}}\:
\begin{pspicture}[shift=-1.5](0,0)(4,4)
\pscircle(2.,2.){.5}\pscircle(2.,2.){1}\pscircle(2.,2.){1.5}\pscircle(2.,2.){2}\rput(0,2.){\proj[.75]}
\end{pspicture}
\nn
&=\TLb\genb{n}-\frac{[x]_n}{[x]_{n+1}}\genb{n}
=[x]_2[x]_{n+1}-\frac{[x]_n}{[x]_{n+1}}[x]_{n+1}
=[x]_{n+2}
\end{align}
Similarly, the twists for general $n$ can be obtained recursively
\begin{align}
\psset{unit=.6cm}
\begin{pspicture}[shift=-.85](-.4,0)(2.4,3)
\rput(0,2){
\psbezier[linewidth=1pt,showpoints=false](-.4,0)(.5,1.25)(1.5,1.25)(2.4,0)
\psbezier[linewidth=1pt,showpoints=false](0,0)(.6,.875)(1.4,.875)(2,0)
\psbezier[linewidth=1pt,showpoints=false](.4,0)(.7,.5)(1.3,.5)(1.6,0)}
\psellipse[linewidth=1pt](0,0)(.4,.15)
\psellipse[linewidth=1pt](2,0)(.4,.15)
\psellipse[linewidth=1pt](0,2)(.4,.15)
\psellipse[linewidth=1pt](2,2)(.4,.15)
\psline[linewidth=1pt](-.4,2)(1.6,0)
\psline[linewidth=1pt](0,2)(2,0)
\psline[linewidth=1pt](.4,2)(2.4,0)
\pspolygon[linewidth=0pt,fillstyle=solid,fillcolor=white,linecolor=white](.35,1.2)(.85,1.6)(1.65,.8)(1.15,.4)
\psline[linewidth=1pt](-.4,0)(1.6,2)
\psline[linewidth=1pt](0,0)(2,2)
\psline[linewidth=1pt](.4,0)(2.4,2) 
\end{pspicture}
\ &=\ 
\psset{unit=.6cm}
\begin{pspicture}[shift=-.85](-.4,0)(2.4,3)
\rput(0,2){
\psbezier[linewidth=1pt,showpoints=false](-.4,0)(.5,1.25)(1.5,1.25)(2.4,0)
\psbezier[linewidth=1pt,showpoints=false](0,0)(.6,.875)(1.4,.875)(2,0)
\psbezier[linewidth=1pt,showpoints=false](.4,0)(.7,.5)(1.3,.5)(1.6,0)}
\psellipse[linewidth=1pt](0,0)(.4,.15)
\psellipse[linewidth=1pt](2,0)(.4,.15)
\psline[linewidth=1pt](-.4,2)(1.6,0)
\psline[linewidth=1pt](0,2)(2,0)
\psline[linewidth=1pt](.4,2)(2.4,0)
\pspolygon[linewidth=0pt,fillstyle=solid,fillcolor=white,linecolor=white](.35,1.2)(.85,1.6)(1.65,.8)(1.15,.4)
\psline[linewidth=1pt](-.4,0)(1.6,2)
\psline[linewidth=1pt](0,0)(2,2)
\psline[linewidth=1pt](.4,0)(2.4,2) 
\end{pspicture}
\ =\ 
\psset{unit=.6cm}
\begin{pspicture}[shift=-.85](-.4,0)(2.4,3)
\rput(0,2){
\psbezier[linewidth=1pt,showpoints=false](0,0)(.6,.875)(1.4,.875)(2,0)
\psbezier[linewidth=1pt,showpoints=false](.4,0)(.7,.5)(1.3,.5)(1.6,0)}
\psline[linewidth=1pt](-.4,.4)(1.6,0)
\psline[linewidth=1pt](0,2)(2,0)
\psline[linewidth=1pt](.4,2)(2.4,0)
\psline[linewidth=4pt,linecolor=white](0,0)(2,2)
\psline[linewidth=4pt,linecolor=white](-.4,0)(1.6,2)
\psline[linewidth=4pt,linecolor=white](.4,0)(.8,.4) 
\psline[linewidth=1pt](-.4,0)(1.6,2)
\psline[linewidth=1pt](0,0)(2,2)
\psline[linewidth=1pt](.4,0)(.8,.4) 
\psarc[linewidth=5pt,linecolor=white](.2,.4){.6}{0}{180}
\psarc[linewidth=1pt](.2,.4){.6}{0}{180}
\psellipse[linewidth=1pt](0,0)(.4,.15)
\psellipse[linewidth=1pt](2,0)(.4,.15)
\end{pspicture}
=(-x)
\begin{pspicture}[shift=-.85](-.4,0)(2.4,3)
\rput(0,2){
\psbezier[linewidth=1pt,showpoints=false](0,0)(.6,.875)(1.4,.875)(2,0)
\psbezier[linewidth=1pt,showpoints=false](.4,0)(.7,.5)(1.3,.5)(1.6,0)}
\psline[linewidth=1pt](.3,.4)(1.6,0)
\psline[linewidth=1pt](0,2)(2,0)
\psline[linewidth=1pt](.4,2)(2.4,0)
\psline[linewidth=4pt,linecolor=white](0,0)(2,2)
\psline[linewidth=4pt,linecolor=white](-.4,0)(1.6,2)
\psline[linewidth=4pt,linecolor=white](.4,0)(.8,.4) 
\psline[linewidth=1pt](-.4,0)(1.6,2)
\psline[linewidth=1pt](0,0)(2,2)
\psline[linewidth=1pt](.4,0)(.8,.4) 
\psarc[linewidth=3.5pt,linecolor=white](.525,.38){.26}{0}{180}
\psarc[linewidth=1pt](.525,.38){.26}{0}{180}
\psellipse[linewidth=1pt](0,0)(.4,.15)
\psellipse[linewidth=1pt](2,0)(.4,.15)
\end{pspicture}
=(-x)^{n-1}
\begin{pspicture}[shift=-.85](-.4,0)(2.4,3)
\rput(0,2){
\psbezier[linewidth=1pt,showpoints=false](0,0)(.6,.875)(1.4,.875)(2,0)
\psbezier[linewidth=1pt,showpoints=false](.4,0)(.7,.5)(1.3,.5)(1.6,0)}
\psline[linewidth=1pt](.7,.3)(1.6,0)
\psline[linewidth=1pt](0,2)(2,0)
\psline[linewidth=1pt](.4,2)(2.4,0)
\psline[linewidth=4pt,linecolor=white](0,0)(2,2)
\psline[linewidth=4pt,linecolor=white](-.4,0)(1.6,2)
\psline[linewidth=1pt](-.4,0)(1.6,2)
\psline[linewidth=1pt](0,0)(2,2)
\psline[linewidth=4pt,linecolor=white](.4,0)(1.3,.3)
\psline[linewidth=1pt](.4,0)(1.3,.3) 
\psarc[linewidth=1pt](1,.3){.3}{0}{180}
\psellipse[linewidth=1pt](0,0)(.4,.15)
\psellipse[linewidth=1pt](2,0)(.4,.15)
\end{pspicture}\nonumber\\
\ &=\ (-x)^{n-1}\omega
\psset{unit=.6cm}
\begin{pspicture}[shift=-.85](-.4,0)(2.4,3)
\rput(0,2){
\psbezier[linewidth=1pt,showpoints=false](0,0)(.6,.875)(1.4,.875)(2,0)
\psbezier[linewidth=1pt,showpoints=false](.4,0)(.7,.5)(1.3,.5)(1.6,0)}
%
\psline[linewidth=1pt](0,2)(2,0)
\psline[linewidth=1pt](.4,2)(2.4,0)
\psline[linewidth=4pt,linecolor=white](0,0)(2,2)
\psline[linewidth=4pt,linecolor=white](-.4,0)(1.6,2)
\psline[linewidth=1pt](-.4,0)(1.6,2)
\psline[linewidth=1pt](0,0)(2,2)
%
\psarc[linewidth=1pt](1,-.4){.73}{28}{152}
\psellipse[linewidth=1pt](0,0)(.4,.15)
\psellipse[linewidth=1pt](2,0)(.4,.15)
\end{pspicture}
\ =\ \omega_n\ 
\begin{pspicture}[shift=-.3](-.4,0)(2.4,1)
\psellipse[linewidth=1pt](0,0)(.4,.125)
\psellipse[linewidth=1pt](2,0)(.4,.125)
\rput(0,0){
\psbezier[linewidth=1pt,showpoints=false](-.4,0)(.5,1.25)(1.5,1.25)(2.4,0)
\psbezier[linewidth=1pt,showpoints=false](0,0)(.6,.875)(1.4,.875)(2,0)
\psbezier[linewidth=1pt,showpoints=false](.4,0)(.7,.5)(1.3,.5)(1.6,0)}
\end{pspicture}\ ,\qquad\ \  \genw{n}=i^{n^2} x^{n(n+2)/2}
\end{align}
by repeatedly using the relation 
\be
 b_j=(-x)b_j^{-1}+ix^{1/2}(x-x^{-1})I
\ee 
and the fact that the second terms do not contribute due to formation of a closed half-arc which is killed by the action of the projector.

In summary, in the general $n\times n$ fused TL algebra~\cite{TL,AkutsuWadati87_I,AkutsuDeguchiWadati87_II,DeguchiAkutsuWadati88_III,AkutsuDeguchiWadati88_IV,DeguchiWadatiAkutsu88_V}, the loop fugacity and twist are
\be
\genb{n}=\sum_{m=-n/2}^{n/2} x^{2m} =[x]_{n+1}= \frac{\sin(n+1)\lambda}{\sin\lambda}, \quad
\genw{n}=i^{n^2} x^{n(n+2)/2}
\ee
where the sum is over half-integers or integers. The braid $B_j$ satisfies a degree $n+1$ polynomial equation~\cite{DeguchiAkutsuWadati88_III,DeguchiWadatiAkutsu88_V}
\be
(B_j-\Genw{n}{1}I)(B_j-\Genw{n}{2}I)\ldots(B_j-\Genw{n}{n+1}I)=0,\qquad 
\Genw{n}{k} = (-1)^{k+1}x^{-k(k-1)} \genw{n}
\ee
and the monoid $E_j$ can be written as a degree $n$ polynomial in the braid operators
\be
E_j=\genb{n}\frac{(B_j-\Genw{n}{2}I)\ldots(B_j-\Genw{n}{n+1}I)}{(\genw{n}-\Genw{n}{2})(\genw{n}-\Genw{n}{3})\ldots(\genw{n}-\Genw{n}{n+1})}
\ee

\subsection{General $n\times n$ fused face operators}

\subsubsection{$m\times n$ fusion and push-through properties}

Let us introduce the following notations that we use throughout the paper
\begin{align}
\lambda&=\frac{(p'-p)\pi}{p'},\quad \mbox{gcd}[p,p']=1,\quad 1\le p<p',\quad p,p'\in \mathbb{N}\\
 &\quad s(u)=\frac{\sin u}{\sin\lambda},\quad u_n=u+n\lambda,\quad s_n(u)=s(u+n\lambda)
\label{notations}
\end{align}
The weights of elementary $1\times 1$ faces of the logarithmic minimal models ${\cal LM}(p,p')$~\cite{PRZ} are expressed in terms of the TL algebra as
\bea
\psset{unit=.9cm}
\begin{pspicture}[shift=-.42](1,1)
\facegrid{(0,0)}{(1,1)}
\rput(.5,.5){$u$}
\psarc[linewidth=1pt,linecolor=red](0,0){.15}{0}{90}
\end{pspicture}\ =\ 
s(\lambda-u)\;\begin{pspicture}[shift=-.45](1,1)
\facegrid{(0,0)}{(1,1)}
\rput[bl](0,0){\loopa}
\end{pspicture}\;+\,s(u)\;
\begin{pspicture}[shift=-.42](1,1)
\facegrid{(0,0)}{(1,1)}
\rput[bl](0,0){\loopb}
\end{pspicture}\;=\;s(\lambda-u) I+s(u) e_j \label{eq:face1}
\eea
where the small arc in the bottom-left corner indicates the orientation of the face operator.
The last equality applies if the planar operators act from the bottom-left to the top-right.
Fusion of $m\times n$ blocks of face operators
\be
\psset{unit=.875cm}
\begin{pspicture}[shift=-2.5](5,5)
\facegrid{(0,0)}{(5,5)}
\rput(4.5,.5){\small $u_0$}
\rput(4.5,1.5){\small $u_1$}
\rput(4.5,2.5){\small $\vdots$}
\rput(4.5,3.5){\small $\vdots$}
\rput[l](4.1,4.5){\small $u_{n\!-\!1}$}
\rput(3.5,.5){\small $u_{-1}$}
\rput(3.5,1.5){\small $u_0$}
\rput(3.5,2.5){\small $\vdots$}
\rput(3.5,3.5){\small $\vdots$}
\rput[r](3.9,4.5){\small $u_{n\!-\!2}$}
\rput(.5,.5){\small $u_{1\!-\!m}$}
\rput(.5,1.5){\small $u_{2\!-\!m}$}
\rput(.5,2.5){\small $\vdots$}
\rput(.5,3.5){\small $\vdots$}
\rput[r](.9,4.5){\small $u_{n\!-\!m}$}
\multiput(0,0)(1,0){2}{\rput(1.5,.5){\small $\cdots$}}
\multiput(0,1)(1,0){2}{\rput(1.5,.5){\small $\cdots$}}
\multiput(0,4)(1,0){2}{\rput(1.5,.5){\small $\cdots$}}
\psarc[linewidth=1pt,linecolor=red](0,0){.15}{0}{90}
\end{pspicture}
\ee
is implemented diagrammatically by applying a restriction or projection, implemented by hand, along each edge. If there is an internal closed arc beginning and ending anywhere along a given edge, then there must be a small internal closed half arc between neighbouring nodes somewhere along that edge. The fusion procedure acts to project out all faces with a closed internal arc along any of the edges.

If there is a closed half arc anywhere in a TL link state~\cite{PRZ}, then there must be a small closed half arc between neighbouring nodes somewhere in the link state. 
Such small external closed half-arcs have a push-through property. 
Specifically, if there is a small half-arc in an in-link state at the top acted upon by a seam, consisting of a $\rho\times 1$ block of fused faces, then by a simple trigonometric identity there must also be a small closed half-arc in the out-link state at the bottom of the seam
\psset{unit=.8cm}
\bea
&&\begin{pspicture}[shift=-.67](-.2,-.3)(2.2,1.8)
\psarc[linewidth=1.5pt,linecolor=blue](1,1){.5}{0}{180}
\psline[linewidth=1.5pt,linecolor=blue](-.3,.5)(2.3,.5)
\psline[linewidth=1.5pt,linecolor=blue](.5,-.3)(.5,.3)
\psline[linewidth=1.5pt,linecolor=blue](1.5,-.3)(1.5,.3)
\facegrid{(0,0)}{(2,1)}
\rput[B](.5,.4){\small$u\!\!-\!\!\lambda$}
\rput[B](1.5,.4){\small $u$}
\psarc[linewidth=1pt,linecolor=red](0,0){.15}{0}{90}
\psarc[linewidth=1pt,linecolor=red](1,0){.15}{0}{90}
\end{pspicture}\ \;=\;
s(2\lambda-u)s(u)\ \ 
\begin{pspicture}[shift=-.67](-.2,-.3)(2.2,1.8)
\psarc[linewidth=1.5pt,linecolor=blue](1,1){.5}{0}{180}
\psline[linewidth=1.5pt,linecolor=blue](-.3,.5)(2.3,.5)
\psline[linewidth=1.5pt,linecolor=blue](.5,-.3)(.5,.3)
\psline[linewidth=1.5pt,linecolor=blue](1.5,-.3)(1.5,.3)
\facegrid{(0,0)}{(2,1)}
\psarc[linewidth=1.5pt,linecolor=blue](1,0){.5}{0}{180}
\psarc[linewidth=1.5pt,linecolor=blue](0,1){.5}{-90}{0}
\psarc[linewidth=1.5pt,linecolor=blue](2,1){.5}{180}{270}
\end{pspicture}\ +
s(2\lambda-u)s(\lambda-u)\ \ 
\begin{pspicture}[shift=-.67](-.2,-.3)(2.2,1.8)
\psarc[linewidth=1.5pt,linecolor=blue](1,1){.5}{0}{180}
\psline[linewidth=1.5pt,linecolor=blue](-.3,.5)(2.3,.5)
\psline[linewidth=1.5pt,linecolor=blue](.5,-.3)(.5,.3)
\psline[linewidth=1.5pt,linecolor=blue](1.5,-.3)(1.5,.3)
\facegrid{(0,0)}{(2,1)}
\psarc[linewidth=1.5pt,linecolor=blue](1,0){.5}{90}{180}
\psarc[linewidth=1.5pt,linecolor=blue](2,0){.5}{90}{180}
\psarc[linewidth=1.5pt,linecolor=blue](0,1){.5}{270}{0}
\psarc[linewidth=1.5pt,linecolor=blue](1,1){.5}{270}{0}
\end{pspicture}\qquad\\
&\!\!+\!\!&
s(u-\lambda)s(\lambda-u)\ \ 
\begin{pspicture}[shift=-.67](-.2,-.3)(2.2,1.8)
\psarc[linewidth=1.5pt,linecolor=blue](1,1){.5}{0}{180}
\psline[linewidth=1.5pt,linecolor=blue](-.3,.5)(2.3,.5)
\psline[linewidth=1.5pt,linecolor=blue](.5,-.3)(.5,.3)
\psline[linewidth=1.5pt,linecolor=blue](1.5,-.3)(1.5,.3)
\facegrid{(0,0)}{(2,1)}
\psarc[linewidth=1.5pt,linecolor=blue](0,0){.5}{0}{90}
\psarc[linewidth=1.5pt,linecolor=blue](2,0){.5}{90}{180}
\psarc[linewidth=1.5pt,linecolor=blue](1,1){.5}{180}{360}
\end{pspicture}\ +
s(u-\lambda)s(u)\ \ 
\begin{pspicture}[shift=-.67](-.2,-.3)(2.2,1.8)
\psarc[linewidth=1.5pt,linecolor=blue](1,1){.5}{0}{180}
\psline[linewidth=1.5pt,linecolor=blue](-.3,.5)(2.3,.5)
\psline[linewidth=1.5pt,linecolor=blue](.5,-.3)(.5,.3)
\psline[linewidth=1.5pt,linecolor=blue](1.5,-.3)(1.5,.3)
\facegrid{(0,0)}{(2,1)}
\psarc[linewidth=1.5pt,linecolor=blue](0,0){.5}{0}{90}
\psarc[linewidth=1.5pt,linecolor=blue](1,1){.5}{180}{270}
\psarc[linewidth=1.5pt,linecolor=blue](1,0){.5}{0}{90}
\psarc[linewidth=1.5pt,linecolor=blue](2,1){.5}{180}{270}
\end{pspicture}\ =s(2\lambda-u)s(u)\ \ 
\begin{pspicture}[shift=-.67](-.2,-.3)(2.2,1.8)
\psarc[linewidth=1.5pt,linecolor=blue](1,1){.5}{0}{180}
\psline[linewidth=1.5pt,linecolor=blue](-.3,.5)(2.3,.5)
\psline[linewidth=1.5pt,linecolor=blue](.5,-.3)(.5,.3)
\psline[linewidth=1.5pt,linecolor=blue](1.5,-.3)(1.5,.3)
\facegrid{(0,0)}{(2,1)}
\psarc[linewidth=1.5pt,linecolor=blue](1,0){.5}{0}{180}
\psarc[linewidth=1.5pt,linecolor=blue](0,1){.5}{-90}{0}
\psarc[linewidth=1.5pt,linecolor=blue](2,1){.5}{180}{270}
\end{pspicture}\nonumber\smallskip
\label{pushthrough}
\eea
Note that the closed loop in the third diagram on the right-side contributes a scalar factor $\beta=s(2\lambda)$. This push-through property means that, if there are no internal closed half loops on the bottom edge, then there must be no closed half loops on the in-link state at the top. This means that the projector can be pushed through from the bottom to the top. If the Wenzl-Jones projector $p_j^{(m)}$ exists, then the action of the projection process described above agrees with the action of the Wenzl-Jones projector $p_j^{(m)}$. However, the projection process described above also makes sense when the Wenzl-Jones projector fails to exist. For $\lambda=\frac{(p'-p)\pi}{p'}$, we will always use the diagrammatic implementation of fusion, including for the construction of fused boundary operators in Section~\ref{BYBE}.

\subsubsection{$2\times 2$ fused face operators}

\psset{unit=.7cm}
\setlength{\unitlength}{.7cm}
For $2\times 2$ fusion, the fused face transfer operator $\face_j(u)$ is defined by
\bea
\psset{unit=.9cm}
\eta_{2,2}(u)\,\face_j(u)\;=\;\begin{pspicture}[shift=-1.27](-.4,-.4)(2.4,2.4)
\multiput(0,0)(0,1){2}{\psline[linewidth=1.5pt,linecolor=blue](-.2,.5)(2.2,.5)}
\psellipse[linewidth=1.5pt,linecolor=blue](-.2,1)(.1,.5)
\psellipse[linewidth=1.5pt,linecolor=blue](2.2,1)(.1,.5)
\multiput(0,0)(1,0){2}{\psline[linewidth=1.5pt,linecolor=blue](.5,-.2)(.5,2.2)}
\psellipse[linewidth=1.5pt,linecolor=blue](1,-.2)(.5,.1)
\psellipse[linewidth=1.5pt,linecolor=blue](1,2.2)(.5,.1)
\facegrid{(0,0)}{(2,2)}
\rput(.5,.5){\small $u\!-\!\lambda$}
\rput(1.5,.5){\small $u$}
\rput(.5,1.5){\small $u$}
\rput(1.5,1.5){\small $u\!+\!\lambda$}
\psarc[linewidth=1pt,linecolor=red](0,0){.15}{0}{90}
\end{pspicture}
\label{eq:fusedFacePic}
\eea
where the projectors are indicated by ovals and $\eta_{2,2}(u)$ is a suitable normalization factor which removes common factors. 
Internally, there are $2^4=16$ loop configurations, 9 of which are killed by the projectors because they have half arcs along the edges. The remaining 7 internal configurations are
\be
\begin{pspicture}(2,2)
\facegrid{(0,0)}{(2,2)}
\rput[bl](0,0){\loopa}
\rput[bl](1,0){\loopa}
\rput[bl](0,1){\loopa}
\rput[bl](1,1){\loopa}
\end{pspicture}\qquad
\begin{pspicture}(2,2)
\facegrid{(0,0)}{(2,2)}
\rput[bl](0,0){\loopa}
\rput[bl](1,0){\loopa}
\rput[bl](0,1){\loopa}
\rput[bl](1,1){\loopb}
\end{pspicture}\qquad
\begin{pspicture}(2,2)
\facegrid{(0,0)}{(2,2)}
\rput[bl](0,0){\loopb}
\rput[bl](1,0){\loopa}
\rput[bl](0,1){\loopa}
\rput[bl](1,1){\loopa}
\end{pspicture}\qquad
\begin{pspicture}(2,2)
\facegrid{(0,0)}{(2,2)}
\rput[bl](0,0){\loopb}
\rput[bl](1,0){\loopa}
\rput[bl](0,1){\loopa}
\rput[bl](1,1){\loopb}
\end{pspicture}\qquad
\begin{pspicture}(2,2)
\facegrid{(0,0)}{(2,2)}
\rput[bl](0,0){\loopb}
\rput[bl](1,0){\loopa}
\rput[bl](0,1){\loopb}
\rput[bl](1,1){\loopb}
\end{pspicture}\qquad
\begin{pspicture}(2,2)
\facegrid{(0,0)}{(2,2)}
\rput[bl](0,0){\loopb}
\rput[bl](1,0){\loopb}
\rput[bl](0,1){\loopa}
\rput[bl](1,1){\loopb}
\end{pspicture}\qquad
\begin{pspicture}(2,2)
\facegrid{(0,0)}{(2,2)}
\rput[bl](0,0){\loopb}
\rput[bl](1,0){\loopb}
\rput[bl](0,1){\loopb}
\rput[bl](1,1){\loopb}
\end{pspicture}
 \label{2by2configs}
\ee
There are three distinct connectivity classes (with respect to the connections between the 8 external nodes) given by the first, the last and the five intermediate planar operators. The first operator is the fused identity $I$ and the last is the fused monoid $E_j$. The other five planar operators combine to give the generalized monoid $X_j$ (\ref{Xoperator})
\bea
\psset{unit=.6cm}
I\;=\ 
\begin{pspicture}[shift=-.9](0,0)(2,2)
\psellipse[linewidth=1pt,linecolor=black](0,1)(.1,.2)
\psellipse[linewidth=1pt,linecolor=black](1,0)(.2,.1)
\psellipse[linewidth=1pt,linecolor=black](2,1)(.1,.2)
\psellipse[linewidth=1pt,linecolor=black](1,2)(.2,.1)
\psarc[linewidth=1pt,linecolor=black](2,0){.8}{90}{180}
\psarc[linewidth=1pt,linecolor=black](2,0){1.2}{90}{180}
\psarc[linewidth=1pt,linecolor=black](0,2){.8}{270}{0}
\psarc[linewidth=1pt,linecolor=black](0,2){1.2}{270}{0}
\end{pspicture}\ =\,
\begin{pspicture}[shift=-1.27](-.4,-.4)(2.4,2.4)
\multiput(0,0)(0,1){2}{\psline[linewidth=1.5pt,linecolor=blue](-.2,.5)(2.2,.5)}
\psellipse[linewidth=1.5pt,linecolor=blue](-.2,1)(.1,.5)
\psellipse[linewidth=1.5pt,linecolor=blue](2.2,1)(.1,.5)
\multiput(0,0)(1,0){2}{\psline[linewidth=1.5pt,linecolor=blue](.5,-.2)(.5,2.2)}
\psellipse[linewidth=1.5pt,linecolor=blue](1,-.2)(.5,.1)
\psellipse[linewidth=1.5pt,linecolor=blue](1,2.2)(.5,.1)
\facegrid{(0,0)}{(2,2)}
\rput[bl](0,0){\loopa}
\rput[bl](1,0){\loopa}
\rput[bl](0,1){\loopa}
\rput[bl](1,1){\loopa}
\end{pspicture}\qquad
E_j\;=\ 
\begin{pspicture}[shift=-.9](0,0)(2,2)
\psellipse[linewidth=1pt,linecolor=black](0,1)(.1,.2)
\psellipse[linewidth=1pt,linecolor=black](1,0)(.2,.1)
\psellipse[linewidth=1pt,linecolor=black](2,1)(.1,.2)
\psellipse[linewidth=1pt,linecolor=black](1,2)(.2,.1)
\psarc[linewidth=1pt,linecolor=black](0,0){.8}{0}{90}
\psarc[linewidth=1pt,linecolor=black](0,0){1.2}{0}{90}
\psarc[linewidth=1pt,linecolor=black](2,2){.8}{180}{270}
\psarc[linewidth=1pt,linecolor=black](2,2){1.2}{180}{270}
\end{pspicture}\ 
=\,\begin{pspicture}[shift=-1.27](-.4,-.4)(2.4,2.4)
\multiput(0,0)(0,1){2}{\psline[linewidth=1.5pt,linecolor=blue](-.2,.5)(2.2,.5)}
\psellipse[linewidth=1.5pt,linecolor=blue](-.2,1)(.1,.5)
\psellipse[linewidth=1.5pt,linecolor=blue](2.2,1)(.1,.5)
\multiput(0,0)(1,0){2}{\psline[linewidth=1.5pt,linecolor=blue](.5,-.2)(.5,2.2)}
\psellipse[linewidth=1.5pt,linecolor=blue](1,-.2)(.5,.1)
\psellipse[linewidth=1.5pt,linecolor=blue](1,2.2)(.5,.1)
\facegrid{(0,0)}{(2,2)}
\rput[bl](0,0){\loopb}
\rput[bl](1,0){\loopb}
\rput[bl](0,1){\loopb}
\rput[bl](1,1){\loopb}
\end{pspicture}\qquad
X_j\;=\ 
\begin{pspicture}[shift=-.9](0,0)(2,2)
\psellipse[linewidth=1pt,linecolor=black](0,1)(.1,.2)
\psellipse[linewidth=1pt,linecolor=black](1,0)(.2,.1)
\psellipse[linewidth=1pt,linecolor=black](2,1)(.1,.2)
\psellipse[linewidth=1pt,linecolor=black](1,2)(.2,.1)
\psarc[linewidth=1pt,linecolor=black](0,0){.8}{0}{90}
\psarc[linewidth=1pt,linecolor=black](2,0){.8}{90}{180}
\psarc[linewidth=1pt,linecolor=black](0,2){.8}{270}{0}
\psarc[linewidth=1pt,linecolor=black](2,2){.8}{180}{270}
\end{pspicture}\ 
=\,\begin{pspicture}[shift=-1.27](-.4,-.4)(2.4,2.4)
\multiput(0,0)(0,1){2}{\psline[linewidth=1.5pt,linecolor=blue](-.2,.5)(2.2,.5)}
\psellipse[linewidth=1.5pt,linecolor=blue](-.2,1)(.1,.5)
\psellipse[linewidth=1.5pt,linecolor=blue](2.2,1)(.1,.5)
\multiput(0,0)(1,0){2}{\psline[linewidth=1.5pt,linecolor=blue](.5,-.2)(.5,2.2)}
\psellipse[linewidth=1.5pt,linecolor=blue](1,-.2)(.5,.1)
\psellipse[linewidth=1.5pt,linecolor=blue](1,2.2)(.5,.1)
\facegrid{(0,0)}{(2,2)}
\psarc[linewidth=1.5pt,linecolor=blue](0,0){.5}{0}{90}
\psarc[linewidth=1.5pt,linecolor=blue](2,0){.5}{90}{180}
\psarc[linewidth=1.5pt,linecolor=blue](2,2){.5}{180}{270}
\psarc[linewidth=1.5pt,linecolor=blue](0,2){.5}{270}{360}
\end{pspicture}
\eea
Explicitly, using trigonometric identities to combine these 5 intermediate weights and removing the common factors $\eta_{2,2}(u)=-s(2\lambda)s(\lambda-u)s(u)$ gives the $2\times 2$ fused face transfer operator
\be
\face_j(u)=\frac{s(\lambda-u)s(2\lambda-u)}{s(2\lambda)}I+s(u)s(\lambda-u)X_j+\frac{s(u)s(u+\lambda)}{s(2\lambda)} E_j
\label{eq:fusedFace}
\ee
With this normalization, we have $\face_j(0)=I$, $\face_j(\lambda)=E_j$.
The face transfer operators of more general $m\times n$ fusions can be calculated similarly. The $n\times n$ fusions are considered explicitly in Section~\ref{nbyn} since these coincide with physical face operators of the logarithmic minimal models ${\cal LM}(P,P';n)$.

\subsubsection{$2\times2$ fused braids}
In this section, we discuss the fused braids to obtain an alternative, but algebraically equivalent, braid-monoid presentation of the generalized monoid form of the face transfer operators (\ref{eq:fusedFace}). 
In fact, the $2\times2$ fused face operator $\face_j(u)$ can be rewritten in terms of the fused braids $B_j, B_j^{-1}$ and the identity~$I$. We first derive expressions for $B_j, B_j^{-1}$ in terms of fused monoids. This is done diagrammatically by taking the diagrams for the braid and inverse braid
\bea
\psset{unit=.7cm}
B_j\;=\;\begin{pspicture}[shift=-1.27](-.4,-.4)(2.4,2.4)
\multiput(0,0)(0,1){2}{\psline[linewidth=1.5pt,linecolor=blue](-.2,.5)(2.2,.5)}
\psellipse[linewidth=1.5pt,linecolor=blue](-.2,1)(.1,.5)
\psellipse[linewidth=1.5pt,linecolor=blue](2.2,1)(.1,.5)
\multiput(0,0)(1,0){2}{\psline[linewidth=1.5pt,linecolor=blue](.5,-.2)(.5,2.2)}
\psellipse[linewidth=1.5pt,linecolor=blue](1,-.2)(.5,.1)
\psellipse[linewidth=1.5pt,linecolor=blue](1,2.2)(.5,.1)
\facegrid{(0,0)}{(2,2)}
\rput[bl](0,0){\braida}
\rput[bl](1,0){\braida}
\rput[bl](0,1){\braida}
\rput[bl](1,1){\braida}
\end{pspicture}\qquad\qquad
B_j^{-1}\;=\;\begin{pspicture}[shift=-1.27](-.4,-.4)(2.4,2.4)
\multiput(0,0)(0,1){2}{\psline[linewidth=1.5pt,linecolor=blue](-.2,.5)(2.2,.5)}
\psellipse[linewidth=1.5pt,linecolor=blue](-.2,1)(.1,.5)
\psellipse[linewidth=1.5pt,linecolor=blue](2.2,1)(.1,.5)
\multiput(0,0)(1,0){2}{\psline[linewidth=1.5pt,linecolor=blue](.5,-.2)(.5,2.2)}
\psellipse[linewidth=1.5pt,linecolor=blue](1,-.2)(.5,.1)
\psellipse[linewidth=1.5pt,linecolor=blue](1,2.2)(.5,.1)
\facegrid{(0,0)}{(2,2)}
\rput[bl](0,0){\braidb}
\rput[bl](1,0){\braidb}
\rput[bl](0,1){\braidb}
\rput[bl](1,1){\braidb}
\end{pspicture}
\eea
and expanding each elementary braid using its definition \eqref{eq:TLbbinv}
\be
b_j=i(x^{1/2}e_j-x^{-1/2} I),\qquad  b_j^{-1}=i(x^{1/2}I-x^{-1/2} e_j) \nn
\ee
After canceling the configurations annihilated by the projectors, the surviving configurations are precisely those in \eqref{2by2configs}. Combining the terms into connectivity classes gives
\be
B_j=x^2E_j-(x+x^{-1})X_j+x^{-2}I, \qquad
B_j^{-1}=x^{-2}E_j-(x+x^{-1})X_j+x^2I 
\ee
which agrees with \eqref{eq:braidEIX}. These expressions can be inverted to give
\be
X_j=\frac{x^{-2} B_j-x^2B_j^{-1}+(x^4-x^{-4})I}{(x^2-x^{-2})(x+x^{-1})}, \quad
E_j=I+\frac{1}{x^2-x^{-2}}(B_j-B_j^{-1})
\ee
where we recognize that the expression for the monoid is the skein relation \eqref{eq:srf}. Finally, using these relations, gives the $2\times2$ face operator in terms of $B_j, B_j^{-1}$ and $I$
\be
\face_j(z) = I + \frac{(z-z^{-1})(x^{-1}z B_j-xz^{-1}B_j^{-1})}{(x-x^{-1})(x^2-x^{-2})}
\ee
This is the one-parameter specialization ($\mu=2\lambda$, $y=x^2$) of the face transfer operator associated with the BMW algebra 
\be
\face_j(z)=I+\frac{(z-z^{-1})(x^{-1}zB_j-xz^{-1}B_j^{-1})}{(x-x^{-1})(y-y^{-1})}\label{eq:BMWface} 
\ee
where 
\be
x=\exp(i\lambda),\quad y=\exp(i\mu),\quad z=\exp(iu)
\ee
More usefully, this can be put into a form which manifestly respects the crossing symmetry (\ref{crossingsymm}) with $u\leftrightarrow \lambda-u$
\be
\face_j(u) =\half\Big[\frac{s(\lambda-u)s(2u)}{s(u)}\,I+\frac{s(\lambda-u)s(u)}{s(2\lambda)}(B_j+B_j^{-1})
+\frac{s(u)s(2\lambda-2u)}{s(\lambda-u)}\,E_j\Big]\label{symmFace}
\ee
Some typical lattice configurations for $2\times 2$ fused logarithmic minimal models are shown in Figure~\ref{configs}.

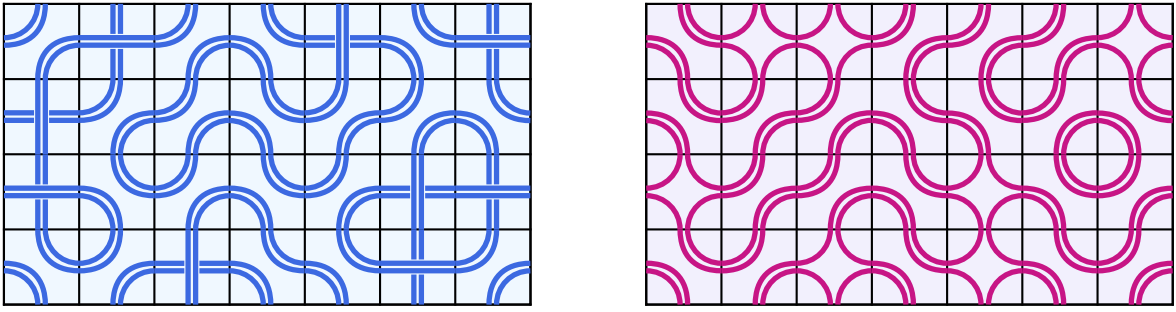
\begin{figure}[htbp]
\begin{center}
\psset{unit=1cm}
\begin{pspicture}[shift=-1.85](0,0)(7,4.5)
\pspolygon[fillstyle=solid,fillcolor=AliceBlue](0,0)(7,0)(7,4)(0,4)
\psgrid[gridlabels=0pt,subgriddiv=1](0,0)(7,4)
\rput(0,0){\TLfoFrb}\rput(1,0){\IfoFrb}\rput(2,0){\BinvfoFrb}\rput(3,0){\TLfoFrb}\rput(4,0){\TLfoFrb}\rput(5,0){\BfoFrb}\rput(6,0){\IfoFrb}
\rput(0,1){\BfoFrb}\rput(1,1){\TLfoFrb}\rput(2,1){\IfoFrb}\rput(3,1){\TLfoFrb}\rput(4,1){\IfoFrb}\rput(5,1){\BinvfoFrb}\rput(6,1){\BfoFrb}
\rput(0,2){\BinvfoFrb}\rput(1,2){\IfoFrb}\rput(2,2){\IfoFrb}\rput(3,2){\TLfoFrb}\rput(4,2){\IfoFrb}\rput(5,2){\IfoFrb}\rput(6,2){\TLfoFrb}
\rput(0,3){\IfoFrb}\rput(1,3){\BfoFrb}\rput(2,3){\IfoFrb}\rput(3,3){\TLfoFrb}\rput(4,3){\BinvfoFrb}\rput(5,3){\TLfoFrb}\rput(6,3){\BfoFrb}
\end{pspicture}\qquad\qquad
\psset{unit=1cm}
\begin{pspicture}[shift=-1.85](0,0)(7,4)
\pspolygon[fillstyle=solid,fillcolor=myboxcolor](0,0)(7,0)(7,4)(0,4)
\psgrid[gridlabels=0pt,subgriddiv=1](0,0)(7,4)
\rput(0,0){\TLfoFp}\rput(1,0){\IfoFp}\rput(2,0){\XfoFp}\rput(3,0){\TLfoFp}\rput(4,0){\XfoFp}\rput(5,0){\TLfoFp}\rput(6,0){\IfoFp}
\rput(0,1){\XfoFp}\rput(1,1){\IfoFp}\rput(2,1){\IfoFp}\rput(3,1){\TLfoFp}\rput(4,1){\XfoFp}\rput(5,1){\TLfoFp}\rput(6,1){\IfoFp}
\rput(0,2){\TLfoFp}\rput(1,2){\IfoFp}\rput(2,2){\IfoFp}\rput(3,2){\TLfoFp}\rput(4,2){\TLfoFp}\rput(5,2){\IfoFp}\rput(6,2){\TLfoFp}
\rput(0,3){\TLfoFp}\rput(1,3){\XfoFp}\rput(2,3){\XfoFp}\rput(3,3){\IfoFp}\rput(4,3){\IfoFp}\rput(5,3){\IfoFp}\rput(6,3){\XfoFp}
\end{pspicture}
\end{center}
\caption{Typical lattice configurations for $2\times 2$ fused models using the algebraically equivalent braid-monoid presentation (\ref{symmFace}) on the left (with tiles $I, B_j, B_j^{-1}, E_j$) and the generalized monoid presentation (\ref{eq:fusedFace}) on the right (with tiles $I, X_j, E_j$). The description on the left is well suited to superconformal dense polymers ${\cal LM}(1,3;2)$ with $\beta_2=0$. The polymers cannot form closed loops but are allowed to cross and to form knots. The planar description on the right is well suited to  superconformal percolation ${\cal LM}(2,4;2)$ with $c=0$ and $\beta_2=1$. The interconnecting webs of cabled connectivities can be localized or can percolate as an infinite connected cluster from one side to the opposite side of a large lattice.\label{configs}}
\end{figure}

\subsubsection{$n\times n$ fused face operators}
\label{nbyn}

The general $n\times n$ face transfer operator is constructed using fusion as
\be
\eta_{n,n}(u)\; \genface{n}_j(u)=
\psset{unit=.75cm}
\begin{pspicture}[shift=-2.75](-.4,-.4)(5.4,5.4)
\multiput(0,0)(0,1){5}{\psline[linewidth=1.5pt,linecolor=blue](-.2,.5)(5.2,.5)}
\psellipse[linewidth=1.5pt,linecolor=blue](-.2,2.5)(.1,2)
\psellipse[linewidth=1.5pt,linecolor=blue](5.2,2.5)(.1,2)
\multiput(0,0)(1,0){5}{\psline[linewidth=1.5pt,linecolor=blue](.5,-.2)(.5,5.2)}
\psellipse[linewidth=1.5pt,linecolor=blue](2.5,-.2)(2,.1)
\psellipse[linewidth=1.5pt,linecolor=blue](2.5,5.2)(2,.1)
\facegrid{(0,0)}{(5,5)}
\rput(4.5,.5){\small $u_0$}
\rput(4.5,1.5){\small $u_1$}
\rput(4.5,2.5){\small $\vdots$}
\rput(4.5,3.5){\small $\vdots$}
\rput[l](4.1,4.5){\small $u_{n\!-\!1}$}
\rput(3.5,.5){\small $u_{-1}$}
\rput(3.5,1.5){\small $u_0$}
\rput(3.5,2.5){\small $\vdots$}
\rput(3.5,3.5){\small $\vdots$}
\rput[r](3.9,4.5){\small $u_{n\!-\!2}$}
\rput(.5,.5){\small $u_{1\!-\!n}$}
\rput(.5,1.5){\small $u_{2\!-\!n}$}
\rput(.5,2.5){\small $\vdots$}
\rput(.5,3.5){\small $\vdots$}
\rput(.5,4.5){\small $u_{0}$}
\multiput(0,0)(1,0){2}{\rput(1.5,.5){\small $\cdots$}}
\multiput(0,1)(1,0){2}{\rput(1.5,.5){\small $\cdots$}}
\multiput(0,4)(1,0){2}{\rput(1.5,.5){\small $\cdots$}}
\psarc[linewidth=1pt,linecolor=red](0,0){.15}{0}{90}
\end{pspicture}
\ee
It is useful to write these in terms of generalized monoids $X_j^{(i)}$~\cite{ZinnJustin}. Explicitly, the face operators take the form
\be
\genface{n}_j(u) = \sum_{i=0}^n S_i(u) X_j^{(i)},\qquad
X_j^{(0)}=I, \quad X_j^{(n)}=E_j
\ee
where
\be
S_i(u)=\frac{(-1)^{i+n} \prod _{k=1}^n s\big(u+(i+k-n-1)\lambda\big)}{\prod _{l=1}^i s(l \lambda)\prod _{m=1}^{n-i} s(m \lambda)}
\ee
The normalization is
\be
\eta_{n,n}(u)=\prod _{k=1}^ns(k \lambda)\prod _{l=1}^{2n-2}s\big(u+(l-n)\lambda\big)^{\text{Min}[l,2 n-1-l]}
\ee
For example, for $n=4$, the intermediate generalized monoid operators $X_j^{(i)}$ are given by
\be
\psset{unit=.7cm}
X_j^{(1)}\!\!=\;\begin{pspicture}[shift=-.9](0,0)(2,2)
\psellipse[linewidth=1pt,linecolor=black](0,1)(.1,.3)
\psellipse[linewidth=1pt,linecolor=black](1,0)(.3,.1)
\psellipse[linewidth=1pt,linecolor=black](2,1)(.1,.3)
\psellipse[linewidth=1pt,linecolor=black](1,2)(.3,.1)
\psarc[linewidth=1pt,linecolor=black](2,0){.7}{90}{180}
\psarc[linewidth=1pt,linecolor=black](2,0){.9}{90}{180}
\psarc[linewidth=1pt,linecolor=black](2,0){1.1}{90}{180}
\psarc[linewidth=1pt,linecolor=black](0,2){.7}{270}{0}
\psarc[linewidth=1pt,linecolor=black](0,2){.9}{270}{0}
\psarc[linewidth=1pt,linecolor=black](0,2){1.1}{270}{0}
\psarc[linewidth=1pt,linecolor=black](0,0){.7}{0}{90}
\psarc[linewidth=1pt,linecolor=black](2,2){.7}{180}{270}
\end{pspicture}\;=
\psset{unit=.45cm}
\begin{pspicture}[shift=-2.25](-.4,-.4)(4.4,4.4)
\multiput(0,0)(0,1){4}{\psline[linewidth=1.5pt,linecolor=blue](-.2,.5)(4.2,.5)}
\psellipse[linewidth=1.5pt,linecolor=blue](-.2,2)(.1,1.5)
\psellipse[linewidth=1.5pt,linecolor=blue](4.2,2)(.1,1.5)
\multiput(0,0)(1,0){4}{\psline[linewidth=1.5pt,linecolor=blue](.5,-.2)(.5,4.2)}
\psellipse[linewidth=1.5pt,linecolor=blue](2,-.2)(1.5,.1)
\psellipse[linewidth=1.5pt,linecolor=blue](2,4.2)(1.5,.1)
\facegrid{(0,0)}{(4,4)}
\multirput(0,0)(0,3){2}{\multirput(0,0)(1,0){4}{\loopa}}
\rput(0,1){\loopa}\rput(1,1){\psarc[linewidth=1.5pt,linecolor=blue](0,0){.5}{0}{90}}\rput(2,1){\psarc[linewidth=1.5pt,linecolor=blue](1,0){.5}{90}{180}}\rput(3,1){\loopa}
\rput(0,2){\loopa}\rput(1,2){\psarc[linewidth=1.5pt,linecolor=blue](0,1){.5}{270}{360}}\rput(2,2){\psarc[linewidth=1.5pt,linecolor=blue](1,1){.5}{180}{270} }\rput(3,2){\loopa}
\end{pspicture}, \ \ 
\psset{unit=.7cm}
X_j^{(2)}\!\!=\;\begin{pspicture}[shift=-.9](0,0)(2,2)
\psellipse[linewidth=1pt,linecolor=black](0,1)(.1,.3)
\psellipse[linewidth=1pt,linecolor=black](1,0)(.3,.1)
\psellipse[linewidth=1pt,linecolor=black](2,1)(.1,.3)
\psellipse[linewidth=1pt,linecolor=black](1,2)(.3,.1)
\psarc[linewidth=1pt,linecolor=black](2,0){.7}{90}{180}
\psarc[linewidth=1pt,linecolor=black](2,0){.9}{90}{180}
\psarc[linewidth=1pt,linecolor=black](0,0){.9}{0}{90}
\psarc[linewidth=1pt,linecolor=black](0,2){.7}{270}{0}
\psarc[linewidth=1pt,linecolor=black](0,2){.9}{270}{0}
\psarc[linewidth=1pt,linecolor=black](2,2){.9}{180}{270}
\psarc[linewidth=1pt,linecolor=black](0,0){.7}{0}{90}
\psarc[linewidth=1pt,linecolor=black](2,2){.7}{180}{270}
\end{pspicture}\;=
\psset{unit=.45cm}
\begin{pspicture}[shift=-2.25](-.4,-.4)(4.4,4.4)
\multiput(0,0)(0,1){4}{\psline[linewidth=1.5pt,linecolor=blue](-.2,.5)(4.2,.5)}
\psellipse[linewidth=1.5pt,linecolor=blue](-.2,2)(.1,1.5)
\psellipse[linewidth=1.5pt,linecolor=blue](4.2,2)(.1,1.5)
\multiput(0,0)(1,0){4}{\psline[linewidth=1.5pt,linecolor=blue](.5,-.2)(.5,4.2)}
\psellipse[linewidth=1.5pt,linecolor=blue](2,-.2)(1.5,.1)
\psellipse[linewidth=1.5pt,linecolor=blue](2,4.2)(1.5,.1)
\facegrid{(0,0)}{(4,4)}
\multirput(0,0)(0,3){2}{\multirput(0,0)(1,0){4}{\loopa}}
\multirput(0,1)(0,1){2}{\multirput(0,0)(3,0){2}{\loopa}}
\multirput(1,1)(0,1){2}{\multirput(0,0)(1,0){2}{\loopb}}
\end{pspicture}, \ \ 
\psset{unit=.7cm}
X_j^{(3)}\!\!=\;\begin{pspicture}[shift=-.9](0,0)(2,2)
\psellipse[linewidth=1pt,linecolor=black](0,1)(.1,.3)
\psellipse[linewidth=1pt,linecolor=black](1,0)(.3,.1)
\psellipse[linewidth=1pt,linecolor=black](2,1)(.1,.3)
\psellipse[linewidth=1pt,linecolor=black](1,2)(.3,.1)
\psarc[linewidth=1pt,linecolor=black](2,0){.7}{90}{180}
\psarc[linewidth=1pt,linecolor=black](0,0){1.1}{0}{90}
\psarc[linewidth=1pt,linecolor=black](0,0){.9}{0}{90}
\psarc[linewidth=1pt,linecolor=black](0,2){.7}{270}{0}
\psarc[linewidth=1pt,linecolor=black](2,2){1.1}{180}{270}
\psarc[linewidth=1pt,linecolor=black](2,2){.9}{180}{270}
\psarc[linewidth=1pt,linecolor=black](0,0){.7}{0}{90}
\psarc[linewidth=1pt,linecolor=black](2,2){.7}{180}{270}
\end{pspicture}\;=
\psset{unit=.45cm}
\begin{pspicture}[shift=-2.25](-.4,-.4)(4.4,4.4)
\multiput(0,0)(0,1){4}{\psline[linewidth=1.5pt,linecolor=blue](-.2,.5)(4.2,.5)}
\psellipse[linewidth=1.5pt,linecolor=blue](-.2,2)(.1,1.5)
\psellipse[linewidth=1.5pt,linecolor=blue](4.2,2)(.1,1.5)
\multiput(0,0)(1,0){4}{\psline[linewidth=1.5pt,linecolor=blue](.5,-.2)(.5,4.2)}
\psellipse[linewidth=1.5pt,linecolor=blue](2,-.2)(1.5,.1)
\psellipse[linewidth=1.5pt,linecolor=blue](2,4.2)(1.5,.1)
\facegrid{(0,0)}{(4,4)}
\multirput(0,0)(0,3){2}{\multirput(0,0)(1,0){4}{\loopb}}
\rput(0,1){\loopb}\rput(1,1){\psarc[linewidth=1.5pt,linecolor=blue](0,0){.5}{0}{90}}\rput(2,1){\psarc[linewidth=1.5pt,linecolor=blue](1,0){.5}{90}{180}}\rput(3,1){\loopb}
\rput(0,2){\loopb}\rput(1,2){\psarc[linewidth=1.5pt,linecolor=blue](0,1){.5}{270}{360}}\rput(2,2){\psarc[linewidth=1.5pt,linecolor=blue](1,1){.5}{180}{270} }\rput(3,2){\loopb}
\end{pspicture}
\ee
Here the $4\times 4$ fused face configurations are just indicative of the connectivity class of configurations contributing to the generalized monoid.

\subsection{$2\times 2$ fused link states and matrix representations}
\label{LinkStatesMatrixReps}
The planar TL operators act on suitable vector spaces ${\cal V}_{r,s,\ell}^{(N)}$ of link states. 
In this section, we introduce \textit{fused or cabled}\/ link states for the case $n=2$ relevant to the logarithmic superconformal minimal models. For simplicity, we restrict to the cases with $r=1$. There are two sectors, Neveu-Schwarz (NS) with $r+s$ even, $\ell=0,2$ and Ramond (R) with $r+s$ odd, $\ell=1$. These correspond to an even and odd number of underlying elementary TL nodes respectively. The number of (single) defects $d$ is directly related to the quantum number $s$ by
\bea
d=s-1=\begin{cases}
2k,&\mbox{NS: \ $2N$ nodes}\\
2k+1,&\mbox{R:\phantom{S}  \ $2N\!+\!1$ nodes}
\end{cases}
\label{defectnumber}
\eea
where the system size $N$ is the number of projected paired nodes and $k$ is the number of cabled defects.

\subsubsection{Neveu-Schwarz}
Let $N$ be the size of the system given by the number of fused pairs of elementary nodes. For $N$ even, corresponding to $\ell=0$, the cabled link states with no (cabled) defects in the Neveu-Schwarz sector $(r,s,\ell)=(1,1,0)$ are generated~\cite{Tartaglia} by acting with $E_j$ and $X_j$ (for $j=1,\ldots,N-1$) on the simplest state in which neighboring pairs of nodes are connected. For $N=2,4$ the basis link states of ${\cal V}_{1,1,0}^{(N)}$ are\\
\psset{unit=.4cm}
$N=2$:\qquad\begin{pspicture}[shift=0](0,0)(2,2)
\multirput(0,0)(2,0){2}{\psellipse[linewidth=1pt](0,0)(.4,.15)}
\psbezier[linewidth=1pt,showpoints=false](.4,0)(.7,.75)(1.3,.75)(1.6,0)
\psbezier[linewidth=1pt,showpoints=false](-.4,0)(.2,1.6)(1.8,1.6)(2.4,0)
\end{pspicture}\\[12pt]
\psset{unit=.4cm}
$N=4$:\qquad\begin{pspicture}[shift=0](0,0)(6,2)
\multirput(0,0)(2,0){4}{\psellipse[linewidth=1pt](0,0)(.4,.15)}
\multirput(0,0)(4,0){2}{\psbezier[linewidth=1pt,showpoints=false](.4,0)(.7,.75)(1.3,.75)(1.6,0)}
\multirput(0,0)(4,0){2}{\psbezier[linewidth=1pt,showpoints=false](-.4,0)(.2,1.6)(1.8,1.6)(2.4,0)}
\end{pspicture}\:\:\;,\qquad
\begin{pspicture}[shift=0](0,0)(6,2)
\multirput(0,0)(2,0){4}{\psellipse[linewidth=1pt](0,0)(.4,.15)}
\rput(0,0){\rput(2,0){\psbezier[linewidth=1pt,showpoints=false](.4,0)(.7,.6)(1.3,.6)(1.6,0)
\psbezier[linewidth=1pt,showpoints=false](-.4,0)(.2,1.4)(1.8,1.4)(2.4,0)}
\psbezier[linewidth=1pt,showpoints=false](-.4,0)(.2,2.7)(5.8,2.7)(6.4,0)
\psbezier[linewidth=1pt,showpoints=false](.4,0)(1,2)(5,2)(5.6,0)}
\end{pspicture}\:\:\;,\qquad
\begin{pspicture}[shift=0](0,0)(6,2)
\multirput(0,0)(2,0){4}{\psellipse[linewidth=1pt](0,0)(.4,.15)}
\cabledFour
\end{pspicture}\:\:\;\\[12pt]
The number of link states, for even $N$, is given by the Riordan numbers $R_N$ with even $N$
\bea
\dim {\cal V}_{1,1,0}^{(N)}=R_N=1,3,15,91,\ldots\qquad N=2,4,6,8,\ldots
\eea
The basis cabled link states in ${\cal V}_{1,1,2}^{(N)}$  in the Neveu-Schwarz sector $(r,s,\ell)=(1,1,2)$ are obtained similarly but with $N$ odd.  For $N=3,5$ the link states are\\[4pt]
\psset{unit=.4cm}
$N=3$:\qquad
\begin{pspicture}[shift=0](0,0)(16,2)
\multirput(0,0)(2,0){3}{\psellipse[linewidth=1pt](0,0)(.4,.15)}
\multirput(0,0)(2,0){2}{\psbezier[linewidth=1pt,showpoints=false](.4,0)(.7,.75)(1.3,.75)(1.6,0)}
\psbezier[linewidth=1pt,showpoints=false](-.4,0)(.2,2.3)(3.8,2.3)(4.4,0)
\end{pspicture}\\[2pt]
\psset{unit=.4cm}
$N=5$:\qquad
\begin{pspicture}[shift=0](0,0)(8,2)
\multirput(0,0)(2,0){5}{\psellipse[linewidth=1pt](0,0)(.4,.15)}
\cabledTwo
\rput(4,0){\cabledThree}
\end{pspicture}\:\:\:,\qquad
\begin{pspicture}[shift=0](0,0)(8,2)
\multirput(0,0)(2,0){5}{\psellipse[linewidth=1pt](0,0)(.4,.15)}
\cabledThree
\rput(6,0){\cabledTwo}
\end{pspicture}\:\:\:,\qquad
\begin{pspicture}[shift=0](0,0)(8,3)
\multirput(0,0)(2,0){5}{\psellipse[linewidth=1pt](0,0)(.4,.15)}
\rput(4,0){\cabledTwo}
\rput(0,0){\psbezier[linewidth=1pt,showpoints=false](.4,0)(.7,.75)(1.3,.75)(1.6,0)
\psbezier[linewidth=1pt,showpoints=false](-.4,0)(.2,4)(7.8,4)(8.4,0)
\psbezier[linewidth=1pt,showpoints=false](2.4,0)(3,2.5)(7,2.5)(7.6,0)}
\end{pspicture}\:\:\:,\\[12pt]
\mbox{}\qquad\qquad\quad
\begin{pspicture}[shift=0](0,0)(8,3)
\multirput(0,0)(2,0){5}{\psellipse[linewidth=1pt](0,0)(.4,.15)}
\rput(2,0){\cabledTwo}
\rput(0,0){\rput(6,0){\psbezier[linewidth=1pt,showpoints=false](.4,0)(.7,.75)(1.3,.75)(1.6,0)}
\psbezier[linewidth=1pt,showpoints=false](-.4,0)(.2,4)(7.8,4)(8.4,0)
\psbezier[linewidth=1pt,showpoints=false](.4,0)(1,2.5)(5,2.5)(5.6,0)}
\end{pspicture}\:\:\:,\qquad
\begin{pspicture}[shift=0](0,0)(8,3)
\multirput(0,0)(2,0){5}{\psellipse[linewidth=1pt](0,0)(.4,.15)}
\rput(2,0){\cabledThree}
\psbezier[linewidth=1pt,showpoints=false](-.4,0)(.2,3.8)(7.8,3.8)(8.4,0)
\psbezier[linewidth=1pt,showpoints=false](.4,0)(1,3)(7,3)(7.6,0)
\end{pspicture}\:\:\:,\qquad
\begin{pspicture}[shift=0](0,0)(8,3)
\multirput(0,0)(2,0){5}{\psellipse[linewidth=1pt](0,0)(.4,.15)}
\rput(0,0){\multirput(0,0)(2,0){4}{\psbezier[linewidth=1pt,showpoints=false](.4,0)(.7,.75)(1.3,.75)(1.6,0)}\psbezier[linewidth=1pt,showpoints=false](-.4,0)(.2,4)(7.8,4)(8.4,0)}
\end{pspicture}\:\:\:\\[12pt]
The number of link states, in ${\cal V}_{1,1,2}^{(N)}$, is given by the Riordan numbers $R_N$ with odd $N$
\bea
\dim {\cal V}_{1,1,2}^{(N)}=R_N=1,6,36,232,\ldots\qquad N=3,5,7,9,\ldots
\eea

There is a simple bijection between Riordan cabled link states and Riordan paths for $N$ even or odd. 
Riordan paths are (spin-1) paths where each step is diagonally up, diagonally down or horizontal, except that the first step must be diagonally up and the last step must be diagonally down. The path must start and finish at the base height 0 and is constrained to lie above the base height. 
The paths are unrestricted in the sense that they can reach arbitrarily large heights by taking $N$ sufficiently large. The number of Riordan paths of length $N$ is given by the Riordan numbers~\cite{Riordan,CatMR}
\be
R_N=0,1,1,3,6,15,36,91,232,\ldots\qquad N=1,2,3,\ldots
\ee
A Riordan path is uniquely labelled by the ordered list of heights at each step. Similarly, a link state is uniquely labelled by the same ordered list of heights where the heights designate the number of doubled strands that pass above the gaps between consecutive (paired elementary) nodes. The matching of the labels gives the bijection. For example,
\be
\psset{unit=.4cm}
\begin{pspicture}[shift=-1.5](-1,-1)(11,3)
\multirput(-1,-.5)(12,0){2}{0}
\multirput(0,0)(6,0){2}{\multirput(1,-.5)(2,0){2}{1}}
\rput(5,-.5){2}
\multirput(0,0)(2,0){6}{\psellipse[linewidth=1pt](0,0)(.4,.15)}
\rput(4,0){\psbezier[linewidth=1pt,showpoints=false](.4,0)(.7,.75)(1.3,.75)(1.6,0)
\psbezier[linewidth=1pt,showpoints=false](-.4,0)(.2,1.6)(1.8,1.6)(2.4,0)}
\rput(0,0){\multirput(0,0)(8,0){2}{\psbezier[linewidth=1pt,showpoints=false](.4,0)(.7,.75)(1.3,.75)(1.6,0)}
\psbezier[linewidth=1pt,showpoints=false](-.4,0)(.2,4)(9.8,4)(10.4,0)
\psbezier[linewidth=1pt,showpoints=false](2.4,0)(3,2.5)(7,2.5)(7.6,0)}
\end{pspicture}\\
\quad\leftrightarrow\quad
\psset{unit=.45cm}
\begin{pspicture}[shift=-1.5](-1,-1)(11,4)
\multirput(-1,-.5)(12,0){2}{0}
\multirput(0,0)(6,0){2}{\multirput(1,-.5)(2,0){2}{1}}
\rput(5,-.5){2}
\psline(-1,0)(1,2)
\psline(1,2)(3,2)
\psline(3,2)(5,4)
\psline(5,4)(7,2)
\psline(7,2)(9,2)
\psline(9,2)(11,0)
\end{pspicture}
\ee

We next form link states with \emph{defects}, that is, cabled strands that connect from the bulk to the boundary. In the Neveu-Schwarz sectors, there is always an even number of underlying elementary nodes. This means that, in the Neveu-Schwarz sectors, the defects consist of an even number $d=2k$ of single strands mutually cabled together in the boundary. In the Ramond sectors with an odd number of underlying elementary nodes, the defects consist of an odd number of single strands cabled together in the boundary. In the following examples, in the NS sector, we take $N$ to be even. The $N$ odd cases are similar. The basis link states for $N=2,4$ with a single ($k=1$) cabled defect are\\[6pt]
\psset{unit=.4cm}
$N=2$:\qquad\begin{pspicture}[shift=0](0,0)(2,2)
\multirput(0,0)(2,0){3}{\psellipse[linewidth=1pt](0,0)(.4,.15)}
\cabledThree
\psline[linestyle=dashed,linecolor=red](3,0)(3,2)
\end{pspicture}\\[12pt]
\psset{unit=.4cm}
$N=4$:\qquad\begin{pspicture}[shift=0](0,0)(8,2)
\multirput(0,0)(2,0){5}{\psellipse[linewidth=1pt](0,0)(.4,.15)}
\cabledTwo
\rput(4,0){\cabledThree}
\psline[linestyle=dashed,linecolor=red](7,0)(7,3)
\end{pspicture}\:\:\:,\qquad
\begin{pspicture}[shift=0](0,0)(8,2)
\multirput(0,0)(2,0){5}{\psellipse[linewidth=1pt](0,0)(.4,.15)}
\cabledThree
\rput(6,0){\cabledTwo}
\psline[linestyle=dashed,linecolor=red](7,0)(7,3)
\end{pspicture}\:\:\:,\qquad
\begin{pspicture}[shift=0](0,0)(8,3)
\multirput(0,0)(2,0){5}{\psellipse[linewidth=1pt](0,0)(.4,.15)}
\rput(4,0){\cabledTwo}
\rput(0,0){\psbezier[linewidth=1pt,showpoints=false](.4,0)(.7,.75)(1.3,.75)(1.6,0)
\psbezier[linewidth=1pt,showpoints=false](-.4,0)(.2,4)(7.8,4)(8.4,0)
\psbezier[linewidth=1pt,showpoints=false](2.4,0)(3,2.5)(7,2.5)(7.6,0)}
\psline[linestyle=dashed,linecolor=red](7,0)(7,3)
\end{pspicture}\:\:\:,\\[12pt]
\mbox{}\qquad\qquad\quad\begin{pspicture}[shift=0](0,0)(8,3)
\multirput(0,0)(2,0){5}{\psellipse[linewidth=1pt](0,0)(.4,.15)}
\rput(2,0){\cabledTwo}
\rput(0,0){\rput(6,0){\psbezier[linewidth=1pt,showpoints=false](.4,0)(.7,.75)(1.3,.75)(1.6,0)}
\psbezier[linewidth=1pt,showpoints=false](-.4,0)(.2,4)(7.8,4)(8.4,0)
\psbezier[linewidth=1pt,showpoints=false](.4,0)(1,2.5)(5,2.5)(5.6,0)}
\psline[linestyle=dashed,linecolor=red](7,0)(7,3)
\end{pspicture}\:\:\:,\qquad
\begin{pspicture}[shift=0](0,0)(8,3)
\multirput(0,0)(2,0){5}{\psellipse[linewidth=1pt](0,0)(.4,.15)}
\rput(2,0){\cabledThree}
\psbezier[linewidth=1pt,showpoints=false](-.4,0)(.2,3.8)(7.8,3.8)(8.4,0)
\psbezier[linewidth=1pt,showpoints=false](.4,0)(1,3)(7,3)(7.6,0)
\psline[linestyle=dashed,linecolor=red](7,0)(7,3)
\end{pspicture}\:\:\:,\qquad
\begin{pspicture}[shift=0](0,0)(8,3)
\multirput(0,0)(2,0){5}{\psellipse[linewidth=1pt](0,0)(.4,.15)}
\rput(0,0){\multirput(0,0)(2,0){4}{\psbezier[linewidth=1pt,showpoints=false](.4,0)(.7,.75)(1.3,.75)(1.6,0)}\psbezier[linewidth=1pt,showpoints=false](-.4,0)(.2,4)(7.8,4)(8.4,0)}
\psline[linestyle=dashed,linecolor=red](7,0)(7,3)
\end{pspicture}\:\:\:\\[12pt]
The (red) dashed line separates the {\it bulk} on the left from the {\it boundary} on the right. The number of these link states with 1 (cabled) defect is 
\bea
\dim {\cal V}_{1,3,0}^{(N)}=R_{N,1}=R_{N+1}=1,6,36,232,\ldots\qquad N=2,4,6,8,\ldots\label{130count}
\eea

 When there are $k>1$ (cabled) defects, each (cabled) defect must be in a separate cluster. In other words, the (cabled) defects are not allowed to be connected in the boundary. For example, in the cases $N=2,4$ with $k=2$, the link states are\\[12pt]
\psset{unit=.4cm}
$N=2$:\qquad \begin{pspicture}[shift=0](0,0)(6,2)
\multirput(0,0)(2,0){2}{\psellipse[linewidth=1pt](0,0)(.4,.15)}
\rput(4,0){\loopProjLonger}
\rput(0,0){\rput(2,0){\psbezier[linewidth=1pt,showpoints=false](.4,0)(.7,.6)(1.3,.6)(1.6,0)
\psbezier[linewidth=1pt,showpoints=false](-.4,0)(.2,1.4)(1.8,1.4)(2.4,0)}
\psbezier[linewidth=1pt,showpoints=false](-.4,0)(.2,2.7)(5.8,2.7)(6.4,0)
\psbezier[linewidth=1pt,showpoints=false](.4,0)(1,2)(5,2)(5.6,0)}
\psline[linestyle=dashed,linecolor=red](3,0)(3,2.5)
\end{pspicture}\\[12pt]
\psset{unit=.4cm}
$N=4$:\qquad \begin{pspicture}[shift=0](0,0)(10,3)
\multirput(0,0)(2,0){4}{\psellipse[linewidth=1pt](0,0)(.4,.15)}
\rput(8,0){\loopProjLonger}
\rput(0,0){\psbezier[linewidth=1pt,showpoints=false](.4,0)(.7,.75)(1.3,.75)(1.6,0)
\psbezier[linewidth=1pt,showpoints=false](-.4,0)(.2,1.6)(1.8,1.6)(2.4,0)}
\rput(4,0){\rput(2,0){\psbezier[linewidth=1pt,showpoints=false](.4,0)(.7,.6)(1.3,.6)(1.6,0)
\psbezier[linewidth=1pt,showpoints=false](-.4,0)(.2,1.4)(1.8,1.4)(2.4,0)}
\psbezier[linewidth=1pt,showpoints=false](-.4,0)(.2,2.7)(5.8,2.7)(6.4,0)
\psbezier[linewidth=1pt,showpoints=false](.4,0)(1,2)(5,2)(5.6,0)}
\psline[linestyle=dashed,linecolor=red](7,0)(7,3)
\end{pspicture}\:\:\:,\qquad
\begin{pspicture}[shift=0](0,0)(10,3)
\multirput(0,0)(2,0){4}{\psellipse[linewidth=1pt](0,0)(.4,.15)}
\rput(8,0){\loopProjLonger}
\multirput(2,0)(4,0){2}{\psbezier[linewidth=1pt,showpoints=false](.4,0)(.7,.75)(1.3,.75)(1.6,0)
\psbezier[linewidth=1pt,showpoints=false](-.4,0)(.2,1.6)(1.8,1.6)(2.4,0)}
\rput(0,0){\psbezier[linewidth=1pt,showpoints=false](-.4,0)(.2,4)(9.8,4)(10.4,0)
\psbezier[linewidth=1pt](.4,0)(1,3.3)(9,3.3)(9.6,0)}
\psline[linestyle=dashed,linecolor=red](7,0)(7,3)
\end{pspicture}\:\:\:,\qquad
\begin{pspicture}[shift=0](0,0)(10,3)
\multirput(0,0)(2,0){4}{\psellipse[linewidth=1pt](0,0)(.4,.15)}
\rput(8,0){\loopProjLonger}
\rput(0,0){\psbezier[linewidth=1pt,showpoints=false](-.4,0)(.2,4)(9.8,4)(10.4,0)
\psbezier[linewidth=1pt](.4,0)(1,3.3)(9,3.3)(9.6,0)}
\rput(2,0){\rput(2,0){\psbezier[linewidth=1pt,showpoints=false](.4,0)(.7,.6)(1.3,.6)(1.6,0)
\psbezier[linewidth=1pt,showpoints=false](-.4,0)(.2,1.4)(1.8,1.4)(2.4,0)}
\psbezier[linewidth=1pt,showpoints=false](-.4,0)(.2,2.7)(5.8,2.7)(6.4,0)
\psbezier[linewidth=1pt,showpoints=false](.4,0)(1,2)(5,2)(5.6,0)}
\psline[linestyle=dashed,linecolor=red](7,0)(7,3)
\end{pspicture}\:\:\:,\qquad\\[12pt]
\mbox{}\qquad\qquad\quad\begin{pspicture}[shift=0](0,0)(10,3)
\multirput(0,0)(2,0){4}{\psellipse[linewidth=1pt](0,0)(.4,.15)}
\rput(8,0){\loopProjLonger}
\rput(4,0){\multirput(0,0)(2,0){2}{\psbezier[linewidth=1pt,showpoints=false](.4,0)(.7,.75)(1.3,.75)(1.6,0)}
\psbezier[linewidth=1pt,showpoints=false](-.4,0)(.2,2.3)(3.8,2.3)(4.4,0)}
\rput(0,0){\psbezier[linewidth=1pt,showpoints=false](.4,0)(.7,.75)(1.3,.75)(1.6,0)
\psbezier[linewidth=1pt,showpoints=false](-.4,0)(.2,4)(9.8,4)(10.4,0)
\psbezier[linewidth=1pt,showpoints=false](2.4,0)(3,3)(9,3)(9.6,0)}
\psline[linestyle=dashed,linecolor=red](7,0)(7,3)
\end{pspicture}\:\:\:,\qquad
\begin{pspicture}[shift=0](0,0)(10,3)
\rput(8,0){\loopProjLonger}
\multirput(0,0)(2,0){4}{\psellipse[linewidth=1pt](0,0)(.4,.15)}
\rput(6,0){\psbezier[linewidth=1pt,showpoints=false](.4,0)(.7,.75)(1.3,.75)(1.6,0)
\psbezier[linewidth=1pt,showpoints=false](-.4,0)(.2,1.6)(1.8,1.6)(2.4,0)}
\rput(0,0){\multirput(0,0)(2,0){2}{\psbezier[linewidth=1pt,showpoints=false](.4,0)(.7,.75)(1.3,.75)(1.6,0)}
\psbezier[linewidth=1pt,showpoints=false](-.4,0)(.2,4)(9.8,4)(10.4,0)
\psbezier[linewidth=1pt,showpoints=false](4.4,0)(5,2.5)(9,2.5)(9.6,0)}
\psline[linestyle=dashed,linecolor=red](7,0)(7,3)
\end{pspicture}\:\:\:,\qquad
\begin{pspicture}[shift=0](0,0)(10,3)
\multirput(0,0)(2,0){4}{\psellipse[linewidth=1pt](0,0)(.4,.15)}
\rput(8,0){\loopProjLonger}
\rput(2,0){\multirput(0,0)(2,0){3}{\psbezier[linewidth=1pt,showpoints=false](.4,0)(.7,.75)(1.3,.75)(1.6,0)}
\psbezier[linewidth=1pt,showpoints=false](-.4,0)(.2,2.7)(5.8,2.7)(6.4,0)}
\rput(0,0){\psbezier[linewidth=1pt,showpoints=false](-.4,0)(.2,4)(9.8,4)(10.4,0)
\psbezier[linewidth=1pt](.4,0)(1,3.3)(9,3.3)(9.6,0)}
\psline[linestyle=dashed,linecolor=red](7,0)(7,3)
\end{pspicture}\:\:\:\\[-2pt]
\bea
\dim {\cal V}_{1,5,0}^{(N)}=R_{N,2}=1,6,40,280,\ldots\qquad N=2,4,6,8,\ldots
\eea

In general, the number of cabled link states, with $k$ (cabled) defects, is given by generalized Riordan numbers $R_{N,k}$ as shown in Table~\ref{Riordan}. Explicitly, these are differences of {\it trinomial coefficients}
\begin{align} 
\dim{\cal V}_{1,2k+1,\ell}^{(N)}&=R_{N,k}=\superTrinomial[N][k]-\superTrinomial[N][k\!+\!1] \nonumber\\
&=\sum_{j=0}^N\left(\trinomial[N][\hf(N\!-\!j\!-\!k)][\hf(N\!-\!j\!+\!k)][j]- \trinomial[N][\hf(N\!-\!j\!-\!k\!-\!1)][\hf(N\!-\!j\!+\!k\!+\!1)][j]\right) \label{eq:RiordanTri}
\end{align}
where $\ell=2N$ mod $4$ and the supertrinomial coefficients are given in terms of trinomial coefficients by
\be
(x+1+x^{-1})^N=\sum_{k=-N}^N\superTrinomial[N][k] x^k,\qquad 
\superTrinomial[N][k]=\sum_{j=0}^N \trinomial[N][\hf(N\!-\!j\!-\!k)][\hf(N\!-\!j\!+\!k)][j]
\ee
and
\be
\trinomial[N][l][m][N\!-\!l\!-\!m]=
\begin{cases}
\frac{N!}{l!m!(N-l-m)!}, &l,m,N-l-m\in \mathbb{Z}_{\ge 0}\\
0,& \text{otherwise}
\end{cases}
\label{trinomial}
\ee

\begin{table}[htbp]
\centering
\subfloat[$R_{N,k}$, $N$ even $(\ell=0)$]{
\begin{tabular}{c|cccc}
\psset{unit=.5cm}
\begin{pspicture}(0,0)(1.5,1.1)
\rput(-.06,-.3){\psline[linewidth=.5pt](-.4,1.4)(2,0)
\rput(.2,.45){\text{$N$}}
\rput(1.5,.9){\text{$k$}}}
\end{pspicture}
&0&1&2&3\\
\hline
2&1&1&1&0\\
4& 3 & 6 & 6 & 3 \\
6& 15 & 36 & 40 & 29 \\
8& 91 & 232 & 280 & 238 \\
10& 603 & 1585 & 2025 & 1890
\end{tabular}
}\hspace{1cm}
\subfloat[$R_{N,k}$, $N$ odd $(\ell=2)$]{
\begin{tabular}{c|cccc}
\psset{unit=.5cm}
\begin{pspicture}(0,0)(1.5,1.1)
\rput(-.06,-.3){\psline[linewidth=.5pt](-.4,1.4)(2,0)
\rput(.2,.45){\text{$N$}}
\rput(1.5,.9){\text{$k$}}}
\end{pspicture}
&0&1&2&3\\
\hline
1&0 & 1 & 0 & 0 \\
 3&1 & 3 & 2 & 1 \\
 5&6 & 15 & 15 & 10 \\
 7&36 & 91 & 105 & 84 \\
 9&232 & 603 & 750 & 672
\end{tabular}
}
\caption{The generalized Riordan numbers $R_{N,k}$ giving the number of link states on $N$ (paired) nodes with $k$ cabled defects in the Neveu-Schwarz sectors with (a) $N$ even ($\ell=0$) and (b) $N$ odd ($\ell=2$). We note that 
the numbers $R_{N,k}$ satisfy the recursion relation $R_{N,k}=R_{N-1,k-1}+R_{N-1,k}+R_{N-1,k+1}$, $N>1$, $k>0$ subject to the initial conditions $R_{1,k}=\delta(1,k)$ and $R_{N,0}=R_{N-1,1}=R_N$.\label{Riordan}}
\end{table}

\subsubsection{Ramond}
In the Ramond sectors, corresponding to $\ell=1$, the link states have an odd number of underlying elementary nodes in the bulk and the number $d=2k+1$ of single-strand cabled defects entering the bulk from the boundary must be odd. We say that the system is of size $N$ if there are $N$ paired nodes plus one single node in the bulk. The number of link states in the Ramond sector is given by generalized Motzkin numbers $M_{N,k}$ as in Table~\ref{Motzkin}.
\begin{table}[tbp]
\begin{center}
\begin{tabular}{c|cccc}
\psset{unit=.5cm}
\begin{pspicture}(0,0)(1.5,1.1)
\rput(-.06,-.3){\psline[linewidth=.5pt](-.4,1.4)(2,0)
\rput(.2,.45){\text{$N$}}
\rput(1.5,.9){\text{$k$}}}
\end{pspicture}
&0&1&2&3\\
\hline
1&1 & 1 & 0 & 0 \\
2& 2 & 2 & 1 & 0 \\
3& 4 & 5 & 3 & 1 \\
4& 9 & 12 & 9 & 4 \\
5& 21 & 30 & 25 & 14 \\
6& 51 & 76 & 69 & 44
\end{tabular}
\end{center}
\caption{Generalized Motzkin numbers $M_{N,k}=R_{N,k}+R_{N,k+1}$ giving the number of link states for system size $N$ with $2k+1$ defects in the Ramond sector. We note that $M_{N,0}=M_N$.\label{Motzkin}}
\end{table}
For example, the Ramond link states for $N=1,2,3$ with one defect ($k=0$) are\\
\psset{unit=.4cm}
$N=1$:\qquad\begin{pspicture}[shift=0](0,0)(2,2)
\loopProj
\cabledTwo
\psline[linestyle=dashed,linecolor=red](2,0)(2,1.4)
\end{pspicture}\nopagebreak\\[6pt]
\psset{unit=.4cm}
$N=2$:\qquad\begin{pspicture}[shift=-.1](0,0)(5,2)
\multirput(0,0)(2,0){2}{\loopProj}
\cabledTwo
\rput(3.2,0){\link}
\psline[linestyle=dashed,linecolor=red](4.2,0)(4.2,2)
\end{pspicture}\!,\qquad
\begin{pspicture}[shift=-.1](0,0)(5,2)
\multirput(0,0)(2,0){2}{\loopProj}
\cabledThree
\psline[linestyle=dashed,linecolor=red](4,0)(4,2)
\end{pspicture}\!\nopagebreak\\[6pt]
\psset{unit=.4cm}
$N=3$:\qquad\begin{pspicture}[shift=-.1](0,0)(7,2)
\multirput(0,0)(2,0){3}{\loopProj}
\cabledThree
\rput(5.2,0){\link}
\psline[linestyle=dashed,linecolor=red](6.2,0)(6.2,2.4)
\end{pspicture}\!,\qquad
\begin{pspicture}[shift=0](0,0)(6,2)
\multirput(0,0)(2,0){3}{\psellipse[linewidth=1pt](0,0)(.4,.15)}
\multirput(0,0)(4,0){2}{\psbezier[linewidth=1pt,showpoints=false](.4,0)(.7,.75)(1.3,.75)(1.6,0)}
\multirput(0,0)(4,0){2}{\psbezier[linewidth=1pt,showpoints=false](-.4,0)(.2,1.6)(1.8,1.6)(2.4,0)}
\psline[linestyle=dashed,linecolor=red](6,0)(6,2.4)
\end{pspicture}\:\:\;,\qquad
\begin{pspicture}[shift=0](0,0)(6,2)
\multirput(0,0)(2,0){3}{\psellipse[linewidth=1pt](0,0)(.4,.15)}
\rput(0,0){\rput(2,0){\psbezier[linewidth=1pt,showpoints=false](.4,0)(.7,.6)(1.3,.6)(1.6,0)
\psbezier[linewidth=1pt,showpoints=false](-.4,0)(.2,1.4)(1.8,1.4)(2.4,0)}
\psbezier[linewidth=1pt,showpoints=false](-.4,0)(.2,2.7)(5.8,2.7)(6.4,0)
\psbezier[linewidth=1pt,showpoints=false](.4,0)(1,2)(5,2)(5.6,0)}
\psline[linestyle=dashed,linecolor=red](6,0)(6,2.4)
\end{pspicture}\:\:\;,\qquad
\begin{pspicture}[shift=0](0,0)(6,2)
\multirput(0,0)(2,0){3}{\psellipse[linewidth=1pt](0,0)(.4,.15)}
\cabledFour
\psline[linestyle=dashed,linecolor=red](6,0)(6,2.4)
\end{pspicture}\:\:\;\\[8pt]
These link states are counted by the usual Motzkin numbers~\cite{Motzkin,CatMR}
\be
\dim{\cal V}_{1,2,1}^{(N)}=M_N=M_{N,0}=R_N+R_{N+1}=1,2,4,9,21,\ldots\qquad N=1,2,3,\ldots\label{eq:motzkinSums}
\ee
This identity follows by partitioning the link states into Riordan link states of size $N$ with a spectator link on the right and Riordan link states of size $N+1$ where the right most pair of nodes is not projected.
Similarly, the Ramond link states for $N=1,2,3$ with three defects ($k=1$) are\\[6pt]
\psset{unit=.4cm}
$N=1$:\qquad\begin{pspicture}[shift=0](0,0)(6,2)
\rput(0,0){\loopProj}
\rput(3,0){\loopProjMed}
\rput(0,0){\linkMeder}
\linkMed
\rput(1,0){\linkLow}
\psline[linestyle=dashed,linecolor=red](2,0)(2,2)
\end{pspicture}\\[12pt]
\psset{unit=.4cm}
$N=2$:\qquad\begin{pspicture}[shift=0](0,0)(6.5,3)
\multirput(0,0)(2,0){2}{\loopProj}
\rput(5,0){\loopProjMed}
\multirput(0,0)(3,0){2}\linkLow
\rput(2,0){\linkMed}
\rput(0,0){\linkLonger}
\psline[linestyle=dashed,linecolor=red](4,0)(4,2.5)
\end{pspicture},\qquad
\begin{pspicture}[shift=0](0,0)(6.5,2)
\multirput(0,0)(2,0){2}{\loopProj}
\rput(4.4,0){\loopProjLong}
\rput(0,0){\rput(2,0){\psbezier[linewidth=1pt,showpoints=false](.4,0)(.7,.6)(1.3,.6)(1.6,0)
\psbezier[linewidth=1pt,showpoints=false](-.4,0)(.2,1.4)(1.8,1.4)(2.4,0)}
\psbezier[linewidth=1pt,showpoints=false](-.4,0)(.2,2.7)(5.8,2.7)(6.4,0)
\psbezier[linewidth=1pt,showpoints=false](.4,0)(1,2)(5,2)(5.6,0)}
\psline[linestyle=dashed,linecolor=red](4,0)(4,2.5)
\end{pspicture}\\[12pt]
\psset{unit=.4cm}
$N=3$:\qquad\begin{pspicture}[shift=0](0,0)(8.5,3)
\multirput(0,0)(2,0){3}{\loopProj}
\rput(7,0){\loopProjMed}
\rput(0,0){\cabledTwo}
\rput(4,0){\linkMeder}
\rput(4,0){\linkMed}
\rput(5,0){\linkLow}
\psline[linestyle=dashed,linecolor=red](6,0)(6,3)
\end{pspicture},\qquad
\begin{pspicture}[shift=0](0,0)(8.5,3)
\multirput(0,0)(2,0){3}{\loopProj}
\rput(7,0){\loopProjMed}
\linkLongX
\linkLongx
\rput(2,0){\cabledTwo}
\rput(5,0){\link}
\psline[linestyle=dashed,linecolor=red](6,0)(6,3)
\end{pspicture},\qquad
\begin{pspicture}[shift=0](0,0)(8.5,3)
\multirput(0,0)(2,0){3}{\loopProj}
\rput(7,0){\loopProjMed}
\linkLongX
\multirput(0,0)(2,0){2}{\link}
\rput(5,0){\linkLow}
\rput(4,0){\linkMed}
\psline[linestyle=dashed,linecolor=red](6,0)(6,3)
\end{pspicture},\\[12pt]
\mbox{}\qquad\quad\qquad\begin{pspicture}[shift=0](0,0)(8,3)
\multirput(0,0)(2,0){3}{\psellipse[linewidth=1pt](0,0)(.4,.15)}
\rput(6.4,0){\loopProjLong}
\rput(4,0){\cabledTwo}
\rput(0,0){\psbezier[linewidth=1pt,showpoints=false](.4,0)(.7,.75)(1.3,.75)(1.6,0)
\psbezier[linewidth=1pt,showpoints=false](-.4,0)(.2,4)(7.8,4)(8.4,0)
\psbezier[linewidth=1pt,showpoints=false](2.4,0)(3,2.5)(7,2.5)(7.6,0)}
\psline[linestyle=dashed,linecolor=red](6,0)(6,3)
\end{pspicture}\:\:\,,\qquad
\begin{pspicture}[shift=0](0,0)(8,3)
\multirput(0,0)(2,0){3}{\psellipse[linewidth=1pt](0,0)(.4,.15)}
\rput(6.4,0){\loopProjLong}
\rput(2,0){\cabledThree}
\psbezier[linewidth=1pt,showpoints=false](-.4,0)(.2,3.8)(7.8,3.8)(8.4,0)
\psbezier[linewidth=1pt,showpoints=false](.4,0)(1,3)(7,3)(7.6,0)
\psline[linestyle=dashed,linecolor=red](6,0)(6,3)
\end{pspicture}\:\:\,\\
\be
\dim{\cal V}_{1,4,1}^{(N)}=M_{N,1} = R_{N,1} + R_{N,2}=1,2,5,12,30,\ldots\qquad N=1,2,3,\ldots\label{eq:riordanSums}
\ee

More generally, the Motzkin link states, for system size $N$ with $2k+1$ single defects, can be partitioned into two sets corresponding to the Riordan link states on $N$ nodes with $k$ cabled defects and separately with $k+1$ cabled defects
\be
M_{N,k} = R_{N,k} + R_{N,k+1}
\ee
Using (\ref{eq:RiordanTri}), the generalized Motzkin numbers $M_{N,k}$, as shown in Table~\ref{Motzkin}, are given explicitly in terms of supertrinomial coefficients and trinomial coefficients by
\begin{align}
\dim{\cal V}_{1,2k+2,1}^{(N)}&=M_{N,k}=\superTrinomial[N][k]-\superTrinomial[N][k+2]\nonumber\\
&=\sum_{j=0}^N\left( \trinomial[N][\hf(N\!-\!j\!-\!k)][\hf(N\!-\!j\!+\!k)][j]- \trinomial[N][\hf(N\!-\!j\!-\!k\!-\!2)][\hf(N\!-\!j\!+\!k\!+\!2)][j]\right)
\end{align}

\subsubsection{Action on cabled link states}
The action of the operator $E_j$ or $X_j$ on a cabled link state is given diagrammatically by gluing the operator underneath the link state and then decomposing any internal projectors to give a decomposition back into the original basis of cabled link states. For example,
\psset{unit=.4cm}
\bea
\begin{pspicture}[shift=-1](-.4,0)(6.4,4)
\rput(0,2){
\multirput(0,0)(2,0){3}{\psbezier[linewidth=1pt,showpoints=false](.4,0)(.7,.75)(1.3,.75)(1.6,0)}
\psbezier[linewidth=1pt,showpoints=false](-.4,0)(.2,2.7)(5.8,2.7)(6.4,0)
\rput(4,-2){\psellipse[linewidth=1pt](0,0)(.4,.15)
\psellipse[linewidth=1pt](2,0)(.4,.15)
\psellipse[linewidth=1pt](0,2)(.4,.15)
\psellipse[linewidth=1pt](2,2)(.4,.15)
\multirput(0,0)(.8,0){2}{\multirput(0,0)(2,0){2}{\psline(-.4,0)(-.4,2)}}}
}
\fusedX
\end{pspicture}
&=&
\begin{pspicture}[shift=-1](-.4,0)(6.4,4)
\rput(0,2){
\multirput(0,0)(2,0){3}{\psbezier[linewidth=1pt,showpoints=false](.4,0)(.7,.75)(1.3,.75)(1.6,0)}
\psbezier[linewidth=1pt,showpoints=false](-.4,0)(.2,2.7)(5.8,2.7)(6.4,0)
\rput(4,-2){\psellipse[linewidth=1pt](0,0)(.4,.15)
\psellipse[linewidth=1pt](2,0)(.4,.15)
\multirput(0,0)(.8,0){2}{\multirput(0,0)(2,0){2}{\psline(-.4,0)(-.4,2)}}}
}
\psellipse[linewidth=1pt](0,0)(.4,.15)
\psellipse[linewidth=1pt](2,0)(.4,.15)
\psbezier[linewidth=1pt](.4,0)(.9,.75)(1.1,.75)(1.6,0)
\psbezier[linewidth=1pt](-.4,0)(.6,.7)(.6,1.3)(-.4,2)
\psbezier[linewidth=1pt](.4,2)(.9,1.25)(1.1,1.25)(1.6,2)
\psbezier[linewidth=1pt](2.4,0)(1.4,.7)(1.4,1.3)(2.4,2)
\end{pspicture}
-\frac{1}{\TLb}
\begin{pspicture}[shift=-1](-.4,0)(6.4,4)
\rput(0,2){
\rput(0,0){\psbezier[linewidth=1pt,showpoints=false](.4,.4)(.7,.75)(1.3,.75)(1.6,0)}
\rput(2,0){\psbezier[linewidth=1pt,showpoints=false](.4,0)(.7,.75)(1.3,.75)(1.6,0)}
\rput(4,0){\psbezier[linewidth=1pt,showpoints=false](.4,0)(.7,.75)(1.3,.75)(1.6,0)}
\psbezier[linewidth=1pt,showpoints=false](-.4,.4)(.2,2.7)(5.8,2.7)(6.4,0)
\rput(4,-2){\psellipse[linewidth=1pt](0,0)(.4,.15)
\psellipse[linewidth=1pt](2,0)(.4,.15)
\multirput(0,0)(.8,0){2}{\multirput(0,0)(2,0){2}{\psline(-.4,0)(-.4,2)}}}
}
\psellipse[linewidth=1pt](0,0)(.4,.15)
\psellipse[linewidth=1pt](2,0)(.4,.15)
\psbezier[linewidth=1pt](.4,0)(.9,.75)(1.1,.75)(1.6,0)
\psbezier[linewidth=1pt](-.4,0)(.6,.7)(.6,1.3)(-.4,1.8)
\psbezier[linewidth=1pt](.4,1.8)(.9,1.25)(1.1,1.25)(1.6,2)
\psbezier[linewidth=1pt](2.4,0)(1.4,.7)(1.4,1.3)(2.4,2)
\psbezier[linewidth=1pt,showpoints=false](-.4,1.8)(-.2,2)(.2,2)(.4,1.8)
\psbezier[linewidth=1pt,showpoints=false](-.4,2.4)(-.2,2.2)(.2,2.2)(.4,2.4)
\end{pspicture}
-\frac{1}{\TLb}
\begin{pspicture}[shift=-1](-.4,0)(6.4,4)
\rput(0,2){
\rput(0,0){\psbezier[linewidth=1pt,showpoints=false](.4,0)(.7,.75)(1.3,.75)(1.6,.4)}
\rput(2,0){\psbezier[linewidth=1pt,showpoints=false](.4,.4)(.7,.75)(1.3,.75)(1.6,0)}
\rput(4,0){\psbezier[linewidth=1pt,showpoints=false](.4,0)(.7,.75)(1.3,.75)(1.6,0)}
\psbezier[linewidth=1pt,showpoints=false](-.4,0)(.2,2.7)(5.8,2.7)(6.4,0)
\rput(4,-2){\psellipse[linewidth=1pt](0,0)(.4,.15)
\psellipse[linewidth=1pt](2,0)(.4,.15)
\multirput(0,0)(.8,0){2}{\multirput(0,0)(2,0){2}{\psline(-.4,0)(-.4,2)}}}
}
\psellipse[linewidth=1pt](0,0)(.4,.15)
\psellipse[linewidth=1pt](2,0)(.4,.15)
\psbezier[linewidth=1pt](.4,0)(.9,.75)(1.1,.75)(1.6,0)
\psbezier[linewidth=1pt](-.4,0)(.6,.7)(.6,1.3)(-.4,2)
\psbezier[linewidth=1pt](.4,2)(.9,1.25)(1.1,1.25)(1.6,1.8)
\psbezier[linewidth=1pt](2.4,0)(1.4,.7)(1.4,1.3)(2.4,1.8)
\psbezier[linewidth=1pt,showpoints=false](1.6,1.8)(1.8,2)(2.2,2)(2.4,1.8)
\psbezier[linewidth=1pt,showpoints=false](1.6,2.4)(1.8,2.2)(2.2,2.2)(2.4,2.4)
\end{pspicture}
+\frac{1}{\TLb^2}
\begin{pspicture}[shift=-1](-.4,0)(6.4,4)
\rput(0,2){
\rput(0,0){\psbezier[linewidth=1pt,showpoints=false](.4,.4)(.7,.75)(1.3,.75)(1.6,.4)}
\rput(2,0){\psbezier[linewidth=1pt,showpoints=false](.4,.4)(.7,.75)(1.3,.75)(1.6,0)}
\rput(4,0){\psbezier[linewidth=1pt,showpoints=false](.4,0)(.7,.75)(1.3,.75)(1.6,0)}
\psbezier[linewidth=1pt,showpoints=false](-.4,.4)(.2,2.7)(5.8,2.7)(6.4,0)
\rput(4,-2){\psellipse[linewidth=1pt](0,0)(.4,.15)
\psellipse[linewidth=1pt](2,0)(.4,.15)
\multirput(0,0)(.8,0){2}{\multirput(0,0)(2,0){2}{\psline(-.4,0)(-.4,2)}}}
}
\psellipse[linewidth=1pt](0,0)(.4,.15)
\psellipse[linewidth=1pt](2,0)(.4,.15)
\psbezier[linewidth=1pt](.4,0)(.9,.75)(1.1,.75)(1.6,0)
\psbezier[linewidth=1pt](-.4,0)(.6,.7)(.6,1.3)(-.4,1.8)
\psbezier[linewidth=1pt](.4,1.8)(.9,1.25)(1.1,1.25)(1.6,1.8)
\psbezier[linewidth=1pt](2.4,0)(1.4,.7)(1.4,1.3)(2.4,1.8)
\psbezier[linewidth=1pt,showpoints=false](1.6,1.8)(1.8,2)(2.2,2)(2.4,1.8)
\psbezier[linewidth=1pt,showpoints=false](1.6,2.4)(1.8,2.2)(2.2,2.2)(2.4,2.4)
\psbezier[linewidth=1pt,showpoints=false](-.4,1.8)(-.2,2)(.2,2)(.4,1.8)
\psbezier[linewidth=1pt,showpoints=false](-.4,2.4)(-.2,2.2)(.2,2.2)(.4,2.4)
\end{pspicture}\nonumber \\[12pt]
&=&
\left(\TLb-2\TLb^{-1}\right)\;\;
\begin{pspicture}[shift=-.5](-.4,0)(6.4,2)
\multirput(0,0)(2,0){4}{\psellipse[linewidth=1pt](0,0)(.4,.15)}
\multirput(0,0)(2,0){3}{\psbezier[linewidth=1pt,showpoints=false](.4,0)(.7,.75)(1.3,.75)(1.6,0)}
\psbezier[linewidth=1pt,showpoints=false](-.4,0)(.2,2.7)(5.8,2.7)(6.4,0)
\end{pspicture}
+\TLb^{-2}
\begin{pspicture}[shift=-.5](-.4,0)(6.4,2)
\multirput(0,0)(2,0){4}{\psellipse[linewidth=1pt](0,0)(.4,.15)}
\multirput(0,0)(4,0){2}{\psbezier[linewidth=1pt,showpoints=false](.4,0)(.7,.75)(1.3,.75)(1.6,0)}
\multirput(0,0)(4,0){2}{\psbezier[linewidth=1pt,showpoints=false](-.4,0)(.2,1.6)(1.8,1.6)(2.4,0)}
\end{pspicture}
\eea
The action of the generators can annihilate certain cabled link states.
For example,
\begin{align}
\begin{pspicture}[shift=-1](-.4,0)(4.4,4)
\rput(0,2){
\rput(0,0){\multirput(0,0)(2,0){2}{\psbezier[linewidth=1pt,showpoints=false](.4,0)(.7,.75)(1.3,.75)(1.6,0)}
\psbezier[linewidth=1pt,showpoints=false](-.4,0)(.2,2.3)(3.8,2.3)(4.4,0)}
}
\multirput(4,0)(.8,0){2}{\psline(-.4,0)(-.4,2)}
\multirput(4,0)(0,2){2}{\psellipse(0,0)(.4,.15)}
\fusedE
\end{pspicture}
&=
\begin{pspicture}[shift=-1](-.4,0)(4.4,4)
\rput(0,2){
\multirput(0,0)(6,0){1}{\multirput(0,0)(2,0){2}{\psbezier[linewidth=1pt,showpoints=false](.4,0)(.7,.75)(1.3,.75)(1.6,0)}
\psbezier[linewidth=1pt,showpoints=false](-.4,0)(.2,2.3)(3.8,2.3)(4.4,0)}
}
\multirput(4,0)(.8,0){2}{\psline(-.4,0)(-.4,2)}
\rput(4,0){\psellipse(0,0)(.4,.15)}
\multirput(0,0)(2,0){2}{\psellipse[linewidth=1pt](0,0)(.4,.15)}
\psellipse(2,2)(.4,.15)
\psbezier[linewidth=1pt,showpoints=false](-.4,0)(.5,1.2)(1.5,1.2)(2.4,0)
\psbezier[linewidth=1pt,showpoints=false](.4,0)(.7,.5)(1.3,.5)(1.6,0)
\psbezier[linewidth=1pt,showpoints=false](-.4,2)(.5,.8)(1.5,.8)(2.4,2)
\psbezier[linewidth=1pt,showpoints=false](.4,2)(.7,1.5)(1.3,1.5)(1.6,2)
\end{pspicture}
=
\begin{pspicture}[shift=-1](-.4,0)(4.4,4)
\multirput(0,0)(2,0){2}{\psellipse[linewidth=1pt](0,0)(.4,.15)}
\rput(0,2){
\rput(0,0){\multirput(0,0)(2,0){2}{\psbezier[linewidth=1pt,showpoints=false](.4,0)(.7,.75)(1.3,.75)(1.6,0)}
\psbezier[linewidth=1pt,showpoints=false](-.4,0)(.2,2.3)(3.8,2.3)(4.4,0)}
}
\multirput(4,0)(.8,0){2}{\psline(-.4,0)(-.4,2)}
\rput(4,0){\psellipse(0,0)(.4,.15)}
\psbezier[linewidth=1pt,showpoints=false](-.4,0)(.5,1.2)(1.5,1.2)(2.4,0)
\psbezier[linewidth=1pt,showpoints=false](.4,0)(.7,.5)(1.3,.5)(1.6,0)
\psbezier[linewidth=1pt,showpoints=false](-.4,2)(.5,.8)(1.5,.8)(2.4,2)
\psbezier[linewidth=1pt,showpoints=false](.4,2)(.7,1.5)(1.3,1.5)(1.6,2)
\end{pspicture}
-\frac{1}{\TLb}
\begin{pspicture}[shift=-1](-.4,0)(4.4,4)
\multirput(0,0)(2,0){2}{\psellipse[linewidth=1pt](0,0)(.4,.15)}
\rput(0,2){\psbezier[linewidth=1pt,showpoints=false](-.4,0)(.2,2.3)(3.8,2.3)(4.4,0)}
\rput(0,2){\psbezier[linewidth=1pt,showpoints=false](.4,0)(.7,.75)(1.3,.75)(1.6,.4)}
\rput(2,2){\psbezier[linewidth=1pt,showpoints=false](.4,.4)(.7,.75)(1.3,.75)(1.6,0)}
\multirput(4,0)(.8,0){2}{\psline(-.4,0)(-.4,2)}
\rput(4,0){\psellipse(0,0)(.4,.15)}
\psbezier[linewidth=1pt,showpoints=false](-.4,0)(.5,1.2)(1.5,1.2)(2.4,0)
\psbezier[linewidth=1pt,showpoints=false](.4,0)(.7,.5)(1.3,.5)(1.6,0)
\psbezier[linewidth=1pt,showpoints=false](-.4,2)(.5,.8)(1.5,.8)(2.4,1.8)
\psbezier[linewidth=1pt,showpoints=false](.4,2)(.7,1.5)(1.3,1.5)(1.6,1.8)
\psbezier[showpoints=false](1.6,2.4)(1.8,2.2)(2.2,2.2)(2.4,2.4)
\psbezier[showpoints=false](1.6,1.8)(1.8,2)(2.2,2)(2.4,1.8)
\end{pspicture}
=
\left(\TLb-\TLb^{-1}\right)
\begin{pspicture}[shift=0](-.4,0)(4.4,3)
\multirput(0,0)(2,0){3}{\psellipse[linewidth=1pt](0,0)(.4,.15)}
\psbezier[linewidth=1pt,showpoints=false](.4,0)(.7,.75)(1.3,.75)(1.6,0)
\psbezier[linewidth=1pt,showpoints=false](-.4,0)(.2,1.6)(1.8,1.6)(2.4,0)
\rput(4,0){\psbezier(-.4,0)(-.3,.6)(.3,.6)(.4,0)}
\end{pspicture}
=0\qquad
\end{align}
since the projector annihilates the single monoid operator \eqref{eq:projProp}.

\subsubsection{Matrix representations}
Matrix representations of the elements in the fused Temperley-Lieb algebra are obtained by acting with the operators on the cabled link states. For example, to find the matrix representation of $E_1$ using the link states on four nodes with no defects, we act with $E_1$ on each of the basis link states, write the result as a linear combination of the basis link states and put the coefficients in the columns of a matrix\\[10pt]
$N=4$: Neveu-Schwarz, no defects ($k=0$)
\be
E_1\!=\!E_3\!=\!\begin{pmatrix}
 \beta _2 & 1 & \frac{\beta _2}{\beta } \\
 0 & 0 & 0 \\
 0 & 0 & 0 \\
\end{pmatrix}\!\!, \;\;
E_2\!=\!\begin{pmatrix}
 0 & 0 & 0 \\
 1 & \beta _2 & \frac{\beta _2}{\beta } \\
 0 & 0 & 0 \\
 \end{pmatrix}\!\!, \;\;
X_1\!=\!X_3\!=\!\begin{pmatrix}
 \frac{\beta _2}{\beta } & 0 & \frac{1}{\beta ^2} \\
 0 & 0 & 0 \\
 0 & 1 & \frac{\beta _3}{\beta ^2} \\
\end{pmatrix}\!\!, \;\;
X_2\!=\!\begin{pmatrix}
\frac{\beta _2}{\beta } & 0 & \frac{1}{\beta ^2} \\
 0 & 0 & 0 \\
 0 & 1 & \frac{\beta _3}{\beta ^2} \\
\end{pmatrix}
\ee
$N=3$: Ramond, one defect ($k=0$)
\be
E_1\!=\!\!\begin{pmatrix}
0 & 0 & 0 & 0 \\
 0 & \beta _2 & 1 & \frac{\beta _2}{\beta } \\
 0 & 0 & 0 & 0 \\
 0 & 0 & 0 & 0 \\
\end{pmatrix}\!\!, \;\;
E_2\!=\!\begin{pmatrix}
 0 & 0 & 0 & 0 \\
 0 & 0 & 0 & 0 \\
 0 & 1 & \beta _2 & \frac{\beta _2}{\beta } \\
 0 & 0 & 0 & 0 \\ \end{pmatrix}\!\!, \;\;
X_1\!=\!\begin{pmatrix}
\frac{\beta _3}{\beta ^2} & 0 & -\frac{1}{\beta } & 0 \\
 0 & \frac{\beta _2}{\beta } & 0 & \frac{1}{\beta ^2} \\
 0 & 0 & 0 & 0 \\
 0 & 0 & 1 & \frac{\beta _3}{\beta ^2} \\
\end{pmatrix}\!\!, \;\;
X_2\!=\!\begin{pmatrix}
 \frac{\beta _3}{\beta ^2} & -\frac{1}{\beta } & 0 & 0 \\
 0 & 0 & 0 & 0 \\
 0 & 0 & \frac{\beta _2}{\beta } & \frac{1}{\beta ^2} \\
 0 & 1 & 0 & \frac{\beta _3}{\beta ^2} \\\end{pmatrix}
\ee
For modest values of the system size $N$, the action of the generators can be implemented in Mathematica~\cite{Wolfram} using the elementary TL algebra to obtain matrix representations of the operators $I$, $E_j$, $X_j$ and hence the double row transfer matrices. In the case of an $(r,s,\ell)$ boundary condition on one side of the strip and the vacuum $(1,1,0)$ on the other side, these matrices can be directly diagonalized to obtain finite-size spectra. In such cases, numerical conformal spectra are obtained by extrapolating the finite-size corrections to the continuum scaling limit. For non-vacuum boundary conditions on both sides of the strip, the transfer matrices need not be diagonalizable and Jordan blocks can occur.

\section{Logarithmic Superconformal Minimal Lattice Models ${\cal LSM}(p,p')$}
\label{Sec:LSM}

In this section, we regard the ${\cal LSM}(p,p')$ models as exactly solvable lattice models~\cite{BaxBook} and discuss their Yang-Baxter integrability. We also discuss duality and the combinatorial structure of finitized Kac characters for the case $n=2$ with $r=1$.

\subsection{Yang-Baxter integrability}

\subsubsection{Generalized Yang-Baxter equations}

The elementary $n=1$ Yang-Baxter Equation (YBE) is expressed~\cite{BaxBook} in the equivalent forms
\begin{align}
\begin{array}{rcl}
X_j(u)X_{j+1}(u+v)X_j(v)&=&X_{j+1}(v)X_j(u+v)X_{j+1}(u)\\[14pt]
\psset{unit=.375cm}
\begin{pspicture}[shift=-1.8](0,0)(7,3.5)
\pspolygon[linewidth=1pt,linecolor=black,fillstyle=solid,fillcolor=lightlightblue](0,2)(2,0)(5,0)(7,2)(5,4)(2,4)(0,2)
\pspolygon[linewidth=1pt,linecolor=black,fillstyle=solid,fillcolor=lightlightblue](0,2)(3,2)(5,0)(7,2)(5,4)(3,2)
\rput(2.5,1){$u$}
\rput(2.5,3){$v$}
\rput(5,2){$v\!-\!u$}
\psarc[linewidth=1pt,linecolor=red](2,0){.3}{0}{135}
\psarc[linewidth=1pt,linecolor=red](0,2){.5}{0}{45}
\psarc[linewidth=1pt,linecolor=red](3,2){.4}{-45}{45}
\end{pspicture}\ &=&\ 
\begin{pspicture}[shift=-1.8](0,0)(7,3.5)
\pspolygon[linewidth=1pt,linecolor=black,fillstyle=solid,fillcolor=lightlightblue](0,2)(2,0)(5,0)(7,2)(5,4)(2,4)(0,2)
\pspolygon[linewidth=1pt,linecolor=black,fillstyle=solid,fillcolor=lightlightblue](7,2)(4,2)(2,0)(0,2)(2,4)(4,2)
\rput(4.5,1){$v$}
\rput(4.5,3){$u$}
\rput(2,2){$v\!-\!u$}
\psarc[linewidth=1pt,linecolor=red](0,2){.4}{-45}{45}
\psarc[linewidth=1pt,linecolor=red](2,0){.5}{0}{45}
\psarc[linewidth=1pt,linecolor=red](4,2){.3}{0}{135}
\end{pspicture}
\end{array}
\label{YBE}
\end{align}
In the diagrammatic representation, the square faces are distorted to rhombi. 
This YBE is satisfied by the elementary face transfer operators (\ref{eq:face1}).
Diagrammatically, starting on the left hand side, the diamond shaped face on the right is pulled through to the left with the effect of interchanging the spectral parameters $u,v$ of the other two faces.

The $2\times 2$ face transfer operators are defined by
\bea
\face_j(u)=p_{j-1}p_{j+1}X_j(u-\lambda)X_{j-1}(u)X_{j+1}(u)X_j(u+\lambda)p_{j-1}p_{j+1}
\eea
where increasing the $j$ in $\face_j(u)$ by one unit equates to increasing the $j$ in $X_j(u)$ of the underlying elementary lattice by two units.
The unprojected $2\times 2$ fused YBE takes the form
\begin{align}
\begin{array}{rcl}
\face_j(u)\face_{j+1}(u+v)\face_j(v)&=&\face_{j+1}(v)\face_j(u+v)\face_{j+1}(u)\\[14pt]
\psset{unit=.7cm}
\begin{pspicture}[shift=-1.8](0,0)(7,4)
\pspolygon[linewidth=1pt,linecolor=black,fillstyle=solid,fillcolor=lightlightblue](0,2)(2,0)(5,0)(7,2)(5,4)(2,4)(0,2)
\pspolygon[linewidth=1pt,linecolor=black,fillstyle=solid,fillcolor=lightlightblue](0,2)(3,2)(5,0)(7,2)(5,4)(3,2)
\psline[linewidth=1pt,linecolor=black](1,1)(4,1)(6,3)
\psline[linewidth=1pt,linecolor=black](1,3)(4,3)(6,1)
\psline[linewidth=1pt,linecolor=black](3.5,0)(1.5,2)(3.5,4)
\rput(2.2,.5){\small$u\!-\!\lambda$}
\rput(3.7,.5){\small$u$}
\rput(1.2,1.5){\small$u$}
\rput(2.7,1.5){\small$u\!+\!\lambda$}
\rput(1.2,2.5){\small$v\!-\!\lambda$}
\rput(2.7,2.5){\small$v$}
\rput(2.2,3.5){\small$v$}
\rput(3.7,3.5){\small$v\!+\!\lambda$}
\rput(5,1){\small$w$}
\rput(4,2){\small$w\!-\!\lambda$}
\rput(6,2){\small$w\!+\!\lambda$}
\rput(5,3){\small$w$}
\psarc[linewidth=1pt,linecolor=red](2,0){.15}{0}{135}
\psarc[linewidth=1pt,linecolor=red](0,2){.25}{0}{45}
\psarc[linewidth=1pt,linecolor=red](3,2){.2}{-45}{45}
\end{pspicture}\ &=&\ 
\psset{unit=.7cm}
\begin{pspicture}[shift=-1.8](0,0)(7,4)
\pspolygon[linewidth=1pt,linecolor=black,fillstyle=solid,fillcolor=lightlightblue](0,2)(2,0)(5,0)(7,2)(5,4)(2,4)(0,2)
\pspolygon[linewidth=1pt,linecolor=black,fillstyle=solid,fillcolor=lightlightblue](7,2)(4,2)(2,0)(0,2)(2,4)(4,2)
\psline[linewidth=1pt,linecolor=black](1,1)(3,3)(6,3)
\psline[linewidth=1pt,linecolor=black](1,3)(3,1)(6,1)
\psline[linewidth=1pt,linecolor=black](3.5,0)(5.5,2)(3.5,4)
\rput(2,1){\small$w$}
\rput(1,2){\small$w\!-\!\lambda$}
\rput(3,2){\small$w\!+\!\lambda$}
\rput(2,3){\small$w$}
\rput(4.2,2.5){\small$u\!-\!\lambda$}
\rput(5.7,2.5){\small$u$}
\rput(3.2,3.5){\small$u$}
\rput(4.7,3.5){\small$u\!+\!\lambda$}
\rput(3.2,.5){\small$v\!-\!\lambda$}
\rput(4.7,.5){\small$v$}
\rput(4.2,1.5){\small$v$}
\rput(5.7,1.5){\small$v\!+\!\lambda$}
\psarc[linewidth=1pt,linecolor=red](0,2){.2}{-45}{45}
\psarc[linewidth=1pt,linecolor=red](2,0){.25}{0}{45}
\psarc[linewidth=1pt,linecolor=red](4,2){.15}{0}{135}
\end{pspicture}
\end{array}
\label{fusedYBE}
\end{align}
where $w=v-u$. The fused YBE is readily proved, starting on the left hand side, by pulling through to the left each of the four elementary faces of the diamond shaped fused face on the right and interchanging spectral parameters appropriately. In this way, the elementary YBE is applied 8 times. Finally, the fusion projector is applied on the six sides of each hexagon and pushed through to obtain the projected $2\times 2$ fused YBE. The scalar normalization factors $\eta_{2,2}(u)$ can be included because they trivially cancel out of the YBE. The same proof easily extends to $n\times n$ fused YBE. Indeed, the faces need not be square. A generalized YBE holds as long as the dimensions on the left and right sides are compatible with the push-through properties. Everything in this section holds generally for the RSOS, vertex and loop representations of TL. In particular, the projected $2\times 2$ fused YBE is satisfied by the face transfer operators $\face_j(u)$ given by (\ref{eq:fusedFace}) or (\ref{symmFace}).

\subsubsection{Boundary Yang-Baxter equations}
\label{BYBE}
To work with lattice models on the strip in the presence of boundaries, we follow \cite{BPO96} and introduce left and right boundary  triangles and the fused crossing parameter $\mu$. Since the left and right boundary triangles and their Boundary Yang-Baxter Equations (BYBEs) are simply related by a reflection in the vertical with the spectral parameter $u$ replaced with $\mu-u$, we only consider right triangles. For the logarithmic superconformal minimal models, the boundary conditions are labelled by the quantum numbers $(r,s,\ell)$. In this paper, we only consider boundary conditions of $s$-type with quantum number $r=1$. More general $r$-type boundary conditions are considered in \cite{Ong}. 
Let us start with the elementary $n=1$ logarithmic minimal models with boundary $K$ operator acting on the last string
\bea
\psset{unit=.6cm}
K_N(u)=
\begin{pspicture}[shift=-.9](1,2)
\psline[linewidth=2pt,linecolor=blue](.5,0)(.5,2)
\pspolygon[linewidth=1pt,fillstyle=solid,fillcolor=lightlightblue](1,0)(1,2)(0,1)(1,0)
\rput(.6,1){\small$u$}
\end{pspicture}
\eea
To ensure integrability, these triangles must satisfy the BYBE
\begin{align}
X_{N-1}(u\!-\!v)K_N(u)X_{N-1}(\lambda\!-\!\mu\!+\!u\!+\!v)K_N(v)
&=K_N(v)X_{N-1}(\lambda\!-\!\mu\!+\!u\!+\!v)K_N(u)X_{N-1}(u\!-\!v)
\label{elemBYBE}
\end{align}
Following \cite{BPO96}, the general $n\times n$ fused transfer matrices satisfy the crossing symmetry $\vec D^{n,n}(u)=\vec D^{n,n}(\mu-(n\!-\!1)\lambda-u)$. To ensure the crossing symmetry $\vec D^{n,n}(u)=\vec D^{n,n}(\lambda-u)$, we need to choose $\mu=n\lambda$ in these equations. For the superconformal loagrithmic minimal models, we choose $\mu=2\lambda$ so that $\vec D(u)=\vec D(\lambda-u)$. Using the commutation relation $[X_j(u),X_j(v)]=0$, it is seen that a simple (vacuum) solution for $n=1$ is given by 
\bea
\psset{unit=.6cm}
K_N(u)=I=
\begin{pspicture}[shift=-.9](1,2)
\pspolygon[linewidth=1pt,fillstyle=solid,fillcolor=lightlightblue](1,0)(1,2)(0,1)(1,0)
\psline[linewidth=1.25pt,linecolor=blue](.5,0)(.5,2)
\end{pspicture}\ =\ 
\begin{pspicture}[shift=-.9](1,2)
\pspolygon[linewidth=1pt,fillstyle=solid,fillcolor=lightlightblue](1,0)(1,2)(0,1)(1,0)
\psarc[linewidth=1.25pt,linecolor=blue](0,1){.65}{-45}{45}
\end{pspicture}
\label{bdytri}
\eea
This boundary condition is conjugate to the identity operator labelled by $(r,s)=(1,1)$.

The fused $2\times 2$ boundary triangles are defined by
\bea
\faceK_N(u)=p_{N-1}K_N(u)X_{N-1}(2u)K_N(u+\lambda)p_{N-1}=\ 
\psset{unit=1cm}
\begin{pspicture}[shift=-.9](1,2)
\psline[linewidth=1.25pt,linecolor=blue](.25,0)(.25,2)
\psline[linewidth=1.25pt,linecolor=blue](.75,0)(.75,2)
\pspolygon[linewidth=1pt,fillstyle=solid,fillcolor=lightlightblue](1,0)(1,2)(0,1)(1,0)
\psline[linewidth=1pt,linecolor=black](.5,.5)(1,1)(.5,1.5)
\rput(.5,1){\small $2u$}
\rput(.82,.5){\small $u$}
\rput(1.13,1.5){\small $u+\!\lambda$}
\psellipse[linewidth=1.25pt,linecolor=blue](.5,0)(.25,.1)
\psellipse[linewidth=1.25pt,linecolor=blue](.5,2)(.25,.1)
\psarc[linewidth=1pt,linecolor=red](.5,.5){.125}{45}{135}
\end{pspicture}
\label{twobdytri}
\eea
The fused BYBE 
\bea
\face_{N-1}(u-v)\faceK_N(u)\face_{N-1}(u+v)\faceK_N(v)=\faceK_N(v)\face_{N-1}(u+v)\faceK_N(u)\face_{N-1}(u-v)
\label{fusedBYBE}
\eea
follows by straightforward algebra, using the elementary YBE (\ref{YBE}) and the elementary BYBE (\ref{elemBYBE}) with $\mu=2\lambda$. 
This relation holds without the projectors so the projectors can be put in at the end. 
Using the commutation relation $[\face_j(u),\face_j(v)]=0$, it is seen that (after normalization) a simple solution is given by 
\bea
\psset{unit=1cm}
\faceK_N(u)=p_{N-1}=
\begin{pspicture}[shift=-.9](1,2)
\pspolygon[linewidth=1pt,fillstyle=solid,fillcolor=lightlightblue](1,0)(1,2)(0,1)(1,0)
\psline[linewidth=1pt,linecolor=black](.5,.5)(1,1)(.5,1.5)
\psline[linewidth=1.25pt,linecolor=blue](.25,0)(.25,2)
\psline[linewidth=1.25pt,linecolor=blue](.75,0)(.75,2)
\psellipse[linewidth=1.25pt,linecolor=blue](.5,0)(.265,.1)
\psellipse[linewidth=1.25pt,linecolor=blue](.5,2)(.265,.1)
\end{pspicture}\ =\ 
\begin{pspicture}[shift=-.9](1,2)
\pspolygon[linewidth=1pt,fillstyle=solid,fillcolor=lightlightblue](1,0)(1,2)(0,1)(1,0)
\psarc[linewidth=1.25pt,linecolor=blue](0,1){.8}{-45}{45}
\psarc[linewidth=1.25pt,linecolor=blue](0,1){.5}{-45}{45}
\psellipse[linewidth=1.25pt,linecolor=blue](.65,1)(.15,.1)
\end{pspicture}
\label{fusedVac}
\eea
This vacuum solution is obtained, up to a scalar, by substituting the elementary vacuum solution (\ref{bdytri}) into (\ref{twobdytri}) and observing that the projector acting on the face $X_{N-1}(2u)$ gives $s_1(-2u)p_{N-1}$. Further solutions to the BYBE are obtained in Section~\ref{commutingDTM} by acting on the vacuum solution (\ref{fusedVac}) with integrable seams to obtain {\em dressed}\/ solutions.

\subsection{Commuting double row transfer matrices}
\label{commutingDTM}

In this section, we construct commuting double row transfer matrices in the Neveu-Schwarz and Ramond sectors. For each sector, we identify the appropriate integrable boundary conditions by the action of integrable seams on the vacuum solution (\ref{fusedVac}) of the BYBE. 
To complete the definition of these transfer matrices we specify, referring to Section~\ref{LinkStatesMatrixReps}, the vector space of link states on which they act. Following the methods of \cite{BPO96}, the  YBE (\ref{fusedYBE}), BYBE (\ref{fusedBYBE}) and inversion relation (\ref{InversionReln}) together suffice to establish commuting transfer matrices and hence integrability at least for boundary conditions of $r=1$ type given by (\ref{fusedVac}).

For the $2\times2$ fused face operators \eqref{eq:fusedFace}, 
we introduce the diagrammatic representations
\be
\psset{unit=.9cm}
\face_j(u)\;=\;
\begin{pspicture}[shift=-.4](0,0)(1,1)
\facegrid{(0,0)}{(1,1)}
\rput(.5,.5){$u$}
\rput(.5,.75){\tiny$(2,2)$}
\psarc[linecolor=red](0,0){.2}{0}{90}
\end{pspicture} \;=\; 
[\eta_{2,2}(u)]^{-1}\,
\begin{pspicture}[shift=-1.27](-.4,-.4)(2.4,2.4)
\multiput(0,0)(0,1){2}{\psline[linewidth=1.5pt,linecolor=blue](-.2,.5)(2.2,.5)}
\psellipse[linewidth=1.5pt,linecolor=blue](-.2,1)(.1,.5)
\psellipse[linewidth=1.5pt,linecolor=blue](2.2,1)(.1,.5)
\multiput(0,0)(1,0){2}{\psline[linewidth=1.5pt,linecolor=blue](.5,-.2)(.5,2.2)}
\psellipse[linewidth=1.5pt,linecolor=blue](1,-.2)(.5,.1)
\psellipse[linewidth=1.5pt,linecolor=blue](1,2.2)(.5,.1)
\facegrid{(0,0)}{(2,2)}
\psarc[linecolor=red](0,0){.2}{0}{90}
\rput(.5,.5){\small $u\!-\!\lambda$}
\rput(1.5,.5){\small $u$}
\rput(.5,1.5){\small $u$}
\rput(1.5,1.5){\small $u\!+\!\lambda$}
\end{pspicture}
\ee
These face operators satisfy the inversion relation 
\bea
\face_j(u)\face_j(-u)=\frac{s(\lambda-u)s(\lambda+u)s(2\lambda-u)s(2\lambda+u)}{s(2\lambda)^2}\,I
\label{InversionReln}
\eea
and crossing symmetry
\be
\face_j(u)\;=\;\face_j(\lambda-u),\qquad\qquad\psset{unit=.9cm}
\begin{pspicture}[shift=-.4](0,0)(1,1)
\facegrid{(0,0)}{(1,1)}
\rput(.5,.5){$u$}
\rput(.5,.75){\tiny$(2,2)$}
\psarc[linecolor=red](0,0){.2}{0}{90}
\end{pspicture} 
\ =\ 
\begin{pspicture}[shift=-.4](0,0)(1,1)
\facegrid{(0,0)}{(1,1)}
\rput(.5,.5){\small$\lambda\!-\!u$}
\rput(.5,.75){\tiny$(2,2)$}
\psarc[linecolor=red](1,0){.2}{90}{180}
\end{pspicture}
\label{Crossing}
\ee

\subsubsection{Neveu-Schwarz sector $(r,s)=(1,1)$}

In the Neveu-Schwarz sector, we define double row transfer matrices with $N$ columns diagrammatically by
\be
\drtm (u) = 
\psset{unit=1.4cm}
\begin{pspicture}[shift=-1.27](0.2,-.4)(6.8,2.4)
\psline[linecolor=red,linewidth=1.5pt,linestyle=dashed](1,-.4)(1,2.4)
\psline[linecolor=red,linewidth=1.5pt,linestyle=dashed](6,-.4)(6,2.4)
\pspolygon[fillstyle=solid,fillcolor=lightlightblue](.25,0)(1,1)(.25,2)
\pspolygon[fillstyle=solid,fillcolor=lightlightblue](6.75,0)(6,1)(6.75,2)
\rput(1,0){
	\multirput(0,0)(3,0){2}{\multirput(0,0)(1,0){2}{
	\psline[linecolor=blue,linewidth=1.5pt](.4,-.4)(.4,2.4) \psline[linecolor=blue,linewidth=1.5pt](.6,-.4)(.6,2.4)
	\multirput(0,0)(0,2.4){2}{\psellipse[linecolor=blue,linewidth=1.5pt](.5,-.2)(.1,.065)}
	}}
	\facegrid{(0,0)}{(5,2)}
	\multirput(0,0)(3,0){2}{\multirput(0,0)(0,1){2}{\multirput(0,0)(1,0){2}{
		\psarc[linecolor=red](0,0){.125}{0}{90}
		\rput(.5,.75){\tiny$(2,2)$}
	}}}
	\multirput(0,0)(3,0){2}{\multirput(.5,1.5)(1,0){2}{\small $\lambda\!-\!u$}}
	\multirput(0,0)(3,0){2}{\multirput(.5,.5)(1,0){2}{\small $u$}}
	\multirput(2.5,.5)(0,1){2}{\ldots}
	\psarc[linecolor=blue,linewidth=1.5pt](0,1){.6}{90}{270}
	\psarc[linecolor=blue,linewidth=1.5pt](0,1){.4}{90}{270}
	\psarc[linecolor=blue,linewidth=1.5pt](5,1){.6}{-90}{90}
	\psarc[linecolor=blue,linewidth=1.5pt](5,1){.4}{-90}{90}
	\multirput(0,0)(6,0){2}{\psellipse[linecolor=blue,linewidth=1.5pt](-.5,1)(.1,.065)}
}	
\end{pspicture}
\label{eq:2fdrtm}
\ee
These transfer matrices act on either the vector space of link states ${\cal V}_{1,1,0}^{(N)}$ or ${\cal V}_{1,1,2}^{(N)}$ depending on whether $N$ is even or odd respectively.

\subsubsection{Ramond sector $(r,s)=(1,2)$}
To move to the Ramond sector with $r=1$ and $s=2$, we change the boundary conditions by adding a seam which consists of a single defect in the bulk which closes on the boundary. By construction, the action of such seams on the vacuum solution (\ref{fusedVac}) of the BYBE produces new {\it dressed}\/ solutions~\cite{Ong} to the BYBE. By convention, we place this (topological) defect seam on the right but by the generalized YBE it can be propagated to any position along the double row. We thus define the double row transfer matrices in the Ramond sector diagrammatically by
\be
\drtm (u) = 
\psset{unit=1.5cm}
\begin{pspicture}[shift=-1.27](0.2,-.4)(8.2,2.4)
\psline[linecolor=red,linewidth=1.5pt,linestyle=dashed](1,-.4)(1,2.4)
\psline[linecolor=red,linewidth=1.5pt,linestyle=dashed](6.9,-.4)(6.9,2.4)
\pspolygon[fillstyle=solid,fillcolor=lightlightblue](.25,0)(1,1)(.25,2)
\pspolygon[fillstyle=solid,fillcolor=lightlightblue](8.24,0)(7.4,1)(8.24,2)
\rput(1,0){
	\multirput(0,0)(3,0){2}{\multirput(0,0)(1,0){2}{
		\psline[linecolor=blue,linewidth=1.5pt](.4,-.4)(.4,2.4) 
		\psline[linecolor=blue,linewidth=1.5pt](.6,-.4)(.6,2.4)
		\multirput(0,0)(0,2.4){2}{\psellipse[linecolor=blue,linewidth=1.5pt](.5,-.2)(.1,.065)}
	}}
	\psline[linecolor=blue,linewidth=1.5pt](5.65,-.4)(5.65,2.4) 
	\multirput(0,0)(0,1){2}{ 
		\psline[linecolor=blue,linewidth=1.5pt](5,.4)(5.5,.4)
		\psline[linecolor=blue,linewidth=1.5pt](5,.6)(5.5,.6)
		\psellipse[linecolor=blue,linewidth=1.5pt](5.2,.5)(.065,.1)
	}
	\facegrid{(0,0)}{(5,2)}
	\multirput(0,0)(3,0){2}{\multirput(0,0)(0,1){2}{\multirput(0,0)(1,0){2}{
		\psarc[linecolor=red](0,0){.125}{0}{90}
		\rput(.5,.75){\tiny$(2,2)$}
	}}}
	\multirput(0,0)(3,0){2}{\multirput(.5,1.5)(1,0){2}{\small $\lambda\!-\!u$}}
	\multirput(0,0)(3,0){2}{\multirput(.5,.5)(1,0){2}{\small $u$}}
	\multirput(2.5,.5)(0,1){2}{\ldots}
	\psarc[linecolor=blue,linewidth=1.5pt](0,1){.6}{90}{270}
	\psarc[linecolor=blue,linewidth=1.5pt](0,1){.4}{90}{270}
	\psarc[linecolor=blue,linewidth=1.5pt](6.4,1){.75}{-90}{90}
	\psarc[linecolor=blue,linewidth=1.5pt](6.4,1){.26}{-90}{90}
	\multirput(0,0)(6,0){1}{\psellipse[linecolor=blue,linewidth=1.5pt](-.5,1)(.1,.065)}
	\psellipse[linecolor=blue,linewidth=1.5pt](6.9,1)(.25,.06)
	\pspolygon[fillstyle=solid,fillcolor=lightlightblue,linewidth=0pt](5.4,0)(6.4,0)(6.4,2)(5.4,2)
	\psline(5.9,0)(5.9,2)
	\multirput(0,0)(0,.5){4}{\psline[linecolor=blue,linewidth=1.5pt](5.9,.25)(6.4,.25)}
	\psline[linecolor=lightlightblue,linewidth=5pt](6.15,0)(6.15,2)
	\psline[linecolor=blue,linewidth=1.5pt](6.15,-.4)(6.15,2.4)
	\multirput(0,0)(0,.5){5}{\psline(5.4,0)(6.4,0)}
	\rput(5.625,.25){\tiny $u\!+\!\xi$}
	\rput(5.625,.825){\tiny $u\!+\!\lambda$}\rput(5.625,.675){\tiny $+\xi$}
	\rput(5.625,1.325){\tiny $\lambda\!-\!u$}\rput(5.625,1.175){\tiny $+\xi$}
	\rput(5.625,1.825){\tiny $\,2\lambda\!-\!u$}\rput(5.625,1.675){\tiny $+\xi$}
	\multirput(0,0)(0,.5){4}{\psarc[linecolor=red](5.4,0){.1}{0}{90}}
	}		
\end{pspicture}
\label{eq:2fdrtm1def}
\ee

The system size $N$ for this transfer matrix is the number of columns in the bulk excluding the Ramond seam. Notice that the single string in the boundary on the right is a spectator since it acts as the identity. 
These transfer matrices act on the vector space of link states ${\cal V}_{1,2,1}^{(N)}$ for $N$ even or odd. The column inhomogeneity $\xi$ is a boundary thermodynamic field. 
If we allow both columns of the Ramond seam to depend on the spectral parameters (with the spectral parameters decreasing by $\lambda$ with each step to the left), fuse the 2 defects and set $\xi=0$, we see that this has the effect of adding an extra column of faces to the double row transfer matrix thus increasing $N$ by 1 and taking us back to the NS sector with the opposite parity of $N$. The width of the fused part of the Ramond seam that is spectral parameter dependent is the quantum number $\ell\in \{0,1,2\}=A_3$ associated to the boundary condition on the right. Since $\ell\in A_3$, it is perhaps not surprising that the Ramond seams and their parfermionic indices satisfy the Ising fusion algebra with $0$ being the identity and $1$ the fundamental. Specifically, $\ell$ satisfies the fusion rules $0\otimes \ell=\ell$, $2\otimes 2=0$, $1\otimes 2=1$ and $1\otimes 1=0\oplus 2$ but where the sum may not be a direct sum. The superconformal picture is generated under the $\mathbb{Z}_2$ orbifolding~\cite{FendGins} of this $A_3$ diagram. 
In the general $n\ge 2$ case, the width of this fused seam is $\ell\in \{0,1,2,\ldots,n\}=A_{n+1}$. In this case, the seams and $\ell$ satisfy the $A_{n+1}$ fusion algebra. 
The parameter $\xi$ associated with an $\ell$-type seam plays a similar role to that of the column inhomogeneity $\xi$ for $r$-type boundary seams. To distinguish the two, we will usually work with the equivalent field $\eta=\lambda-\xi\in(0,\pi)$. Numerically, the continuum scaling limit is found to be independent of the value chosen for $\eta$ provided $\eta$ is restricted to certain subintervals of $(0,\pi)$. In contrast, taking $\eta$ to have an imaginary part, suitably scaled with $\log N$, will induce a boundary renormalization group flow between different conformal boundary conditions in the continuum scaling limit~\cite{FPR}.

\subsubsection{Neveu Schwarz sectors $(1,s)$, $s\ge 3$}

To move to sectors with $r=1$, $s\ge 3$ and $s$ odd, we increase the number of defects $d=s-1$ in the transfer matrices and link states by adding an $s$-type seam on the right side in accord with (\ref{defectnumber}). The action of the $s$-type seams on the vacuum solution (\ref{fusedVac}) of the BYBE produces new solutions to the BYBE known as $s$-type boundary conditions. The $s$-type seams are obtained by fusing face operators to form a seam with spectral parameters, introducing an inhomogeneity $\xi$ and taking the braid limit $\xi\to \pm i\infty$. 
As in Section~\ref{LinkStatesMatrixReps}, the quantum number $s-1$ is the number of single strands, in the link state, that connect the bulk to the right boundary. These defects then just propagate along the right boundary. In this way, the number of defects in the bulk is controlled by the boundary condition. 
The double row transfer matrices in the NS sector for $s=3$ are defined diagrammatically by
\be
\drtm (u) = 
\psset{unit=1.4cm}
\begin{pspicture}[shift=-1.27](0.2,-.4)(6.8,2.4)
\psline[linecolor=red,linewidth=1.5pt,linestyle=dashed](1,-.4)(1,2.4)
\psline[linecolor=red,linewidth=1.5pt,linestyle=dashed](5,-.4)(5,2.4)
\pspolygon[fillstyle=solid,fillcolor=lightlightblue](.25,0)(1,1)(.25,2)
\pspolygon[fillstyle=solid,fillcolor=lightlightblue](6.75,0)(6,1)(6.75,2)
\rput(1,0){
	\multirput(0,0)(3,0){2}{\multirput(0,0)(1,0){1}{
	\psline[linecolor=blue,linewidth=1.5pt](.4,-.4)(.4,2.4) \psline[linecolor=blue,linewidth=1.5pt](.6,-.4)(.6,2.4)
	\multirput(0,0)(0,2.4){2}{\psellipse[linecolor=blue,linewidth=1.5pt](.5,-.2)(.1,.065)}
	}}
	\multirput(1,0)(3,0){1}{\multirput(0,0)(1,0){1}{
	\psline[linecolor=blue,linewidth=1.5pt](.4,-.4)(.4,2.4) \psline[linecolor=blue,linewidth=1.5pt](.6,-.4)(.6,2.4)
	\multirput(0,0)(0,2.4){2}{\psellipse[linecolor=blue,linewidth=1.5pt](.5,-.2)(.1,.065)}
	}}
	\facegrid{(0,0)}{(5,2)}
	\multirput(0,0)(3,0){2}{\multirput(0,0)(0,1){2}{\multirput(0,0)(1,0){1}{
		\psarc[linecolor=red](0,0){.125}{0}{90}
		\rput(.5,.75){\tiny$(2,2)$}
	}}}
         \multirput(1,0)(3,0){1}{\multirput(0,0)(0,1){2}{\multirput(0,0)(1,0){1}{
		\psarc[linecolor=red](0,0){.125}{0}{90}
		\rput(.5,.75){\tiny$(2,2)$}
	}}}
	\multirput(0,0)(3,0){2}{\multirput(.5,1.5)(1,0){1}{\small $\lambda\!-\!u$}}
	\multirput(0,0)(3,0){2}{\multirput(.5,.5)(1,0){1}{\small $u$}}
	\multirput(2.5,.5)(0,1){2}{\ldots}
	\multirput(1,0)(3,0){1}{\multirput(.5,1.5)(1,0){1}{\small $\lambda\!-\!u$}}
	\multirput(1,0)(3,0){1}{\multirput(.5,.5)(1,0){1}{\small $u$}}
	\psarc[linecolor=blue,linewidth=1.5pt](0,1){.6}{90}{270}
	\psarc[linecolor=blue,linewidth=1.5pt](0,1){.4}{90}{270}
	\psarc[linecolor=blue,linewidth=1.5pt](5,1){.6}{-90}{90}
	\psarc[linecolor=blue,linewidth=1.5pt](5,1){.4}{-90}{90}
	\multirput(0,0)(6,0){2}{\psellipse[linecolor=blue,linewidth=1.5pt](-.5,1)(.1,.065)}
	\multirput(4,0)(3,0){1}{\multirput(0,0)(1,0){1}{
	\psline[linecolor=blue,linewidth=1.5pt](.4,-.4)(.4,2.4) \psline[linecolor=blue,linewidth=1.5pt](.6,-.4)(.6,2.4)
	\multirput(0,0)(0,2.4){2}{\psellipse[linecolor=blue,linewidth=1.5pt](.5,-.2)(.1,.065)}
	}}
	\multirput(4,.4)(0,.2){2}{\psline[linecolor=blue,linewidth=1.5pt](0,0)(1,0)}
	\multirput(4,1.4)(0,.2){2}{\psline[linecolor=blue,linewidth=1.5pt](0,0)(1,0)}
	\multirput(4.4,.0)(.2,0){2}{\psline[linecolor=lightlightblue,linewidth=4.5pt](0,0)(0,2)}
	\multirput(4.4,.0)(.2,0){2}{\psline[linecolor=blue,linewidth=1.5pt](0,0)(0,2)}
	\psline[linewidth=1pt,linecolor=black](4,1)(5,1)
}	
\end{pspicture}
\label{s3}
\ee
These transfer matrices act on either the vector space of link states ${\cal V}_{1,3,0}^{(N)}$ or ${\cal V}_{1,3,2}^{(N)}$ depending on whether the number of columns $N$ in the bulk is even or odd respectively. The cabled defect enters the bulk through the link state and propagates through the bulk system. Transfer matrices with more defects in the Neveu-Schwarz and Ramond sectors are defined similarly.

\subsection{Hamiltonian Limit}
\label{appHam}

In this section, we derive expressions for the quantum Hamiltonians associated with the matrix representations $\vec D(u)$ of the double row transfer tangles of the previous section. For the numerical calculations of spectra, it is more efficient to work with the Hamiltonians rather than the double row transfer matrices.

\subsubsection{Neveu-Schwarz sector $(r,s)=(1,1)$}

Let us define a normalised double row transfer matrix 
\be
 \vec d(u)=\begin{cases}
\beta_2^{-1} \vec D(u),&\ \beta_2\neq0\\[4pt]
\frac{3\sin 2\lambda}{2\sin 3u}\,\vec D(u),&\ \beta_2=0
 \end{cases}
\label{db}
\ee
such that
\be
 \lim_{u\to0}\vec d(u)=I,\qquad \vec d(u)=\vec d(\lambda-u)
\ee
for all $\beta_2$.
For small $u$, the double row transfer matrices then admit a series expansion of the form
\be
 \vec d(u) = I-\frac{2 u}{\sin 2\lambda} \ham + O(u^2) 
\label{eq:drtmExpansion}
\ee
The quantum Hamiltonian ${\cal H}$ is thus given by the logarithmic derivative of the double row transfer matrices
\be
\ham = \left. -\frac{\sin 2\lambda}{2} \frac{\partial}{\partial u} \log \vec d(u)\right|_{u=0},\qquad \lambda\in(0,\pi), \quad \lambda\ne \tfrac{\pi}{2}
\ee

To obtain the Hamiltonian, let us assume initially that $\beta_2\neq0$ and expand the double row transfer matrix $\vec D(u)$ \eqref{eq:2fdrtm} to order $O(u^2)$. We do this diagrammatically. First, we open the double row transfer matrix and use the crossing symmetry property to rotate the faces in the upper row
\be
\drtm(u) = 
\psset{unit=1.1cm}
\begin{pspicture}[shift=-3](.6,0)(5.6,6)
\psellipticarc[linecolor=blue,linewidth=1.5pt](1.75,3)(1.1,2.55){85}{275}
\psellipticarc[linecolor=blue,linewidth=1.5pt](1.75,3)(.9,2.37){85}{275}
\psellipticarc[linecolor=blue,linewidth=1.5pt](4.9,3)(.42,.57){-90}{90}
\psellipticarc[linecolor=blue,linewidth=1.5pt](4.9,3)(.6,.75){-90}{90}

\pscurve[linecolor=blue,linewidth=1.5pt](2,5.1)(1.65,4)(1.65,2)(2,.9)
\pscurve[linecolor=blue,linewidth=1.5pt](2.2,5.1)(1.85,4)(1.85,2)(2.2,.9)
\pscurve[linecolor=blue,linewidth=1.5pt](2.8,4.45)(2.62,3.8)(2.62,2.2)(2.8,1.55)
\pscurve[linecolor=blue,linewidth=1.5pt](3.,4.45)(2.82,3.8)(2.82,2.1)(3.,1.55)
\pscurve[linecolor=blue,linewidth=1.5pt](3.7,4)(3.55,3.6)(3.55,2.4)(3.7,2)
\pscurve[linecolor=blue,linewidth=1.5pt](3.9,4)(3.75,3.6)(3.75,2.4)(3.9,2)
\pscurve[linecolor=blue,linewidth=1.5pt,showpoints=false](4.6,3.5)(4.44,3.2)(4.44,2.8)(4.6,2.5)
\pscurve[linecolor=blue,linewidth=1.5pt](4.8,3.5)(4.64,3.2)(4.64,2.8)(4.8,2.5)
\psellipse[linecolor=blue,linewidth=1.5pt](.78,3)(.1,.07)
\psellipse[linecolor=blue,linewidth=1.5pt](1.68,3)(.1,.07)
\psellipse[linecolor=blue,linewidth=1.5pt](2.66,3)(.1,.07)
\psellipse[linecolor=blue,linewidth=1.5pt](3.6,3)(.1,.07)
\psellipse[linecolor=blue,linewidth=1.5pt](4.52,3)(.1,.07)
\psellipse[linecolor=blue,linewidth=1.5pt](5.39,3)(.1,.07)
\rput{150}(5.5,3.866){
	\facegrid{(0,0)}{(4,1)}
	\multirput(0,0)(1,0){4}{
		\psarc[linecolor=red](1,1){.15}{180}{270}
		\psline[linecolor=blue,linewidth=1.5pt](.4,0)(.4,-.4)
		\psline[linecolor=blue,linewidth=1.5pt](.6,0)(.6,-.4)
		\psellipse[linecolor=blue,linewidth=1.5pt](.5,-.2)(.1,.07)
	}
}
\rput{210}(5,3){
	\facegrid{(0,0)}{(4,1)}
	\multirput(0,0)(1,0){4}{
		\psarc[linecolor=red](1,1){.15}{180}{270}
		\psline[linecolor=blue,linewidth=1.5pt](.4,1)(.4,1.4)
		\psline[linecolor=blue,linewidth=1.5pt](.6,1)(.6,1.4)
		\psellipse[linecolor=blue,linewidth=1.5pt](.5,1.2)(.1,.07)
	}
}
	\multirput(2.2,5.2)(.866,-.5){4}{
		\rput(.4,.3){\tiny$(2,2)$}
		\small $\lambda\!-\!u$
	}
	\multirput(2.2,.8)(.866,.5){4}{
		\rput(.15,.3){\tiny$(2,2)$}
		\small $u$
	}

\end{pspicture}
=
\psset{unit=1.1cm}
\begin{pspicture}[shift=-3](.6,0)(5.6,6)
\psellipticarc[linecolor=blue,linewidth=1.5pt](1.75,3)(1.1,2.55){85}{275}
\psellipticarc[linecolor=blue,linewidth=1.5pt](1.75,3)(.9,2.37){85}{275}
\psellipticarc[linecolor=blue,linewidth=1.5pt](4.9,3)(.42,.57){-90}{90}
\psellipticarc[linecolor=blue,linewidth=1.5pt](4.9,3)(.6,.75){-90}{90}

\pscurve[linecolor=blue,linewidth=1.5pt](2,5.1)(1.65,4)(1.65,2)(2,.9)
\pscurve[linecolor=blue,linewidth=1.5pt](2.2,5.1)(1.85,4)(1.85,2)(2.2,.9)
\pscurve[linecolor=blue,linewidth=1.5pt](2.8,4.45)(2.62,3.8)(2.62,2.2)(2.8,1.55)
\pscurve[linecolor=blue,linewidth=1.5pt](3.,4.45)(2.82,3.8)(2.82,2.1)(3.,1.55)
\pscurve[linecolor=blue,linewidth=1.5pt](3.7,4)(3.55,3.6)(3.55,2.4)(3.7,2)
\pscurve[linecolor=blue,linewidth=1.5pt](3.9,4)(3.75,3.6)(3.75,2.4)(3.9,2)
\pscurve[linecolor=blue,linewidth=1.5pt,showpoints=false](4.6,3.5)(4.44,3.2)(4.44,2.8)(4.6,2.5)
\pscurve[linecolor=blue,linewidth=1.5pt](4.8,3.5)(4.64,3.2)(4.64,2.8)(4.8,2.5)
\psellipse[linecolor=blue,linewidth=1.5pt](.78,3)(.1,.07)
\psellipse[linecolor=blue,linewidth=1.5pt](1.68,3)(.1,.07)
\psellipse[linecolor=blue,linewidth=1.5pt](2.66,3)(.1,.07)
\psellipse[linecolor=blue,linewidth=1.5pt](3.6,3)(.1,.07)
\psellipse[linecolor=blue,linewidth=1.5pt](4.52,3)(.1,.07)
\psellipse[linecolor=blue,linewidth=1.5pt](5.39,3)(.1,.07)
\rput{150}(5.5,3.866){
	\facegrid{(0,0)}{(4,1)}
	\multirput(0,0)(1,0){4}{
		\psarc[linecolor=red](0,1){.15}{-90}{0}
		\psline[linecolor=blue,linewidth=1.5pt](.4,0)(.4,-.4)
		\psline[linecolor=blue,linewidth=1.5pt](.6,0)(.6,-.4)
		\psellipse[linecolor=blue,linewidth=1.5pt](.5,-.2)(.1,.07)
	}
}
\rput{210}(5,3){
	\facegrid{(0,0)}{(4,1)}
	\multirput(0,0)(1,0){4}{
		\psarc[linecolor=red](1,1){.15}{180}{270}
		\psline[linecolor=blue,linewidth=1.5pt](.4,1)(.4,1.4)
		\psline[linecolor=blue,linewidth=1.5pt](.6,1)(.6,1.4)
		\psellipse[linecolor=blue,linewidth=1.5pt](.5,1.2)(.1,.07)
	}
}
	\multirput(2.2,5.2)(.866,-.5){4}{
		\rput(.15,.3){\tiny$(2,2)$}
		\small $u$
	}
	\multirput(2.2,.8)(.866,.5){4}{
		\rput(.15,.3){\tiny$(2,2)$}
		\small $u$
	}
\end{pspicture}
\ee
In the diagrams we show $N=4$ but the general $N$ case works similarly. We expand to $O(u^2)$ using the series expansion of the face operator
\be
\face_j(u) = I-\frac{\beta_2}{\sin2\lambda}u I + \frac{u}{\sin \lambda} X_j + \frac{u}{\sin 2\lambda} E_j + O(u^2),\qquad \beta_2=2 \cos2\lambda+1
\ee
The term with the identity on all faces is
\be
\psset{unit=1.1cm}
\vec D(0)\ =\ \begin{pspicture}[shift=-3](.6,0)(5.6,6)
\psellipticarc[linecolor=blue,linewidth=1.5pt](1.75,3)(1.1,2.55){85}{275}
\psellipticarc[linecolor=blue,linewidth=1.5pt](1.75,3)(.9,2.37){85}{275}
\psellipticarc[linecolor=blue,linewidth=1.5pt](4.9,3)(.42,.57){-90}{90}
\psellipticarc[linecolor=blue,linewidth=1.5pt](4.9,3)(.6,.75){-90}{90}

\pscurve[linecolor=blue,linewidth=1.5pt](2,5.1)(1.65,4)(1.65,2)(2,.9)
\pscurve[linecolor=blue,linewidth=1.5pt](2.2,5.1)(1.85,4)(1.85,2)(2.2,.9)
\pscurve[linecolor=blue,linewidth=1.5pt](2.8,4.45)(2.62,3.8)(2.62,2.2)(2.8,1.55)
\pscurve[linecolor=blue,linewidth=1.5pt](3.,4.45)(2.82,3.8)(2.82,2.1)(3.,1.55)
\pscurve[linecolor=blue,linewidth=1.5pt](3.7,4)(3.55,3.6)(3.55,2.4)(3.7,2)
\pscurve[linecolor=blue,linewidth=1.5pt](3.9,4)(3.75,3.6)(3.75,2.4)(3.9,2)
\pscurve[linecolor=blue,linewidth=1.5pt,showpoints=false](4.6,3.5)(4.44,3.2)(4.44,2.8)(4.6,2.5)
\pscurve[linecolor=blue,linewidth=1.5pt](4.8,3.5)(4.64,3.2)(4.64,2.8)(4.8,2.5)
\psellipse[linecolor=blue,linewidth=1.5pt](.78,3)(.1,.07)
\psellipse[linecolor=blue,linewidth=1.5pt](1.68,3)(.1,.07)
\psellipse[linecolor=blue,linewidth=1.5pt](2.66,3)(.1,.07)
\psellipse[linecolor=blue,linewidth=1.5pt](3.6,3)(.1,.07)
\psellipse[linecolor=blue,linewidth=1.5pt](4.52,3)(.1,.07)
\psellipse[linecolor=blue,linewidth=1.5pt](5.39,3)(.1,.07)
\rput{150}(5.5,3.866){
	\facegrid{(0,0)}{(4,1)}
	\multirput(0,0)(1,0){4}{
		\psline[linecolor=blue,linewidth=1.5pt](.4,0)(.4,-.4)
		\psline[linecolor=blue,linewidth=1.5pt](.6,0)(.6,-.4)
		\psellipse[linecolor=blue,linewidth=1.5pt](.5,-.2)(.1,.07)
	}
}
\rput{210}(5,3){
	\facegrid{(0,0)}{(4,1)}
	\multirput(0,0)(1,0){4}{
		\psline[linecolor=blue,linewidth=1.5pt](.4,1)(.4,1.4)
		\psline[linecolor=blue,linewidth=1.5pt](.6,1)(.6,1.4)
		\psellipse[linecolor=blue,linewidth=1.5pt](.5,1.2)(.1,.07)
	}
}
\multirput{210}(2.4,1.5)(.866,.5){4}{
	\rput(0,0){\TLtile}
}
\multirput{240}(2.034,5.873)(.866,-.5){4}{
	\TLtile
}
\end{pspicture}
= \fTLb I
\label{IdTerm}
\ee
Since we are expanding to $O(u^2)$ and the operators $X_j$ and $E_j$ have a coefficient of $u$, we only need configurations that have at most one of the operators $X_j$ and $E_j$. This gives terms such as
\begin{align}
\psset{unit=1.1cm}
\begin{pspicture}[shift=-3](.6,0)(5.6,6)
\psellipticarc[linecolor=blue,linewidth=1.5pt](1.75,3)(1.1,2.55){85}{275}
\psellipticarc[linecolor=blue,linewidth=1.5pt](1.75,3)(.9,2.37){85}{275}
\psellipticarc[linecolor=blue,linewidth=1.5pt](4.9,3)(.42,.57){-90}{90}
\psellipticarc[linecolor=blue,linewidth=1.5pt](4.9,3)(.6,.75){-90}{90}
\pscurve[linecolor=blue,linewidth=1.5pt](2,5.1)(1.65,4)(1.65,2)(2,.9)
\pscurve[linecolor=blue,linewidth=1.5pt](2.2,5.1)(1.85,4)(1.85,2)(2.2,.9)
\pscurve[linecolor=blue,linewidth=1.5pt](2.8,4.45)(2.62,3.8)(2.62,2.2)(2.8,1.55)
\pscurve[linecolor=blue,linewidth=1.5pt](3.,4.45)(2.82,3.8)(2.82,2.1)(3.,1.55)
\pscurve[linecolor=blue,linewidth=1.5pt](3.7,4)(3.55,3.6)(3.55,2.4)(3.7,2)
\pscurve[linecolor=blue,linewidth=1.5pt](3.9,4)(3.75,3.6)(3.75,2.4)(3.9,2)
\pscurve[linecolor=blue,linewidth=1.5pt,showpoints=false](4.6,3.5)(4.44,3.2)(4.44,2.8)(4.6,2.5)
\pscurve[linecolor=blue,linewidth=1.5pt](4.8,3.5)(4.64,3.2)(4.64,2.8)(4.8,2.5)
\psellipse[linecolor=blue,linewidth=1.5pt](.78,3)(.1,.07)
\psellipse[linecolor=blue,linewidth=1.5pt](1.68,3)(.1,.07)
\psellipse[linecolor=blue,linewidth=1.5pt](2.66,3)(.1,.07)
\psellipse[linecolor=blue,linewidth=1.5pt](3.6,3)(.1,.07)
\psellipse[linecolor=blue,linewidth=1.5pt](4.52,3)(.1,.07)
\psellipse[linecolor=blue,linewidth=1.5pt](5.39,3)(.1,.07)
\rput{150}(5.5,3.866){
	\facegrid{(0,0)}{(4,1)}
	\multirput(0,0)(1,0){4}{
		\psline[linecolor=blue,linewidth=1.5pt](.4,0)(.4,-.4)
		\psline[linecolor=blue,linewidth=1.5pt](.6,0)(.6,-.4)
		\psellipse[linecolor=blue,linewidth=1.5pt](.5,-.2)(.1,.07)
	}
}
\rput{210}(5,3){
	\facegrid{(0,0)}{(4,1)}
	\multirput(0,0)(1,0){4}{
		\psline[linecolor=blue,linewidth=1.5pt](.4,1)(.4,1.4)
		\psline[linecolor=blue,linewidth=1.5pt](.6,1)(.6,1.4)
		\psellipse[linecolor=blue,linewidth=1.5pt](.5,1.2)(.1,.07)
	}
}
\multirput{210}(2.4,1.5)(.866,.5){4}{
	\TLtile
}
\rput{240}(2.034,5.873){\Xtile}
\multirput{240}(2.9,5.373)(.866,-.5){3}{
	\TLtile
}
\end{pspicture}
&= \frac{\fTLb}{\TLb} I
&
\psset{unit=1.1cm}
\begin{pspicture}[shift=-3](.6,0)(5.6,6)
\psellipticarc[linecolor=blue,linewidth=1.5pt](1.75,3)(1.1,2.55){85}{275}
\psellipticarc[linecolor=blue,linewidth=1.5pt](1.75,3)(.9,2.37){85}{275}
\psellipticarc[linecolor=blue,linewidth=1.5pt](4.9,3)(.42,.57){-90}{90}
\psellipticarc[linecolor=blue,linewidth=1.5pt](4.9,3)(.6,.75){-90}{90}
\pscurve[linecolor=blue,linewidth=1.5pt](2,5.1)(1.65,4)(1.65,2)(2,.9)
\pscurve[linecolor=blue,linewidth=1.5pt](2.2,5.1)(1.85,4)(1.85,2)(2.2,.9)
\pscurve[linecolor=blue,linewidth=1.5pt](2.8,4.45)(2.62,3.8)(2.62,2.2)(2.8,1.55)
\pscurve[linecolor=blue,linewidth=1.5pt](3.,4.45)(2.82,3.8)(2.82,2.1)(3.,1.55)
\pscurve[linecolor=blue,linewidth=1.5pt](3.7,4)(3.55,3.6)(3.55,2.4)(3.7,2)
\pscurve[linecolor=blue,linewidth=1.5pt](3.9,4)(3.75,3.6)(3.75,2.4)(3.9,2)
\pscurve[linecolor=blue,linewidth=1.5pt,showpoints=false](4.6,3.5)(4.44,3.2)(4.44,2.8)(4.6,2.5)
\pscurve[linecolor=blue,linewidth=1.5pt](4.8,3.5)(4.64,3.2)(4.64,2.8)(4.8,2.5)
\psellipse[linecolor=blue,linewidth=1.5pt](.78,3)(.1,.07)
\psellipse[linecolor=blue,linewidth=1.5pt](1.68,3)(.1,.07)
\psellipse[linecolor=blue,linewidth=1.5pt](2.66,3)(.1,.07)
\psellipse[linecolor=blue,linewidth=1.5pt](3.6,3)(.1,.07)
\psellipse[linecolor=blue,linewidth=1.5pt](4.52,3)(.1,.07)
\psellipse[linecolor=blue,linewidth=1.5pt](5.39,3)(.1,.07)
\rput{150}(5.5,3.866){
	\facegrid{(0,0)}{(4,1)}
	\multirput(0,0)(1,0){4}{
		\psline[linecolor=blue,linewidth=1.5pt](.4,0)(.4,-.4)
		\psline[linecolor=blue,linewidth=1.5pt](.6,0)(.6,-.4)
		\psellipse[linecolor=blue,linewidth=1.5pt](.5,-.2)(.1,.07)
	}
}
\rput{210}(5,3){
	\facegrid{(0,0)}{(4,1)}
	\multirput(0,0)(1,0){4}{
		\psline[linecolor=blue,linewidth=1.5pt](.4,1)(.4,1.4)
		\psline[linecolor=blue,linewidth=1.5pt](.6,1)(.6,1.4)
		\psellipse[linecolor=blue,linewidth=1.5pt](.5,1.2)(.1,.07)
	}
}
\multirput{210}(2.4,1.5)(.866,.5){4}{
	\TLtile
}
\rput{150}(2.9,5.373){\TLtile}
\multirput{240}(2.9,5.373)(.866,-.5){3}{
	\TLtile
}
\end{pspicture}
&= I\label{Iterms}\\
\psset{unit=1.1cm}
\begin{pspicture}[shift=-3](.6,0)(5.6,6)
\psellipticarc[linecolor=blue,linewidth=1.5pt](1.75,3)(1.1,2.55){85}{275}
\psellipticarc[linecolor=blue,linewidth=1.5pt](1.75,3)(.9,2.37){85}{275}
\psellipticarc[linecolor=blue,linewidth=1.5pt](4.9,3)(.42,.57){-90}{90}
\psellipticarc[linecolor=blue,linewidth=1.5pt](4.9,3)(.6,.75){-90}{90}
\pscurve[linecolor=blue,linewidth=1.5pt](2,5.1)(1.65,4)(1.65,2)(2,.9)
\pscurve[linecolor=blue,linewidth=1.5pt](2.2,5.1)(1.85,4)(1.85,2)(2.2,.9)
\pscurve[linecolor=blue,linewidth=1.5pt](2.8,4.45)(2.62,3.8)(2.62,2.2)(2.8,1.55)
\pscurve[linecolor=blue,linewidth=1.5pt](3.,4.45)(2.82,3.8)(2.82,2.1)(3.,1.55)
\pscurve[linecolor=blue,linewidth=1.5pt](3.7,4)(3.55,3.6)(3.55,2.4)(3.7,2)
\pscurve[linecolor=blue,linewidth=1.5pt](3.9,4)(3.75,3.6)(3.75,2.4)(3.9,2)
\pscurve[linecolor=blue,linewidth=1.5pt,showpoints=false](4.6,3.5)(4.44,3.2)(4.44,2.8)(4.6,2.5)
\pscurve[linecolor=blue,linewidth=1.5pt](4.8,3.5)(4.64,3.2)(4.64,2.8)(4.8,2.5)
\psellipse[linecolor=blue,linewidth=1.5pt](.78,3)(.1,.07)
\psellipse[linecolor=blue,linewidth=1.5pt](1.68,3)(.1,.07)
\psellipse[linecolor=blue,linewidth=1.5pt](2.66,3)(.1,.07)
\psellipse[linecolor=blue,linewidth=1.5pt](3.6,3)(.1,.07)
\psellipse[linecolor=blue,linewidth=1.5pt](4.52,3)(.1,.07)
\psellipse[linecolor=blue,linewidth=1.5pt](5.39,3)(.1,.07)
\rput{150}(5.5,3.866){
	\facegrid{(0,0)}{(4,1)}
	\multirput(0,0)(1,0){4}{
		\psline[linecolor=blue,linewidth=1.5pt](.4,0)(.4,-.4)
		\psline[linecolor=blue,linewidth=1.5pt](.6,0)(.6,-.4)
		\psellipse[linecolor=blue,linewidth=1.5pt](.5,-.2)(.1,.07)
	}
}
\rput{210}(5,3){
	\facegrid{(0,0)}{(4,1)}
	\multirput(0,0)(1,0){4}{
		\psline[linecolor=blue,linewidth=1.5pt](.4,1)(.4,1.4)
		\psline[linecolor=blue,linewidth=1.5pt](.6,1)(.6,1.4)
		\psellipse[linecolor=blue,linewidth=1.5pt](.5,1.2)(.1,.07)
	}
}
\multirput{210}(2.4,1.5)(.866,.5){4}{
	\TLtile
}
\rput{240}(2.034,5.873){\TLtile}
\rput{240}(2.9,5.373){\Xtile}
\multirput{240}(3.766,4.873)(.866,-.5){2}{
	\TLtile
}
\end{pspicture}
&= \fTLb X_1
&
\psset{unit=1.1cm}
\begin{pspicture}[shift=-3](.6,0)(5.6,6)
\psellipticarc[linecolor=blue,linewidth=1.5pt](1.75,3)(1.1,2.55){85}{275}
\psellipticarc[linecolor=blue,linewidth=1.5pt](1.75,3)(.9,2.37){85}{275}
\psellipticarc[linecolor=blue,linewidth=1.5pt](4.9,3)(.42,.57){-90}{90}
\psellipticarc[linecolor=blue,linewidth=1.5pt](4.9,3)(.6,.75){-90}{90}
\pscurve[linecolor=blue,linewidth=1.5pt](2,5.1)(1.65,4)(1.65,2)(2,.9)
\pscurve[linecolor=blue,linewidth=1.5pt](2.2,5.1)(1.85,4)(1.85,2)(2.2,.9)
\pscurve[linecolor=blue,linewidth=1.5pt](2.8,4.45)(2.62,3.8)(2.62,2.2)(2.8,1.55)
\pscurve[linecolor=blue,linewidth=1.5pt](3.,4.45)(2.82,3.8)(2.82,2.1)(3.,1.55)
\pscurve[linecolor=blue,linewidth=1.5pt](3.7,4)(3.55,3.6)(3.55,2.4)(3.7,2)
\pscurve[linecolor=blue,linewidth=1.5pt](3.9,4)(3.75,3.6)(3.75,2.4)(3.9,2)
\pscurve[linecolor=blue,linewidth=1.5pt,showpoints=false](4.6,3.5)(4.44,3.2)(4.44,2.8)(4.6,2.5)
\pscurve[linecolor=blue,linewidth=1.5pt](4.8,3.5)(4.64,3.2)(4.64,2.8)(4.8,2.5)
\psellipse[linecolor=blue,linewidth=1.5pt](.78,3)(.1,.07)
\psellipse[linecolor=blue,linewidth=1.5pt](1.68,3)(.1,.07)
\psellipse[linecolor=blue,linewidth=1.5pt](2.66,3)(.1,.07)
\psellipse[linecolor=blue,linewidth=1.5pt](3.6,3)(.1,.07)
\psellipse[linecolor=blue,linewidth=1.5pt](4.52,3)(.1,.07)
\psellipse[linecolor=blue,linewidth=1.5pt](5.39,3)(.1,.07)
\rput{150}(5.5,3.866){
	\facegrid{(0,0)}{(4,1)}
	\multirput(0,0)(1,0){4}{
		\psline[linecolor=blue,linewidth=1.5pt](.4,0)(.4,-.4)
		\psline[linecolor=blue,linewidth=1.5pt](.6,0)(.6,-.4)
		\psellipse[linecolor=blue,linewidth=1.5pt](.5,-.2)(.1,.07)
	}
}
\rput{210}(5,3){
	\facegrid{(0,0)}{(4,1)}
	\multirput(0,0)(1,0){4}{
		\psline[linecolor=blue,linewidth=1.5pt](.4,1)(.4,1.4)
		\psline[linecolor=blue,linewidth=1.5pt](.6,1)(.6,1.4)
		\psellipse[linecolor=blue,linewidth=1.5pt](.5,1.2)(.1,.07)
	}
}
\multirput{210}(2.4,1.5)(.866,.5){4}{
	\TLtile
}
\rput{240}(2.034,5.873){\TLtile}
\rput{150}(3.766,4.873){\TLtile}
\multirput{240}(3.766,4.873)(.866,-.5){2}{
	\TLtile
}
\end{pspicture}
&= \fTLb E_1
\label{XEterms}
\end{align}

The expansion for the double row transfer matrix is thus
\begin{align}
\drtm(u) &=\Big[ \beta_2 I-2N\Big(\frac{\beta_2^2}{\sin2\lambda}\Big)u I + 2\left( \frac{u}{\sin \lambda} \frac{\fTLb}{\TLb}I + \frac{u}{\sin 2\lambda} I \right) + 2\sum_{j=1}^{N-1}\left( \frac{u}{\sin \lambda} \fTLb X_j + \frac{u}{\sin 2\lambda} \fTLb E_j \right)\Big] +O(u^2)\nn
&=\beta_2\Bigg\{ I + \frac{2 u}{\sin 2\lambda}\bigg[ \Big(\!\!-\!N\beta_2 +  \frac{\beta^2}{\beta_2}\Big)I + 
\sum_{j=1}^{N-1}\big(\beta X_j + E_j\big)  \bigg] \Bigg\}+ O(u^2),\qquad \beta_2\ne 0
\end{align}
Comparing this with the expansion \eqref{eq:drtmExpansion} gives the Neveu-Schwarz Hamiltonian
\be
-\ham 
= \sum_{j=1}^{N-1}\left(\TLb X_j + E_j\right)
\label{HamNS}
\ee
where we have shifted the zero of energy (bulk and boundary free energies) by $-N\beta_2+\beta_2^{-1} \beta^2$ to remove the constant terms in the Hamiltonian. 

To obtain the Hamiltonian for superconformal dense polymers, we fix $\beta_2=0$ and expand $\vec d(u)$ to order $O(u^2)$. Because of the normalisation factor $\sin 3u$ in (\ref{db}), this is equivalent to expanding $\vec D(u)$ to order $O(u^3)$. 
In this case, the contributions (\ref{IdTerm}) to (\ref{XEterms}) all vanish except for the term on the right of (\ref{Iterms}) which gives the constant term in the Hamiltonian. 
At the next order in $u$, the only surviving terms are diagrammatically as in the right side of (\ref{Iterms}) but with an additional $X_j$ or $E_j$ for $j=1,2,\ldots,N-1$. 
After shifting the zero of energy, we again obtain the Hamiltonian (\ref{HamNS}).
In all cases, the Hamiltonian (\ref{HamNS}) acts on the vector space of link states ${\cal V}_{1,1,0}^{(N)}$ or ${\cal V}_{1,1,2}^{(N)}$ depending on whether $N$ is even or odd.

\subsubsection{Ramond sector $(r,s)=(1,2)$}

Applying the same method used in the Neveu-Schwarz sector $(r,s)=(1,1)$ to the Ramond sector $(r,s)=(1,2)$, gives the Ramond Hamiltonian
\be
-\ham = \sum_{j=1}^{N-1}\left(\TLb X_j + E_j\right)+h_N(\eta)\,p_{2 N-1} e_{2N}p_{2N-1},\quad h_N(\eta)=\frac{2 \sin^2 2\lambda}{\sin \eta \sin(3\lambda-\eta)},\quad \eta\in(0,\pi)
\label{HamR}
\ee
The extra boundary term with trigonometric coefficient $h_N(\eta)$ is due to the presence of the Ramond seam introducing the single defect in the bulk. It acts at position $N$ in the fused Temperley-Lieb algebra and has the diagrammatic representation
\be
p_{2 N-1} e_{2N}p_{2N-1} = 
\psset{unit=1.1cm}
\begin{pspicture}[shift=-.4](0,0)(1,1)
\multirput(0,0)(0,.6){2}{\psellipse[linecolor=blue,linewidth=1.5pt](.15,.2)(.15,.075)}
\psarc[linecolor=blue,linewidth=1.5pt](.5,.2){.2}{0}{180}
\psarc[linecolor=blue,linewidth=1.5pt](.5,.8){.2}{180}{360}
\end{pspicture}
\ee
The Ramond Hamiltonian acts on the vector space of link states ${\cal V}_{1,2,1}^{(N)}$.

The trigonometric coefficient $h_N(\eta)$ has singularities at the endpoints $\eta=0,\pi$. If $\lambda=\pi/3$ or $2\pi/3$, there are no further singularities  on $\eta\in(0,\pi)$ and $h_N(\eta)$ is always positive if $\lambda=\pi/3$ and $h_N(\eta)$ is always negative if 
$\lambda=2\pi/3$. 
Otherwise, there is a third singularity at $\eta=\eta_{sing}$ and so we restrict the parameter $\eta$ to 
\be
\eta\in(0,\eta_{sing}) \cup (\eta_{sing},\pi), \qquad \eta_{sing} = \mbox{$3\lambda$ mod $\pi$}
\ee
In these cases, the field $h_N(\eta)$ is positive (ferromagnetic) on one subinterval and negative (antiferromagnetic) on the other subinterval. The midpoints of the two subintervals are
\be
\eta_{\pm} = \frac{\eta_{sing}}{2}, \frac{\eta_{sing}}{2}+\frac{\pi}{2},\quad
\pm =\mbox{sgn}(h_N(\eta)),\quad \mbox{sgn}(h_N(\eta_+))>0,\quad \mbox{sgn}(h_N(\eta_-))<0
\ee
More explicitly, the midpoint of the positive interval of $h_N(\eta)$ is
\bea
\eta_+=\eta_+(\lambda)=\begin{cases}
\frac{\eta_{sing}}{2},&\lambda\in(0,\frac{\pi}{3})\cup(\frac{2\pi}{3},\pi)\\[2pt]
\frac{\eta_{sing}}{2}+\frac{\pi}{2},&\lambda\in(\frac{\pi}{3},\frac{2\pi}{3})
\end{cases};\qquad
\eta_+=\tfrac{\pi}{2},\quad \lambda = \tfrac{\pi}{3}
\eea
Numerical investigations indicate that the conformal properties in the continuum limit depend only on the choice of the sign of $h_N(\eta)$, that is which subinterval $\eta$ lies in, and that they are otherwise independent of the choice of the value for $\eta$. For the purpose of generating numerical estimates with $\lambda<\frac{\pi}{2}$, it is therefore convenient to fix the value of $\eta$ to $\eta=\eta_+$.

\subsubsection{Neveu-Schwarz sectors $(1,s)$, $s\ge 3$}

The Hamiltonians, in the NS and R sectors, with $r=1$ and $s\ge 3$ are given by the same expressions (\ref{HamNS}) and (\ref{HamR}) depending on the parity of $s$. The only difference is that they act on the vector space of link states 
${\cal V}_{1,s,0}^{(N)}$ ($s$ odd, $N$ even), ${\cal V}_{1,s,2}^{(N)}$ ($s$ odd, $N$ even) or ${\cal V}_{1,s,1}^{(N)}$ ($s$ even) with $s\ge 3$. In these cases, the generators $X_j$, $E_j$ in the Hamiltonian do not act on the $s-1$ unpaired sites on the boundary. 
Specifically, for $s=3$, the Hamiltonian acting on the space of link states ${\cal V}_{1,3,0}^{(N)}$ is
\be
-\ham 
= \sum_{j=1}^{N-1}\left(\TLb X_j + E_j\right)
\label{HamNSs=3}
\ee
where the total number of underlying TL sites is $2N+2$.

\subsection{Duality}
\label{Duality}

In this section, we observe that the logarithmic superconformal minimal models exhibit an exact duality for finite systems under the involution
\be
 \lambda\leftrightarrow\pi-\lambda\quad \mbox{or}\quad p\leftrightarrow p'-p
\label{dual}
\ee 
Since the spectra are invariant under this duality, as argued below, it therefore suffices to restrict our study to models satisfying
\bea
\lambda<\frac{\pi}{2},\qquad p>\frac{p'}{2}
\eea

Under the duality (\ref{dual}), we see that
\be
 x\leftrightarrow-x^{-1},\qquad \beta\leftrightarrow-\beta
\ee
and observe that the TL algebra is invariant provided we supplement the map (\ref{dual}) with
\be
 I\leftrightarrow I,\qquad e_j\leftrightarrow-e_j
\ee
The WJ projectors are then readily seen to be invariant ($p_j\leftrightarrow p_j$), while the generators of the fused TL algebra
are mapped as
\be
 I\leftrightarrow I,\qquad X_j\leftrightarrow-X_j,\qquad E_j\leftrightarrow E_j,\qquad B_j\leftrightarrow B_j^{-1}
\ee
As a consequence, the fused face operators, the double row transfer matrices \eqref{eq:2fdrtm} and the Hamiltonians (\ref{HamNS}) 
are all invariant under the duality transformation if we map
\be
 u\leftrightarrow-u\quad \mbox{or}\quad z\leftrightarrow z^{-1}
\ee
It follows that, in the Neveu-Schwarz sectors, the matrix representatives $\vec D(u)=\vec D(u,\lambda)$ and ${\cal H}={\cal H}(\lambda)$ satisfy
\be
\drtm(u,\lambda)\sim \drtm(-u,\pi-\lambda),\qquad {\cal H}(\lambda)\sim {\cal H}(\pi-\lambda)
\ee
where $\sim$ denotes matrix similarity.

The same duality transformation holds in the Ramond sectors with the additional mapping $\eta \leftrightarrow \pi-\eta$. The double row transfer and Hamiltonian matrix representatives then satisfy
\be
\drtm(u,\lambda,\eta) \sim \drtm(-u,\pi-\lambda,\pi-\eta),\qquad {\cal H}(\lambda,\eta)\sim {\cal H}(\pi-\lambda,\pi-\eta)
\ee
We note that $h_N(\eta)$ changes sign under the duality mapping $\eta \leftrightarrow \pi-\eta$, $\lambda\leftrightarrow \pi-\lambda$ so the roles of $\eta_+$ and $\eta_-$ are interchanged 
under duality and
\be
 \eta_{\pm}(\lambda)=\pi-\eta_{\mp}(\pi-\lambda),\qquad \lambda\ne\tfrac{\pi}{3}, \tfrac{2\pi}{3}
\ee

\subsection{Free energies}

As argued in \cite{PRZ}, although we consider models with crossing parameter $\lambda$ given by rational fractions of $\pi$, these values are dense on the real line and so the bulk free energies, the boundary free energies and central charges are all given as continuous functions of $\lambda$. In the next two subsections, we calculate the bulk and  boundary free energies analytically for $0<\lambda<\frac{\pi}{2}$. The central charges and first few conformal weights are determined numerically in Section~\ref{numericalSection}.

\subsubsection{Bulk free energies}
\def\Re{\mathop{\mbox{Re}}}

The bulk free energies are calculated using the inversion method of Baxter~\cite{BaxInv82}. 
The $2\times 2$ fused face operators satisfy the inversion (\ref{InversionReln}) and crossing relation (\ref{Crossing}).
It follows that the partition function per face $\kappa(u)=\exp(-f_{bulk}(u))$ satisfies the inversion and crossing  relations
\bea
\kappa(u)\kappa(u+\lambda)=\frac{s(\lambda-u)s(\lambda+u)s(2\lambda-u)s(2\lambda+u)}{s(2\lambda)^2},\qquad 
\kappa(u)=\kappa(\lambda-u)
\eea
Using the identities
\be
 \frac{d^2}{du^2}\,\log\frac{\sin u}{\sin\lambda}=-\int_{-\infty}^{\infty} \frac{2t\,\cosh(\pi-2u)t}{\sinh \pi t}\,dt,
\qquad 0<\Re u<\pi
\label{logsinId1}
\ee
and
\be
\frac{d^2}{du^2}\,\log\frac{\sin(\lambda-u)\sin(\lambda+u)}{\sin^2\lambda}=
-\int_{-\infty}^{\infty} \frac{4t\,\cosh(\pi-2\lambda)t}{\sinh\pi t}\,e^{2ut}\,dt,
\qquad |\Re u|<\lambda
\label{logsinId2}
\ee
it follows that
\be
\frac{d^2}{du^2}\log\frac{s(\lambda-u)s(\lambda+u)s(2\lambda-u)s(2\lambda+u)}{s(2\lambda)^2}
=-\!\int_{-\infty}^{\infty}
\frac{8t\cosh(\pi-3\lambda)t \cosh{\lambda t}}{\sinh\pi t}\,e^{2ut}dt
\ee
WritingÊ
\be
 \frac{d^2}{du^2} \log \kappa(u)=\int_{-\infty}^{\infty} c(t) e^{2ut}\,dt
\ee
we conclude that
\bea
c(t)=e^{-2\lambda t}c(-t),\qquad\quad 
(1+e^{2\lambda t})c(t)=-\frac{8t\cosh(\pi-3\lambda)t \cosh{\lambda t}}{\sinh\pi t}
\eea
so that the solution is
\be
c(t)= -\frac{4t\cosh(\pi-3\lambda)t}{\sinh\pi t}\,e^{-\lambda t}
\ee
Integrating twice and evaluating the integration constants gives
\begin{align}
\log\kappa(u)&=\int_{-\infty}^\infty 
\frac{\cosh(\pi-3\lambda)t\,[\cosh(\lambda-2u)t-\cosh\lambda t]}{t\,\sinh\pi t}\,dt+Au\nonumber\\
&=\!2\int_{-\infty}^\infty 
\frac{\cosh(\pi-3\lambda)t\sinh ut\sinh(\lambda-u)t}{t\,\sinh\pi t}\,dt,
\quad  -\frac{\lambda}{2}<\mbox{Re}(u)<\frac{3\lambda}{2}
\end{align}
where applying the crossing symmetry and setting $u=\lambda$ implies $A=0$.
Since the solution without zeros and poles in the relevant (physical) analyticity strip is unique, this solution can be rewritten as
\bea
\kappa(u)=\frac{s(\lambda+u)s(2\lambda-u)}{s(2\lambda)},\qquad -\frac{\lambda}{2}<\Re u<\frac{3\lambda}{2},\qquad 0<\lambda<\frac{\pi}{2}
\eea

Let $E_0$ be the lowest energy eigenvalue of the quantum Hamiltonian ${\cal H}$ in the vacuum sector $(r,s,\ell)=(1,1,0)$. Then taking the Hamiltonian limit, as in Section~\ref{appHam}, and allowing for the fact that there are two faces in each column gives
\bea
\kappa(u)\sim \exp\Big[-\frac{u}{\sin2\lambda}\Big(\frac{E_0}{N}+\beta^2\Big)\Big]
\sim \frac{s(\lambda+u)s(2\lambda-u)}{s(2\lambda)},\qquad N\to\infty
\eea
Taking the logarithmic derivative, or equivalently equating the linear terms in $u$ for the small $u$ expansion
and using
\bea
\frac{s(\lambda+u)s(2\lambda-u)}{s(2\lambda)}\sim \frac{u}{\sin 2\lambda}
\eea
 gives the energy eigenvalue $E_0$ of the quantum Hamiltonian ${\cal H}$ as 
\bea
-\frac{E_0}{N}\sim \beta_2+1=\beta^2,\qquad N\to\infty
\eea
This result is confirmed to high precision by our numerics in both the NS and R sectors.

\subsubsection{Boundary free energies}

\def\Re{\mathop{\!\mbox{Re}}}

The boundary free energies are calculated using the boundary inversion relation methods of \cite{OPB95}. 
Diagrammatically, following \cite{UnivTBA}, the boundary partition function $\kappa_0(u)=\exp(-f_{bdy}(u))$ satisfies the boundary inversion relation
\bea
\psset{unit=.8cm}
\kappa_0(u)\kappa_0(u+\lambda)=
\frac{s(2\lambda)^2}{s(\lambda\!-\!2u)s(\lambda\!+\!2u)s(2\lambda\!-\!2u)s(2\lambda\!+\!2u)}\quad
\begin{pspicture}[shift=-2](0,0)(4,4)
\multirput(0,3.4)(0,.2){2}{\psline[linewidth=1pt](0,0)(4,0)}
\multirput(0,2.4)(0,.2){2}{\psline[linewidth=1pt](0,0)(4,0)}
\psarc[linewidth=1pt](1,1){.4}{-135}{45}
\psarc[linewidth=1pt](1,1){.6}{-135}{45}
\psarc[linewidth=1pt](3,1){.4}{-225}{-45}
\psarc[linewidth=1pt](3,1){.6}{-225}{-45}
\pspolygon[linewidth=1pt,fillstyle=solid,fillcolor=lightlightblue](0,0)(2,2)(0,4)(0,0)
\pspolygon[linewidth=1pt,fillstyle=solid,fillcolor=lightlightblue](4,0)(2,2)(4,4)(4,0)
\pspolygon[linewidth=1pt,fillstyle=solid,fillcolor=lightlightblue](0,2)(1,1)(2,2)(1,3)(0,2)
\pspolygon[linewidth=1pt,fillstyle=solid,fillcolor=lightlightblue](4,2)(3,1)(2,2)(3,3)(4,2)
\rput(1,2.4){\tiny $(2,2)$}
\rput(1,1.94){$2u$}
\rput(3,2.4){\tiny $(2,2)$}
\rput(3,1.94){$-2u$}
\rput(3.7,1){$u$}
\rput(3.8,3){$u\!+\lambda$}
\rput(.2,1){$\lambda-\!u$}
\rput(.4,3){$-u$}
\psarc[linewidth=1pt,linecolor=red](0,2){.175}{-45}{45}
\psarc[linewidth=1pt,linecolor=red](2,2){.175}{-45}{45}
\end{pspicture}
\label{diagInvReln}
\eea
where the faces with spectral parameters $\pm2u$ are put in using the inversion relation (\ref{InversionReln}) and compensated by the scalar prefactors.
The identity acting internally at the bottom is decomposed into the complementary fusion projectors
\bea
I=\frac{E_j}{\beta_2}+\big(I-\frac{E_j}{\beta_2}\big),\qquad \beta_2\ne 0
\label{IdDecomp}
\eea
and only the first projector is kept. 
Replacing the four boundary triangles in (\ref{diagInvReln}) with the vacuum triangles (\ref{fusedVac}), rotating the central faces and using the push-through relations
\bea
\face_j(\lambda\!-\!2u)E_j=\frac{s(3\lambda\!-\!2u)s(2\lambda\!-\!2u)}{s(2\lambda)}\,E_j,\qquad 
\face_j(\lambda\!+\!2u)E_j=\frac{s(3\lambda\!+\!2u)s(2\lambda\!+\!2u)}{s(2\lambda)}\,E_j
\eea
gives the scalar contributions of the central faces and the scalar boundary inversion relation
\bea
\kappa_0(u)\kappa_0(u+\lambda)=\frac{s(3\lambda\!-\!2u)s(3\lambda\!+\!2u)}{s(\lambda\!-\!2u)s(\lambda\!+\!2u)}
\label{bdyInv}
\eea
Applying the decomposition (\ref{IdDecomp}) to the full double row transfer matrices, as in (\ref{diagInvReln}),  gives an inversion identity~\cite{PearceInv} of the form
\bea
\vec D(u)\vec D(u+\lambda)=\frac{s(3\lambda\!-\!2u)s(3\lambda\!+\!2u)}{s(\lambda\!-\!2u)s(\lambda\!+\!2u)}\,I
+\phi(u) \vec D^{(1,2)}(u)
\eea
where $\phi(u)$ is a scalar function and $\vec D^{(1,2)}(u)$ is a fused double row transfer matrix. The scalar  inversion relation ensues because the second term on the right of this inversion identity is exponentially small for $u$ in an appropriate strip in the complex $u$ plane. 

We first calculate the boundary free energies for the transfer matrices $\vec D(u)$. Taking the Hamiltonian limit gives the boundary free energies of the Hamiltonians valid on the interval $0<\lambda<\frac{\pi}{3}$. We then analytically continue these results to the interval $0<\lambda<\frac{\pi}{2}$.

In its analyticity strip $-\frac{\lambda}{2}<\Re  u<\frac{3\lambda}{2}$, the vacuum boundary contribution 
$\tilde\kappa_0(u)=\kappa_0(u)/\beta_2$ satisfies the inversion relation and crossing relations
\be
\tilde\kappa_0(u)\tilde\kappa_0(u+\lambda)
 =\frac{\sin^2\lambda\,\sin(3\lambda-2u)\sin(3\lambda+2u)}{\sin^2 3\lambda\,\sin(\lambda-2u)\sin(\lambda+2u)},\qquad \tilde\kappa_0(u)=\tilde\kappa_0(\lambda-u),
\qquad \tilde\kappa(0)=1
\ee
\be
\log\tilde\kappa_0(u)+\log\tilde\kappa_0(u+\lambda)=
\log\frac{\sin^3\lambda\,\sin(3\lambda-2u)\sin(3\lambda+2u)}{\sin^2 3\lambda\,\sin(\lambda-2u)\sin(\lambda+2u)},
\qquad -\frac{\lambda}{2}<\Re u<\frac{\lambda}{2}
\ee
But now using the identities (\ref{logsinId1}) and (\ref{logsinId2}), 
it follows that in the strip $|\Re u|<\lambda/2$
\be
\frac{d^2}{du^2}\log\frac{\sin(3\lambda-2u)\sin(3\lambda+2u)}{\sin(\lambda-2u)\sin(\lambda+2u)}
=\!\int_{-\infty}^{\infty}
\frac{2t\sinh\frac{(\pi-4\lambda)t}{2}\sinh{\lambda t}}{\sinh\frac{\pi t}{2}}\,e^{2ut}dt
\ee
We conclude that
\be
c(t)=e^{-2\lambda t}c(-t),\qquad\quad 
(1+e^{2\lambda t})c(t)=\frac{2t\sinh\frac{(\pi-4\lambda)t}{2}\sinh{\lambda t}}{\sinh\frac{\pi t}{2}}
\ee
so that the solution is
\be
c(t)=\frac{t\sinh\frac{(\pi-4\lambda)t}{2}\sinh{\lambda t}}{\sinh\frac{\pi t}{2}\cosh \lambda t}\,e^{-\lambda t}
\ee
Integrating twice and evaluating the integration constants gives
\begin{align}
\log\tilde\kappa_0(u)&=\int_{-\infty}^\infty 
\frac{\sinh\frac{(\pi-4\lambda)t}{2}\sinh{\lambda t}}{4t\,\sinh\frac{\pi t}{2}\cosh\lambda t}\,e^{-(\lambda-2u)t}\,dt+Au+B\nonumber\\
&=\int_{-\infty}^\infty 
\frac{\sinh\frac{(\pi-4\lambda)t}{2}\sinh{\lambda t}\,[\cosh(\lambda-2u)t-\cosh\lambda t]}{4t\,\sinh\frac{\pi t}{2}\cosh\lambda t}\,dt+Au\nonumber\\
&=\!\int_{-\infty}^\infty 
\frac{\sinh\frac{(\pi-4\lambda)t}{2}\sinh{\lambda t}\sinh ut\sinh(\lambda-u)t}{2t\,\sinh\frac{\pi t}{2}\cosh\lambda t}\,dt,
\quad  -\frac{\lambda}{2}<\mbox{Re}(u)<\frac{3\lambda}{2}
\label{DTMbdy}
\end{align}
where applying the crossing symmetry and setting $u=\lambda$ implies $A=0$. 
This result applies in the interval $0<\lambda<\frac{\pi}{3}$. This result can be analytically continued to the interval 
$0<\lambda<\frac{\pi}{2}$, but since we do not need it, we do not give the expression here.

\begin{figure}[t]
   \centering
   \includegraphics[width=5in]{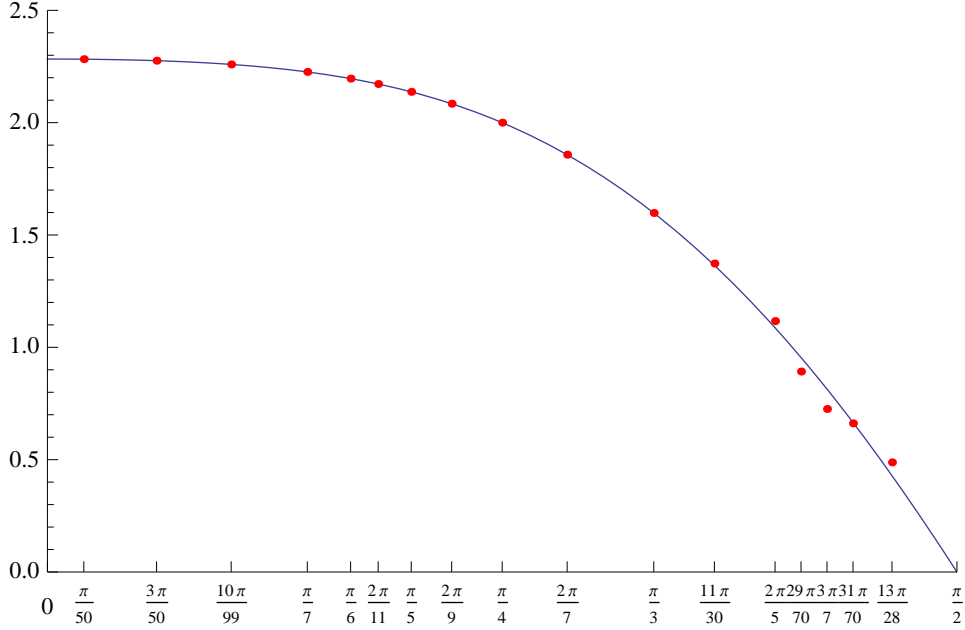} 
\caption{Plot of the Hamiltonian boundary free energy (\ref{analBdyFree}) against $\lambda$. 
The agreement with numerical estimates is good for $\lambda<\frac{\pi}{3}$. There is some scatter of numerical data points for 
$\lambda>\frac{\pi}{3}$ due to larger errors arising from slow convergence of the sequences of approximants.
\label{superBdyPlot}}
\end{figure}

Taking the Hamiltonian limit of (\ref{DTMbdy}) gives the boundary free energy for the Hamiltonian (\ref{HamNS})
\begin{align}
f_{bdy}=f^{2,2}_{bdy}&=\frac{\beta^2}{\beta_2}+\sin 2\lambda 
\int_{-\infty}^\infty \frac{\sinh\frac{(\pi-4\lambda)t}{2} \sinh\lambda t\tanh\lambda t}{\sinh\frac{\pi t}{2}}\,dt\nonumber\\
&=\sin 2\lambda \Big[\frac{2\cos\lambda}{\sin 3\lambda}+
\int_{-\infty}^\infty \frac{\sinh\frac{(\pi-4\lambda)t}{2} \sinh\lambda t\tanh\lambda t}{\sinh\frac{\pi t}{2}}\,dt\Big],
\qquad 0<\lambda<\tfrac{\pi}{3}
\label{superFbdy}
\end{align}
For $\lambda>\frac{\pi}{3}$, the integral diverges so we analytically continue this result to $0<\lambda<\frac{\pi}{2}$ by combining the two terms in square brackets into one integral. We start with the identity
\bea
\log\sin\lambda=-\int_{-\infty}^\infty \frac{\sinh^2\frac{(\pi-2\lambda)t}{2}}{t\sinh\pi t}\,dt,\qquad 0<\lambda<\pi
\eea
It follows that
\begin{align}
\frac{2\cos\lambda}{\sin 3\lambda}&=\frac{1}{3}\frac{d}{d\lambda}\big[3\log\sin\lambda-\log\sin3\lambda\big]
=\frac{1}{3}\frac{d}{d\lambda}\int_{-\infty}^\infty 
\frac{\sinh^2\frac{(\pi-6\lambda)t}{2}-3\sinh^2\frac{(\pi-2\lambda)t}{2}}{t\sinh\pi t}\,dt\nonumber\\
&=\int_{-\infty}^\infty 
\frac{\sinh(\pi\!-\!2\lambda)t-\sinh(\pi\!-\!6\lambda)t}{\sinh\pi t}\,dt
=\int_{-\infty}^\infty \frac{\cosh\frac{(\pi-4\lambda)t}{2}\sinh\lambda t}{\sinh\frac{\pi t}{2}}\,dt
\end{align}
Substituting into (\ref{superFbdy}) and simplifying the single integrand gives
\bea
f_{bdy}=\sin2\lambda\,\int_{-\infty}^\infty \frac{\cosh\frac{(\pi-2\lambda)t}{2}\tanh\lambda t}{\sinh\frac{\pi t}{2}}\,dt,
\qquad 0<\lambda<\tfrac{\pi}{2}
\label{analBdyFree}
\eea
A plot of the analytic boundary free energy $f_{bdy}$ against $\lambda$ is shown in Figure~\ref{superBdyPlot}. This analytic result is well confirmed numerically.

\subsection{Finitized $r=1$ Kac characters}

\label{FinStrFn+Char}

In this section, we use combinatorial arguments to conjecture the $r=1$ finitized Kac characters of the logarithmic superconformal ($n=2$) minimal models ${\cal LSM}(p,p')$. Our arguments generalize those applied to the logarithmic minimal models ${\cal LM}(p,p')$~\cite{PRZ} by replacing the role of $q$-binomials with $q$-trinomials. The quantum numbers $r,s,\ell$ determining the sector are fixed by boundary conditions. In each sector, we impose the requirements that (i) as $q\to1$, the finitized characters give the correct counting of the link states with $s\!-\!1$ single defects, (ii) as $N\to\infty$, the finitized characters converge to the corresponding full character (up to the leading power of $q$) and (iii) the finitized characters are fermionic in the sense that, for each $N$, they admit a $q$-expansion with nonnegative integer coefficients. The manner in which sums of $q$-trinomial coefficients enter is fixed by the trinomial coefficients appearing in the counting formulas of Section~\ref{LinkStatesMatrixReps}. These $q$-trinomial coefficients are multiplied by suitable powers of $q$ with quadratic exponents to ensure condition (ii) is satisfied. That such finitized characters exist in each sector (at least with $r=1$), satisfying all of the required properties (i)--(iii), gives remarkable confirmation of the consistency of the lattice approach.

To obtain a finitized form $\chi^{P,P';2,(N)}_{r,s,\ell}(q)$ of $\chi^{P,P';2}_{r,s,\ell}(q)$, we use {\it $q$-trinomial coefficients} defined in terms of $q$-factorials (\ref{qfactorials}) by
\be
\qTrinomial[n][l][m][n-l-m]=\begin{cases}
\frac{(q)_n}{(q)_l(q)_m(q)_{n\!-\!l\!-\!m}}& l,m,n\!-\!l\!-\!m\in \mathbb{Z}_{\ge 0}\\
0,&\mbox{otherwise}\end{cases}
\ee
In the limit $q\to1$, the $q$-trinomials reduce to the trinomial coefficients (\ref{trinomial})
\be
\lim_{q\to1}\qTrinomial[n][l][m][n\!-\!l\!-\!m]=\trinomial[n][l][m][n\!-\!l\!-\!m] \label{eq:qto1}
\ee
In the limit $N\to\infty$, the $q$-trinomials satisfy
\begin{align}
\lim_{N\to\infty}\qTrinomial[N][\hf(N\!-\!j\!-\!k)][\hf(N\!-\!j\!+\!k)][j]
&=\;\lim_{N\to\infty}\frac{(q)_N}{(q)_{\hf(N-j-k)}(q)_{\hf(N-j+k)}(q)_{j}}\nn
&=\;\frac{(q)_\infty}{(q)_{\infty}(q)_{\infty}(q)_{j}}
\;=\;\frac{1}{(q)_{\infty}(q)_{j}}\label{qtriN}
\end{align}

We present our conjectured finitized superconformal Kac characters separately in the Neveu-Schwarz and Ramond sectors. Since the form of the finitized characters for $r>1$ is more complicated, we restrict our attention in this paper to $r=1$. 
Observe that, as in the $n=1$ case, the dependence on $P,P'$ or $p,p'$ in the Kac characters only enters in the fractional power of $q$ in the prefactor. 
To remove these fractional powers of $q$, it is convenient to write the superconformal Kac characters (\ref{superKac}) and $\mathbb{Z}_2$ string functions (\ref{string}) as
\bea
\chi^{P,P'\!,2}_{r,s,\ell}(q)=q^{-\frac{c^{P,P'\!;2}}{24}+\Delta_{r,s,\ell}^{P,P'\!;2}}\,
\hat\chi_{r,s,\ell}(q),\qquad c_1^1=q^{\frac{1}{24}}\,\hat{c}_1^1(q),\qquad c_m^\ell=q^{-\frac{1}{48}}\,\hat{c}_m^\ell(q),\quad\ell=0,2
\label{chihat}
\eea
where $\hat{c}_0^2(q)=\hat{c}_2^0(q)=q^{\frac{1}{2}}(1+\cdots)$ is the only hatted quantity with a $q$-expansion not starting with $1$. We recall that $s\!-\!1$ is the number of single defects and, since $r=1$, we must have $s+\ell$ odd in the NS and R sectors. 
In the NS sector with $\ell=0,2$, $r=1$ and $s$ odd, we also define
\bea
\Delta_s^\ell=\Delta_{1-s}^{\ell;2}+\mbox{Max}[\tfrac{1}{2}(\ell\!-\!s\!+\!1),0]=\begin{cases}
\frac{3}{2},&\mbox{$s=1$, $\ell=2$}\\
\frac{1}{2},&\mbox{$\frac{s+\ell-1}{2}$ odd,\ $s>1$}\\
0,&\mbox{$\frac{s+\ell-1}{2}$ even}
\end{cases}
\eea

\subsubsection{Neveu-Schwarz sectors}

NS: $s$ odd, $N$ even ($\ell=0$) or $N$ odd ($\ell=2$)
\be
q^{\Delta_s^\ell}\,\hat\chi^{(N)}_{1,s,\ell}(q)
=\sum_{j=0}^N q^{\frac{j^2}{2}}\Bigg(\!\!\qTrinomial[N][\hf(N\!-\!j-\frac{s-1}{2})][\hf(N\!-\!j\!+\!\frac{s-1}{2})][j]\!\!\!-\!q^{\frac{s}{2}}\qTrinomial[N][\hf(N\!-\!j\!-\!\frac{s+1}{2})][\hf(N\!-\!j\!+\!\frac{s+1}{2})][j]\!\Bigg)
\label{NSfinchi}
\ee
Taking the limit as $q\to1$ and using (\ref{eq:qto1}) gives
\begin{align}
\lim_{q\to1}\hat\chi^{(N)}_{1,s,\ell}(q)
&=\sum_{j=0}^N \Bigg(\!\!\trinomial[N][\hf(N\!-\!j-\frac{s-1}{2})][\hf(N\!-\!j\!+\!\frac{s-1}{2})][j]\!-\!\trinomial[N][\hf(N\!-\!j\!-\!\frac{s+1}{2})][\hf(N\!-\!j\!+\!\frac{s+1}{2})][j]\!\!\Bigg)\nn
&=\superTrinomial[N][\frac{s-1}{2}]-\superTrinomial[N][\frac{s-1}{2}+1]=R_{N,\frac{s-1}{2}}
\end{align}
which is the correct number of Riordan link states with $s-1$ single or $\frac{1}{2}(s-1)$ cabled defects. 
We check that the limit $N\to\infty$ gives $\hat\chi_{1,s,\ell}(q)$ separately for $\ell=0$ and $\ell=2$.\\[6pt]
NS: $s$ odd, $N$ even ($\ell=0$)\\[4pt]
We separate the two terms in (\ref{NSfinchi}) and sum over only the nonzero terms (in which the arguments of the $q$-trinomial coefficients are integers). This gives two cases:  $s=1$ mod $4$ and $s=3$ mod $4$.\\[6pt]
NS: $s=1$ mod $4$, $N$ even ($\ell=0$)
\begin{align}
\lim_{N\to\infty}\hat\chi^{(N)}_{1,s,0}(q)
&=\lim_{N\to\infty}\Bigg\{\sum_{{j=0\atop j \text{ even}}}^\infty q^{\frac{j^2}{2}}\qTrinomial[N][\hf(N\!-\!j-\frac{s-1}{2})][\hf(N\!-\!j\!+\!\frac{s-1}{2})][j]\nn
&\hspace{3cm}\mbox{}
-q^{\frac{s}{2}}\sum_{j=1 \atop j\text{ odd}}^\infty q^{\frac{j^2}{2}}\qTrinomial[N][\hf(N\!-\!j\!-\!\frac{s+1}{2})][\hf(N\!-\!j\!+\!\frac{s+1}{2})][j]\Bigg\}\nn
&=\frac{1}{(q)_{\infty}}\sum_{j=0\atop j \text{ even}}^\infty \frac{q^{\frac{j^2}{2}}}{(q)_{j}}
-\frac{q^{\frac{s}{2}}}{(q)_{\infty}}\sum_{j=1 \atop j \text{ odd}}^\infty \frac{q^{\frac{j^2}{2}}}{(q)_{j}}
\;=\;\hat c_0^0-q^{\frac{s}{2}} \hat c_2^0
\end{align}
where we have used (\ref{qtriN}) and the fermionic forms (\ref{Isingfermionic}) of the string functions. Indeed, these fermionic forms were used to fix the quadratic exponents of $q$ in the factors multiplying the $q$-trinomial coefficients. We notice that such fermionic forms are known~\cite{HKOTT2002} more generally for $n\ge 2$ parafermionic string functions.\\[6pt]
NS: $s=3$ mod $4$, $N$ even ($\ell=0$)
\begin{align}
\lim_{N\to\infty}q^{\frac{1}{2}}\hat\chi^{(N)}_{1,s,0}(q)
&=\lim_{N\to\infty}\Bigg\{\sum_{{j=1\atop j \text{ odd}}}^\infty q^{\frac{j^2}{2}}\qTrinomial[N][\hf(N\!-\!j-\frac{s-1}{2})][\hf(N\!-\!j\!+\!\frac{s-1}{2})][j]\nn
&\hspace{3cm}\mbox{}-q^{\frac{s}{2}}\sum_{j=0 \atop j\text{ even}}^\infty q^{\frac{j^2}{2}}\qTrinomial[N][\hf(N\!-\!j\!-\!\frac{s+1}{2})][\hf(N\!-\!j\!+\!\frac{s+1}{2})][j]\Bigg\}\nn
&=\frac{1}{(q)_{\infty}}\sum_{j=1\atop j \text{ odd}}^\infty \frac{q^{\frac{j^2}{2}}}{(q)_{j}}
-\frac{q^{\frac{s}{2}}}{(q)_{\infty}}\sum_{j=0 \atop j \text{ even}}^\infty \frac{q^{\frac{j^2}{2}}}{(q)_{j}}
\;=\;\hat c_2^0-q^{\frac{s}{2}} \hat c_0^0
\end{align}
NS: $s=1$ mod $4$, $N$ odd ($\ell=2$)
\begin{align}
\lim_{N\to\infty}q^{\frac{1}{2}+\delta(s,1)}\hat\chi^{(N)}_{1,s,2}(q)
&=\frac{1}{(q)_{\infty}}\sum_{j=1\atop j \text{ odd}}^\infty \frac{q^{\frac{j^2}{2}}}{(q)_{j}}
-\frac{q^{\frac{s}{2}}}{(q)_{\infty}}\sum_{j=0 \atop j \text{ even}}^\infty \frac{q^{\frac{j^2}{2}}}{(q)_{j}}
\;=\;\hat c_0^2-q^{\frac{s}{2}} \hat c_2^2
\end{align}
NS: $s=3$ mod $4$, $N$ odd ($\ell=2$)
\begin{align}
\lim_{N\to\infty}\hat\chi^{(N)}_{1,s,2}(q)
&=\frac{1}{(q)_{\infty}}\sum_{j=0\atop j \text{ even}}^\infty \frac{q^{\frac{j^2}{2}}}{(q)_{j}}
-\frac{q^{\frac{s}{2}}}{(q)_{\infty}}\sum_{j=1 \atop j \text{ odd}}^\infty \frac{q^{\frac{j^2}{2}}}{(q)_{j}}
\;=\;\hat c_2^2-q^{\frac{s}{2}} \hat c_0^2
\end{align}
In the Neveu-Schwarz sectors with $r=1$, these four expressions for the $N\to\infty$ limit of the finitized characters combine to give the required result in accord with (\ref{superKac}):\\[6pt]
NS: $s$ odd, $\ell=0,2$
\be
\lim_{N\to\infty}\hat\chi^{(N)}_{1,s,\ell}(q)
=q^{-\Delta_s^\ell}(\hat c_{m_-}^\ell\!-q^\frac{s}{2} \hat c_{m_+}^\ell)
\ee
where $m_-=s-1$ mod $4$, $m_+=s+1$ mod $4$ and $m_+=2-m_-$ mod $4$.

\subsubsection{Ramond sectors}

In the Ramond sectors, we have $\ell=1$ and $s$ even.\\[6pt]
Ramond: $s$ even, $N$ even or odd ($\ell=1$)
\be
\hat\chi^{(N)}_{1,s,1}(q)
=\!\sum_{j=0}^N q^{\frac{j^2\!-\!j}{2}}\!\Bigg(\!\!\qTrinomial[N][\hf(N\!-\!j\!-\!\frac{s-2}{2})][\hf(N\!-\!j\!+\!\frac{s-2}{2})][j]\!\!\!-\!q^{\frac{s}{2}}\qTrinomial[N][\hf(N\!-\!j\!-\!\frac{s+2}{2})][\hf(N\!-\!j\!+\!\frac{s+2}{2})][j]\!\Bigg)\quad
\ee
Taking the limit $q\to 1$ gives
\begin{align}
\lim_{q\to1}\hat\chi^{(N)}_{1,s,1}(q)
&=\!\sum_{j=0}^N \Bigg(\!\!\trinomial[N][\hf(N\!-\!j-\frac{s-2}{2})][\hf(N\!-\!j\!+\!\frac{s-2}{2})][j]\!-\!\trinomial[N][\hf(N\!-\!j\!-\!\frac{s+2}{2})][\hf(N\!-\!j\!+\!\frac{s+2}{2})][j]\!\!\Bigg)\nn
&=\superTrinomial[N][\frac{s-2}{2}]-\superTrinomial[N][\frac{s-2}{2}+2]
\;=\;M_{N,\frac{s-2}{2}}
\end{align}
which is the correct number of Motzkin link states for system size $N$ with $s-1$ single defects.
For the limit $N\to\infty$, the cases $N$ odd or even and $s=0$ or $2$ mod $4$ must be considered separately\\[6pt]
Ramond: $s=0$ mod $4$, $N$ even ($\ell=1$)
\begin{align}
\mbox{}\hspace{-6pt}\lim_{N\to\infty}\!\hat\chi^{(N)}_{1,s,1}(q)\!
&=\!\!\lim_{N\to\infty}\!\sum_{j=1\atop j \text{ odd}}^\infty\! q^{\frac{j^2\!-\!j}{2}}\!\Bigg(\!\!\qTrinomial[N][\!\hf(N\!-\!j\!-\!\frac{s-2}{2})][\!\hf(N\!-\!j\!+\!\frac{s-2}{2})][j]\!\!\!-q^{\frac{s}{2}}\!\!\qTrinomial[N][\!\hf(N\!-\!j\!-\!\frac{s+2}{2})][\!\hf(N\!-\!j\!+\!\frac{s+2}{2})][j]\!\Bigg)\quad\nn
&=\frac{(1-q^{\frac{s}{2}})}{(q)_{\infty}}\sum_{j=1\atop j \text{ odd}}^\infty \frac{q^{\frac{j^2-j}{2}}}{(q)_{j}}
=(1-q^{\frac{s}{2}})\,\hat c_1^1(q)
\end{align}
Ramond: $s=2$ mod $4$, $N$ even ($\ell=1$)
\begin{align}
\mbox{}\hspace{-6pt}\lim_{N\to\infty}\!\hat\chi^{(N)}_{1,s,1}(q)\!
&=\!\!\lim_{N\to\infty}\!\sum_{j=0\atop j \text{ even}}^\infty\! q^{\frac{j^2\!-\!j}{2}}\!\Bigg(\!\!\qTrinomial[N][\!\hf(N\!-\!j\!-\!\frac{s-2}{2})][\!\hf(N\!-\!j\!+\!\frac{s-2}{2})][j]\!\!\!-q^{\frac{s}{2}}\!\!\qTrinomial[N][\!\hf(N\!-\!j\!-\!\frac{s+2}{2})][\!\hf(N\!-\!j\!+\!\frac{s+2}{2})][j]\!\Bigg)\quad\nn
&=\frac{(1-q^{\frac{s}{2}})}{(q)_{\infty}}\sum_{j=0\atop j \text{ even}}^\infty \frac{q^{\frac{j^2-j}{2}}}{(q)_{j}}
=(1-q^{\frac{s}{2}})\,\hat c_1^1(q)
\end{align}
Ramond: $s=0$ mod $4$, $N$ odd ($\ell=1$)
\begin{align}
\lim_{N\to\infty}\hat\chi^{(N)}_{1,s,1}(q)
&=\frac{(1-q^{\frac{s}{2}})}{(q)_{\infty}}\sum_{j=0\atop j \text{ even}}^\infty \frac{q^{\frac{j^2-j}{2}}}{(q)_{j}}
\;=\;(1-q^{\frac{s}{2}})\,\hat c_1^1(q)
\end{align}
Ramond: $s=2 \mod 4$, $N$ odd ($\ell=1$)
\begin{align}
\lim_{N\to\infty}\hat\chi^{(N)}_{1,s,1}(q)
&=\frac{(1-q^{\frac{s}{2}})}{(q)_{\infty}}\sum_{j=1\atop j \text{ odd}}^\infty \frac{q^{\frac{j^2-j}{2}}}{(q)_{j}}
\;=\;(1-q^{\frac{s}{2}})\,\hat c_1^1(q)
\end{align}
In the Ramond sector with $r=1$, these four expressions for the $N\to\infty$ limit of the finitized characters all give the required result in accord with (\ref{superKac}):\\[6pt]
Ramond: $s$ even, $N$ even or odd ($m_+\!=m_-\!=\ell=1$)
\be
\lim_{N\to\infty}\hat\chi^{(N)}_{1,s,1}(q)
=(1-q^{\frac{s}{2}})\,\hat c_1^1(q)
\ee

\section{Numerical Strip Partition Functions for ${\cal LSM}(p,p')$}
\label{numericalSection}

\subsection{Finite-size corrections}
\label{FiniteSize}
In this section, we use finite size scaling to numerically estimate the eigenvalue spectra of the logarithmic superconformal minimal Hamiltonians in the NS and R sectors.

\subsubsection{Transfer matrices}
For given $p,p'$ or the related $P,P'$, the partition function, on a strip with $N$ columns and $N'$ double rows, of the $n\times n$ fused lattice models ${\cal LM}(p,p')_{n\times n}$ with double row transfer matrix $\vec D^{n,n}(u)$ of fused faces is defined by
\be
  Z^{P,P';n,(N)}_{(r_1,s_1,\ell_1)|(r_2,s_2,\ell_2)}\;=\;\mathop{\rm Tr}\vec D^{n,n}(u)^{N'}\;=\;\sum_j D^{n,n}_j(u)^{N'}
   \;=\;\sum_j e^{-N'E^{n,n}_j(u)}
\label{ZNM}
\ee
where the sum over $j=0,1,2,\ldots$ is over all eigenvalues of the transfer matrix $\vec D^{n,n}(u)$ including possible
multiplicities and $E^{n,n}_j(u)$ is the energy associated to the eigenvalue
$D^{n,n}_j(u)$. The form of the partition function (\ref{ZNM}) holds in all sectors where $(r_1,s_1,\ell_1)$ and $(r_2,s_2,\ell_2)$ are the boundary conditions on the left and right of the strip respectively in the Virasoro picture. 

We usually work in the $(r,s,\ell)$ sector by taking $(r_2,s_2,\ell_2)=(r,s,\ell)$ and $(r_1,s_1,\ell_1)=(1,1,0)$ which is the vacuum boundary condition conjugate to the identity operator. In these cases, it is found numerically that the double row transfer matrices $\vec D^{n,n}(u)$ are diagonalizable with real eigenvalues. 
Conformal invariance of the model in the continuum scaling limit dictates 
\cite{BCN,Aff} that the leading finite-size corrections for large $N$ in the $(r,s,\ell)$ sector
are of the form
\be
 E^{n,n}_j(u)=-\ln D^{n,n}_j(u)\simeq Nf^{n,n}_{bulk}(u)+f^{n,n}_{bdy}(u)
   +\frac{2\pi\sin\vartheta}{N}\Big(\!\!-\!\frac{c}{24}+\Delta^{P,P';n}_{r,s,\ell}+k\Big)+\cdots
\label{logLa}
\ee
for some $k=0,1,2,\ldots$ where the anisotropy angle $\vartheta$~\cite{KimPearce87} and modular nome $q$ are
\bea
\vartheta=\frac{\pi u}{\lambda},\qquad \lambda=\frac{(p'-p)\pi}{p'},\qquad 
q=\exp\Big(\!\!-2\pi\,\frac{N'}{N}\sin\vartheta\Big)
\eea
The central charge of the CFT is $c$ while the 
spectrum of conformal weights is given by the possible values of $\Delta^{P,P';n}_{r,s,\ell}$
with excitations or descendants labelled by the non-negative integers $k$. Here $f^{n,n}_{bulk}(u)$ is the limiting bulk free energy per fused column (which is twice the free energy per fused face) and $f^{n,n}_{bdy}(u)$ is the 
boundary free energy. The bulk free energy is independent of the boundary parameters $r,s,\ell$ and the boundary free energy is independent of $s$. 
\subsubsection{Hamiltonians}

Similarly, the finite-size corrections for the eigenenergies of the Hamiltonian $\calH$  are given by
\be
E^{n,n}_j=Nf^{n,n}_{bulk}+f^{n,n}_{bdy}
  +\frac{\pi v_s}{N}\left(-\frac{c}{24}+\Delta^{P,P';n}_{r,s,\ell}+k\right)+\cdots, \quad k=0,1,2,\ldots
\label{HamBulkFS}
\ee
where $v_s$ is the velocity of sound. For $2\times 2$ fusion, the velocity of sound is given explicitly by
\bea
v_s=\frac{\pi\sin2\lambda'}{\lambda'},\qquad \lambda'=\mbox{Min}[\lambda,\pi-\lambda]
\label{vsound}
\eea
in accord with duality. For the superconformal $n=2$ case of primary interest, we will usually drop the superscripts $(n,n)$. Again the bulk free energy is independent of the boundary parameters $r,s,\ell$ and the boundary free energy is independent of $s$. In particular, for the logarithmic superconformal theories with $n=2$, the Hamiltonian finitized conformal partition functions are
\bea
Z^{P,P';2,(N)}_{(1,1,0)|(r,s,\ell)}(q)=\chi_{r,s,\ell}^{P,P';2,(N)}(q)
\eea
where the $r=1$ finitized characters are given in Section~\ref{FinStrFn+Char} with the modular nome 
\bea
q=\exp\Big(\!\!-\frac{N'\pi v_s}{N}\Big)
\eea

\subsection{Numerical central charges and conformal weights}
\label{numerical}

In this section, we present the results of numerical calculations for the central charges and conformal weights  of the 
${\cal LSM}(p,p'):={\cal LM}(p,p')_{2\times 2}$ models, in the NS ($r+s$ even) and R ($r+s$ odd) sectors, to confirm the identification with the $n=2$ logarithmic minimal coset models
\bea
{\cal LSM}(p,p')\equiv {\cal LM}(P,P';2),\qquad P=|2p-p'|,\ P'=p'
\eea 
Since it is numerically more efficient, we calculate the central charges and conformal weights using the Hamiltonians and not the double row transfer matrices. Because of duality, we can restrict ourselves to the case $0<\lambda<\frac{\pi}{2}$. The numerics support well our theoretical arguments. Ultimately, of course, these models are exactly solvable. So, in principle, it should be possible to obtain all of the conformal data analytically by extending the analysis of Kl\"umper and Pearce~\cite{KlumperPearce1992} based on $T$- and $Y$-systems. Indeed, these functional equations are derived for the logarithmic minimal models ${\cal LM}(P,P';n)$ in \cite{MDPR13}.

\subsubsection{Neveu-Schwarz sector $(r,s)=(1,1)$}

\begin{table}[p]
\center
\begin{tabular}{|c| c | c|c c | c |c|c|c|}
\hline
\multicolumn{9}{|c|}{\rule{0pt}{14pt}Central charges $c$ and $\Delta_{1,1,2}=\frac{3}{2}$ of ${\cal LSM}(p,p')$ with $(r,s)=(1,1)$}\\[3pt]
\hline
\multirow{2}{*}{$(p,p')$} & \multirow{2}{*}{$(P,P')$}& \multirow{2}{*}{$\frac{\lambda}{\pi}$} & \multicolumn{4}{c|}{NS, $\ell=0$ ($N$ even)}&\multicolumn{2}{c|}{NS, $\ell=2$ ($N$ odd)}  \\
	  &   & &\multicolumn{2}{c}{Exact} & \multicolumn{1}{c}{Estimated}& Error &\multicolumn{1}{c}{Estimated}&Error\\
\hline
 $(6,7)$ & $(5,7)$ & 0.143  & $81/70$  & $1.15714$  & $1.15709 $ & $.005\%$&$1.4996$ & $.03\%$\\
  $(5,6)$ & $(4,6)$ & 0.167  & $1 $  & $1 $  & $1.0004$ & $.04\%$ & $1.5008 $ &$.05\%$\\
$(4,5)$ & $(3,5)$ & 0.2  &$7/10$  & $0.7$  & $0.6998$ & $.02\%$ &$1.4995$& $.03\%$\\
$(3,4)$ & $(2,4)$ & 0.25&0 & 0  & 0 &0&$1.498$ & $.14\%$\\
$(5,7)$ & $(3,7)$ &0.286&$-11/14$ & $-0.7857$ &$-0.7854$ &.04\%&$1.499$ & $.07\%$\\
$(2,3)$ & $(1,3)$ &0.333&$-5/2$ & $-2.5$ & $-2.502$&.08\%& $1.504$&$.27\%$\\
$(3,5)$ & $(1,5)$ & 0.4  &$-81/10$ & $-8.1$ &$-7.97$  &1.6\%&$1.45$&3.4\%\\
$(4,7)$ & $(1,7)$ &0.429 &$-195/14$ & $-13.93$ &$-14.12$ &1.4\% &$1.46$ &2.5\%\\
\hline
\end{tabular}
\caption{Numerical estimates of central charges $c$ and the conformal weight $\Delta_{1,1,2}=\frac{3}{2}$ in the Neveu-Schwarz sector with $(r,s)=(1,1)$ for different logarithmic superconformal minimal models ${\cal LSM}(p,p')$. The models are listed in increasing order of the crossing parameter $\lambda=\frac{(p'-p)\pi}{p'}$ with $0<\frac{\lambda}{\pi}<\frac{1}{2}$: (i)~Comparison of some exact (\ref{supercentralcharges}) and estimated values (\ref{groundstateE}) of $c$  with $(r,s,\ell)=(1,1,0)$ and $N$ even. There are no finite size corrections for $c$ in the case of superconformal percolation ${\cal LSM}(3,4)$ with the largest eigenvalue of $-\cal H$ given exactly by $2(N-1)$.
(ii)~Some estimates of $\Delta_{1,1,2}$ with $\ell=2$ and $N$ odd. The exact value $\Delta_{1,1,2}=\frac{3}{2}$ is independent of $(p,p')$ corresponding to the supersymmetric partner of the energy-momentum tensor. The agreement of the numerical values with the theoretical prediction is good for $\lambda<\frac{\pi}{3}$. For $\Delta_{1,1,2}$, the convergence is slower and the errors greater for $\lambda>\frac{\pi}{3}$.
The results in this table, along with the additional results in the plot shown in Figure~\ref{centralPlot}, confirm the identifications $P=|2p-p'|$, $P'=p'$.\label{centralTab}}
\end{table}
\begin{figure}[p]
   \centering
   \includegraphics[width=6in]{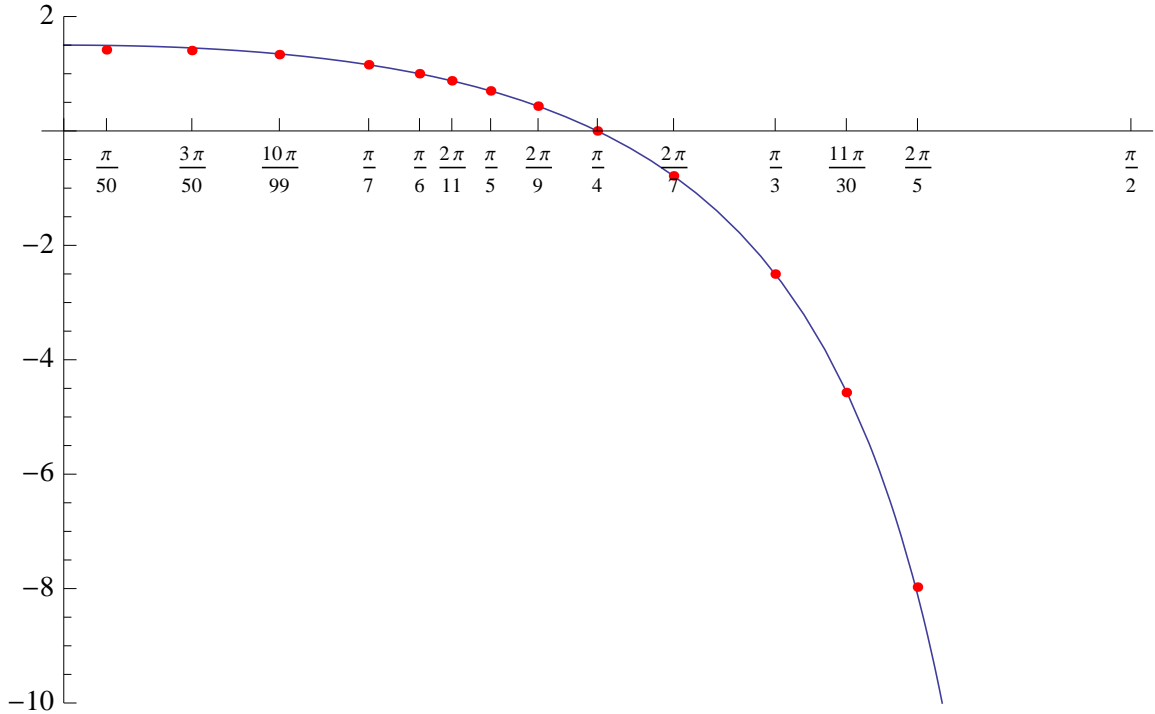} 
   \caption{Plot of the central charges $c=\frac{3}{2}-\frac{12\lambda^2}{\pi(\pi-2\lambda)}$, $0<\lambda<\frac{\pi}{2}$ of the Hamiltonian (\ref{HamNS}) in the Neveu-Schwarz sector $(r,s,\ell)=(1,1,0)$. For $\lambda<\frac{2\pi}{5}$, the agreement with numerics is good. As $\lambda$ approaches $\frac{\pi}{2}$, the central charges diverge to $-\infty$ and the errors in the numerical estimates are larger.\label{centralPlot}}
\end{figure}

\noindent
Estimates for the central charge $c$ and conformal weights $\Delta_j$ are obtained by applying the finite-size corrections (\ref{HamBulkFS}) with the velocity of sound $v_s$ (\ref{vsound}) to the ground state and excited states of the Hamiltonians (\ref{HamNS}) and (\ref{HamR}). The matrices representing these Hamiltonians are constructed and diagonalized in Mathematica~\cite{Wolfram} out to system sizes $N=13$. Since we work in the $2\times 2$ fused TL algebra, this means we work with chains of up to 26 sites in the underlying TL algebra. Let $E_0$ be the lowest energy in the vacuum sector with (1,1,0) boundary conditions on both sides of the strip. For the quantum chains (\ref{HamNS}) and (\ref{HamR}), 
the precise sequences of approximants we use for $E_0$ and excited state energies $E_j$ are
\bea
&\disp  \frac{E_0}{n}\to f_{bulk},\qquad 
 (E_0-N f_{bulk})\to f_{bdy},\qquad
-\frac{24 n}{\pi v_s}(E_0-nf_{bulk}-f_{bdy})\to c&\qquad
\label{groundstateE}\\
&\disp \frac{c}{24}+ \frac{n}{\pi v_s}(E_j-nf_{bulk}-f_{bdy})\to \Delta_j+k,\qquad k=0,1,2,\ldots\qquad n=N-1&
\label{excitedE}
\eea
These sequences are extrapolated using Vanden Broeck-Schwartz~\cite{vandenBroeckSchwartz79} acceleration. 
The lowest energy $E_0$ in the vacuum Neveu-Schwarz sector with $(r,s,\ell)=(1,1,0)$ and $\Delta_{1,1,0}=0$ gives the central charge. 
Our numerical results are shown as a plot in Figure~\ref{centralPlot} and tabulated in Table~\ref{centralTab}.
For the special case of superconformal percolation with $(p,p')=(3,4)$, the eigenenergies of the Hamiltonian in this sector are given by $-E_0(N)=2(N-1)=2,6,10,14,\ldots$ for $N=2,4,6,8,\ldots$ . In this case, we have $\lambda=\frac{\pi}{4}$, $-f_{bulk}=2=\beta^2$, $f_{bdy}=0$ and there are no finite-size corrections so that $c=0$.

Some results for the numerical estimates to the conformal weight $\Delta_{1,1,2}=\frac{3}{2}$ in the Neveu-Schwarz sector with $N$ odd are also shown in Table~\ref{centralTab}.

\subsubsection{Neveu-Schwarz sector $(r,s)=(1,3)$}

The Hamiltonian in the Neveu-Schwarz sectors with $(r,s)=(1,3)$ is given by (\ref{HamNSs=3}). The quantum number $\ell$ takes the values $\ell=0$ ($N$ even) or $\ell=2$ ($N$ odd). Some numerical estimates for the conformal weights $\Delta_{1,3,0}=1-\frac{2\lambda}{\pi}$ and $\Delta_{1,3,2}=\frac{1}{2}-\frac{2\lambda}{\pi}$ are shown in Table~\ref{s3Table} for $0<\lambda<\frac{\pi}{3}$. For $\lambda>\frac{\pi}{3}$, the extrapolations are less reliable.
\begin{table}[ht]
\center
\begin{tabular}{|c| c | c|c c | c |c| cc|c| c|}
\hline
\multicolumn{11}{|c|}{\rule{0pt}{14pt}Conformal weights $\Delta_{1,3,0}$ and $\Delta_{1,3,2}$ of ${\cal LSM}(p,p')$ with $(r,s)=(1,3)$}\\[3pt]
\hline
\multirow{2}{*}{$(p,p')$} & \multirow{2}{*}{$(P,P')$}& \multirow{2}{*}{$\frac{\lambda}{\pi}$} & \multicolumn{4}{c|}{$\Delta_{1,3,0}$ ($N$ even)}&\multicolumn{4}{c|}{$\Delta_{1,3,2}$ ($N$ odd)}  \\
	  &   & &\multicolumn{2}{c}{Exact} & \multicolumn{1}{c}{Est.}& Error &\multicolumn{2}{c}{Exact}& \multicolumn{1}{c}{Est.}  & Error\\
\hline
$(6,7)$ & $(5,7)$ & $0.143$ & $5/7$ & $0.7143$ & $0.7148$ & $.07\%$ &3/14&0.2143& $0.2142$ & $.05\%$ \\
$(5,6)$ & $(4,6)$  & $0.167$ & $2/3$ & $0.6667$ & $0.6673$ & $.10\%$ &1/6&0.1667& $0.1666$ & $.01$\% \\
$(4,5)$ & $(3,5)$  & $0.2$ & $3/5$ & $0.6$ & $0.6001$ & $.01\%$ &1/10&0.1& $0.0999$ & $.06\%$ \\
$(3,4)$ & $(2,4)$ & 0.25 & 1/2 & 0.5 &0.5001& $.03\%$ &0&0&0& $0$ \\
$(5,7)$ & $(3,7)$ & $0.286$ & $3/7$ & $0.429$ & $0.4289$ & $.07\%$ &$-1/14$&$-0.07143$& $-0.07136$ & $.09\%$ \\
$(2,3)$ & $(1,3)$ & $0.333$ & $1/3$ & $0.3333$ & $0.3336$ & $.09\%$ &$-1/6$&$-0.1667$& $-0.1665$ & $.09\%$ \\
\hline
\end{tabular}
\caption{Some numerical estimates of the conformal weights $\Delta_{1,3,0}=1-\frac{2\lambda}{\pi}$ and $\Delta_{1,3,2}=\frac{1}{2}-\frac{2\lambda}{\pi}$ in the Neveu-Schwarz sector with $(r,s)=(1,3)$ and $\ell=0$ ($N$ even), $\ell=2$ ($N$ odd) for different logarithmic superconformal minimal models ${\cal LSM}(p,p')$. The models are listed in increasing order of the crossing parameter $\lambda=\frac{(p'-p)\pi}{p'}$ with $0<\frac{\lambda}{\pi}<\frac{1}{3}$. In this range, the agreement of the numerical values with the theoretical prediction is good. For $\lambda>\frac{\pi}{3}$, the estimates are less reliable. There are no finite size corrections for $\Delta_{1,3,2}$ in the case of superconformal percolation ${\cal LSM}(3,4)$ with the largest eigenvalue of $-\cal H$ given exactly by $2(N-1)$.\label{s3Table}}
\end{table}

\subsubsection{Ramond sector $(r,s)=(1,2)$}
\begin{figure}[p]
   \centering
   \includegraphics[width=6in]{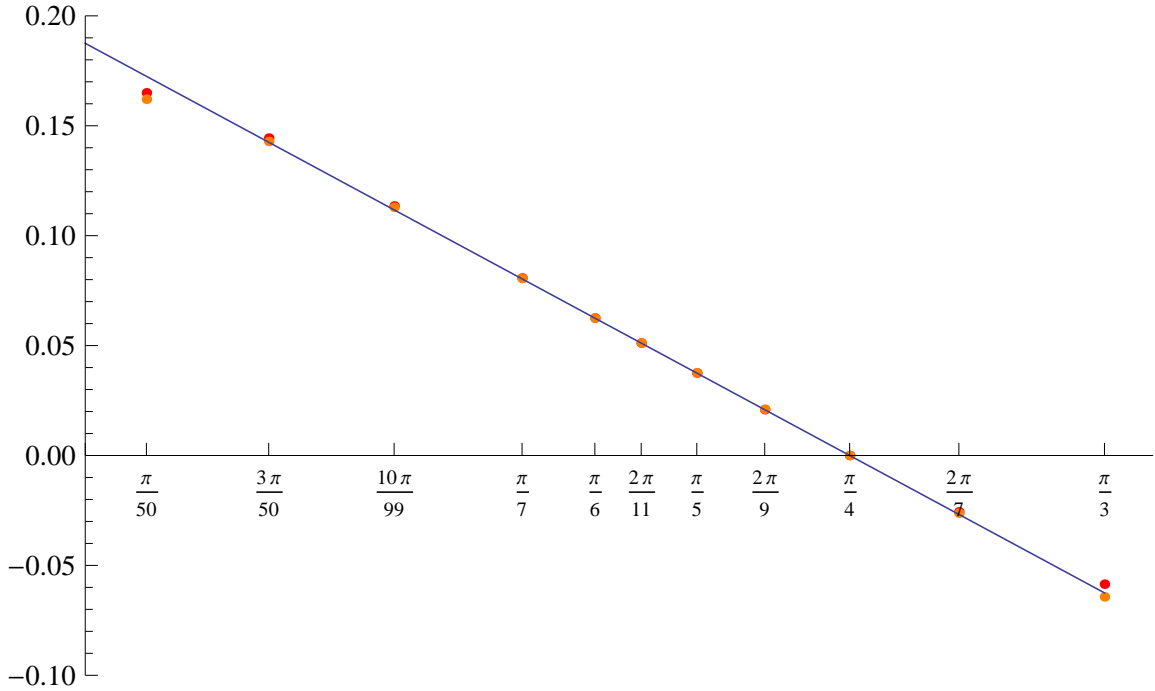} 
   \caption{Plot of the conformal weight $\Delta_{1,2,1}=\frac{3}{16}\big(1-\frac{4\lambda}{\pi}\big)$ for $0<\lambda<\frac{\pi}{3}$ in the Ramond sector against numerical estimates for even (red) and odd (orange) $N$.\label{RamondPlot}}
\end{figure}

The boundary free energies have not been calculated analytically in the Ramond sectors, so they are estimated by numerical extrapolation. Since the extrapolations of conformal quantities are sensitive to the value of the boundary free energy, the numerical precision of the estimates is reduced in the Ramond sectors. Some numerical estimates of the conformal weight $\Delta_{1,2,1}=\frac{3}{16}\big(1-\frac{4\lambda}{\pi}\big)$ are shown in Table~\ref{RamondTable} for $0<\lambda<\frac{\pi}{3}$. For $\lambda>\frac{\pi}{3}$, the extrapolations are unreliable. A plot of the conformal weight $\Delta_{1,2,1}$ for $0<\lambda<\frac{\pi}{3}$ against the numerical estimates for even and odd $N$ is shown in Figure~\ref{RamondPlot}.

\begin{table}[p]
\center
\begin{tabular}{|c| c | c|c c | c |c| c | c|}
\hline
\multicolumn{9}{|c|}{\rule{0pt}{14pt}Conformal weight $\Delta_{1,2,1}$ of ${\cal LSM}(p,p')$ with $(r,s)=(1,2)$}\\[3pt]
\hline
\multirow{2}{*}{$(p,p')$} & \multirow{2}{*}{$(P,P')$}& \multirow{2}{*}{$\frac{\lambda}{\pi}$} & \multicolumn{6}{c|}{Ramond, $\ell=1$}  \\
	  &   & &\multicolumn{2}{c|}{Exact} & {Est. ($N$ even)}& Error & Est. ($N$ odd)  & Error\\
\hline
 $(6,7)$ & $(5,7)$ & $0.143$ & $9/112$ & $0.0804$ & $0.0808$ & $0.49\%$ & $0.0806$ & $0.25\%$ \\
  $(5,6)$ & $(4,6)$  & $0.167$ & $1/16$ & $0.0625$ & $0.0626$ & $0.10\%$ & $0.0625$ & $0.05$\% \\
$(4,5)$ & $(3,5)$  & $0.2$ & $3/80$ & $0.0375$ & $0.0375$ & $0.04\%$ & $0.0376$ & $0.39\%$ \\
$(3,4)$ & $(2,4)$ & 0.25 & 0 & 0 & 0 & $0$ & 0 & $0$ \\
$(5,7)$ & $(3,7)$ & $0.286$ & $-3/112$ & $-0.0268$ & $-0.0256$ & $4.3\%$ & $-0.0262$ & $2.1\%$ \\
$(2,3)$ & $(1,3)$ & $0.333$ & $-1/16$ & $-0.0625$ & $-0.0585$ & $6.3\%$ & $-0.0643$ & $2.9\%$ \\
\hline
\end{tabular}
\caption{Some numerical estimates of the conformal weight $\Delta_{1,2,1}=\frac{3}{16}\big(1-\frac{4\lambda}{\pi}\big)$ in the Ramond sector with $(r,s)=(1,2)$, $\ell=1$ and $N$ even or odd for different logarithmic superconformal minimal models ${\cal LSM}(p,p')$. The models are listed in increasing order of the crossing parameter $\lambda=\frac{p'-p}{p'}$ with $0<\frac{\lambda}{\pi}<\frac{1}{3}$. In this range, the agreement of the numerical values with the theoretical prediction is good. For $\lambda>\frac{\pi}{3}$, the convergence is unreliable. The errors are greater in the Ramond sector since, for this sector, the boundary free energy is not known analytically and the extrapolations are sensitive to the value of the boundary free energy. There are no finite size corrections for the case of superconformal percolation ${\cal LSM}(3,4)$ with the largest eigenvalue of $-\cal H$ given exactly by $4\sqrt{2}+2(N-3)$ for both even and odd $N$.\label{RamondTable}}
\end{table}

\subsection{Excitations}

It is possible to numerically estimate the first few excitations in the conformal towers in each sector. 
Although the finite sizes we can handle are too small to go very far, the results we do obtain are consistent with the $q$ expansion of the $r=1$ Kac characters (\ref{superKac}) for all $P,P'$
\begin{align}
Z_{(1,1,0)|(1,1,0)}^{P,P';2}(q)&=q^{-\frac{c}{24}}(1+q^2+q^3+\cdots)\label{qexp1}\\
Z_{(1,1,0)|(1,1,2)}^{P,P';2}(q)&=q^{-\frac{c}{24}+\frac{3}{2}}(1+q+2q^2+\cdots)\label{qexp2}\\
Z_{(1,1,0)|(1,2,1)}^{P,P';2}(q)&=q^{-\frac{c}{24}+\Delta_{1,2,1}^{P,P';2}}(1+q+2q^2+\cdots)\label{qexp3}\\
Z_{(1,1,0)|(1,3,0)}^{P,P';2}(q)&=q^{-\frac{c}{24}+\Delta_{1,3,0}^{P,P';2}}(1+q+3q^2+\cdots)\label{qexp4}\\
Z_{(1,1,0)|(1,3,2)}^{P,P';2}(q)&=q^{-\frac{c}{24}+\Delta_{1,3,2}^{P,P';2}}(1+q+2q^2+\cdots)\label{qexp5}
\end{align}

\section{Jordan Cells and Representation Theory}
\label{Sec:JordanRep}

In this section, we work with exact transfer matrices for small system sizes $N$ to give some examples of the occurrence of 
Jordan cells in the NS sector. 
We also show, in a couple of examples, how this relates to the representation theory of the $N=1$ superconformal algebra.

\subsection{Example rank-2 Jordan cells for ${\cal LSM}(2,3)$ and ${\cal LSM}(3,4)$}
\label{Sec:ExRank2}

As already mentioned in Section~\ref{FiniteSize}, if the double row transfer matrix has a ($1,s,\ell)$ boundary on one side and the vacuum on the other, then the matrix is found to be diagonalizable with real eigenvalues. Although well supported by numerics, this observation is highly non-trivial because these transfer matrices are not normal matrices. Typically, in the Hamiltonian limit, the eigenvalues are all distinct but this is not always the case. We conjecture that, in general, these matrices are diagonalizable in the Hamiltonian limit $u\to 0$ and assume this in the following discussion.

Following~\cite{PRZ}, let us consider fusing boundary conditions by placing non-trivial boundary conditions on both the left and right edges of the strip. Specifically, let us consider fusion of the boundary conditions $(r,s,\ell)=(1,3,\ell)$ with $\ell=0,2$ in the NS sectors. If $\ell_1=\ell_2=0$, the number of bulk sites $N$ is even. To obtain $N$ odd, we add a seam with $\ell=2$ on the left or the right. These boundary conditions correspond to one cabled defect entering the bulk from the left and one from the right. Since $s=3$, this is described by the fusion of spin-1 representations given by the $su(2)$ fusion rule
\bea
(1,3,\ell_1)\otimes (1,3,\ell_2)=(1,1,\ell)\oplus (1,3,\ell)\oplus (1,5,\ell),\qquad
\ell_1,\ell_2=0,2,\quad \ell=\mbox{$\ell_1+\ell_2$ mod 4}
\label{su2}
\eea
where the decompositions are direct sums if
\bea
\Delta^{P,P';2}_{1,3,\ell}-\Delta^{P,P';2}_{1,1,\ell}\notin \mathbb{Z},\qquad 
\Delta^{P,P';2}_{1,5,\ell}-\Delta^{P,P';2}_{1,3,\ell}\notin \mathbb{Z},\qquad 
\Delta^{P,P';2}_{1,5,\ell}-\Delta^{P,P';2}_{1,1,\ell}\notin \mathbb{Z}
\label{differences}
\eea
In case of integer differences, Jordan cells can be formed thereby rendering the corresponding sums non-direct. 
In terms of link states, the decomposition (\ref{su2}) says that $2, 1$ or 0 single defects can close between the left and right boundaries with the remaining $0, 2$ or 4 single defects entering and propagating in the bulk. If the link states are partitioned in this way, the double row transfer matrices become upper block triangular since defects can only be annihilated in pairs and not created by the action of the TL algebra. In all cases, the blocks on the diagonal are diagonalizable with real eigenvalues. The finitized conformal partition functions are
\bea
Z^{P,P';2,(N)}_{(1,3,\ell_1)|(1,3,\ell_2)}(q)=q^{-\frac{c}{24}}\mbox{Tr}\, q^{L_0^{(N)}}=\chi^{P,P';2,(N)}_{(1,1,\ell)}(q)+\chi^{P,P';2,(N)}_{(1,3,\ell)}(q)+\chi^{P,P';2,(N)}_{(1,5,\ell)}(q)
\eea
where $L_0^{(N)}$ is a finitized Virasoro dilatation operator (related to the Hamiltonian by a shift in energy) with $\lim_{N\to\infty} L_0^{(N)}=L_0$.

If the conditions (\ref{differences}) are satisfied, the double row transfer matrices, corresponding to fusing the two cabled defects, are expected to be diagonalizable and this is confirmed numerically. If the conditions (\ref{differences}) are not satisfied, then it is possible to form rank-2 Jordan cells between two blocks whose conformal weights differ by integers. The formation of such non-trivial Jordan cells is a hallmark of a logarithmic CFT and, indeed, we explicitly find such non-trivial Jordan cells for small system sizes between diagonal blocks with $\Delta^{P,P';2}_{1,s,\ell}-\Delta^{P,P';2}_{1,s',\ell}\in \mathbb{Z}$. We illustrate this with two examples. Explicitly, we consider superconformal dense polymers ${\cal LSM}(2,3)$ with $(P,P')=(1,3)$, $\ell_1=0$, $\ell_2=2$, $N=3$, $\beta=1$, $\beta_2=0$, $c=-\frac{5}{2}$ and superconformal percolation ${\cal LSM}(3,4)$ with $(P,P')=(2,4)$, $\ell_1=\ell_2=0$, $N=4$, $\beta=\sqrt{2}$, $\beta_2=1$, $c=0$. The relevant conformal weights in these two case are respectively
\begin{align}
{\cal LSM}(2,3):&\qquad \Delta^{1,3;2}_{1,1,2}=\tfrac{3}{2},\ \  \Delta^{1,3;2}_{1,3,2}=-\tfrac{1}{6},\ \ 
\Delta^{1,3;2}_{1,5,2}=\tfrac{1}{2},\qquad \Delta^{1,3;2}_{1,5,2}-\Delta^{1,3;2}_{1,1,2}\in\mathbb{Z}\\[3pt]
{\cal LSM}(3,4):&\qquad \Delta^{2,4;2}_{1,1,0}=0,\ \  \Delta^{2,4;2}_{1,3,0}=\tfrac{1}{2},\ \ \ \ \,
\Delta^{2,4;2}_{1,5,0}=\tfrac{1}{2},\qquad \Delta^{2,4;2}_{1,5,0}-\Delta^{2,4;2}_{1,3,0}\in\mathbb{Z}
\end{align}
In the first case, we expect Jordan cells to appear between the $s=1$ and $s=5$ blocks in (\ref{su2}). In the second case, we expect Jordan cells to appear between the $s=3$ and $s=5$ blocks, as discussed in Section~\ref{Sec:rep}. These expectations are confirmed by Jordan decompositions of the exact finite-size double row transfer matrices. In the examples we consider, only rank-2 Jordan cells appear.

For the example of superconformal dense polymers, the Jordan decomposition of the $6\times 6$ (energy-shifted) Hamiltonian matrix gives
\bea
\tfrac{1}{2}(\sqrt{17}-1)I-{\cal H}\sim 0\oplus\tfrac{1}{2}(\sqrt{17}-1)\oplus \tfrac{1}{2}(\sqrt{17}-1)\oplus \begin{pmatrix}\frac{1}{2}(\sqrt{17}+3)&1\\ 0&\frac{1}{2}(\sqrt{17}+3)\end{pmatrix}\oplus \sqrt{17}
\eea
where the diagonal blocks are ordered with increasing energy eigenvalues. The 6 link states, in the NS sector, are the same as the $N=4$, $(r,s,\ell)=(1,3,0)$ link states shown above (\ref{130count}) but with a dashed red line inserted to separate off the cabled defect on the left. The states in the $s=1,3,5$ blocks appear in positions $\{5\}$; $\{3,4,6\}$ and $\{1,2\}$ respectively. The $N=3$ finitized conformal partition function is
\begin{align}
Z^{1,3;2,(3)}_{(1,3,0)|(1,3,2)}(q)&=\chi^{1,3;2,(3)}_{1,1,2}(q)+\chi^{1,3;2,(3)}_{1,3,2}(q)+\chi^{1,3;2,(3)}_{1,5,2}(q)
\nonumber\\
&=q^{-\frac{c}{24}}[q^{\frac{3}{2}}+q^{-\frac{1}{6}}(1+q+q^2)+q^{\frac{1}{2}}(1+q)]
=q^{-\frac{c}{24}}[q^{-\frac{1}{6}}+q^{\frac{1}{2}}+q^{\frac{5}{6}}+2q^{\frac{3}{2}}+q^{\frac{11}{6}}]
\end{align}
corresponding to
\bea
L_0^{(3)}=\big(\!-\!\tfrac{1}{6}\big)\oplus\tfrac{1}{2}\oplus\tfrac{5}{6}\oplus\begin{pmatrix}\frac{3}{2}&1\\ 0&\frac{3}{2}\end{pmatrix}\oplus\tfrac{11}{6}
\eea
Matching the ordering of the energies, confirms that the single rank-2 Jordan cell indeed forms between the two degenerate energy eigenvalues $\frac{1}{2}(\sqrt{17}+3)$ in the $s=1$ and $s=5$ blocks of the Hamiltonian
exactly coinciding with the degenerate conformal energy eigenvalues $E=\frac{3}{2}$ between the $s=1$ and $s=5$ finitized characters. Moreover, the appearance of such Jordan cells is robust as the system size $N$ is increased~\cite{AMDYSA2011}. The Hamiltonian energy eigenvalues will, of course, only approach the exact conformal energies in the limit $N\to\infty$. 

For the example of superconformal percolation, the Jordan decomposition of the $15\times 15$ Hamiltonian matrix gives
\bea
-{\cal H}\sim(-6)\!\oplus\! \vec J_1\!\oplus\!\mbox{\small $\begin{pmatrix}-\!1\!-\!\sqrt{3}&1\\ \!0&\!\!-\!1\!-\!\sqrt{3}\end{pmatrix}$}\!\oplus\!(-\sqrt{2})\!\oplus\!\vec J_2 \!\oplus\!\vec J_3\!\oplus\!\mbox{\small $\begin{pmatrix}\sqrt{3}\!-\!1&1\\ \!0&\!\!\sqrt{3}\!-\!1\end{pmatrix}$}\!\oplus\!\sqrt{2}\oplus\!\vec J_4
\eea
where the diagonal blocks are ordered with increasing energy eigenvalues, $x=x_j$ are the four real roots of the quartic equation $x^4+4x^3-6x^2-12x-4=0$ and
\bea
\vec J_j=\begin{pmatrix}x_j&1\\0&x_j\end{pmatrix},\quad j=1,2,3,4;\qquad x_1<x_2<x_3<x_4
\eea
The $N=4$ finitized conformal partition function is
\begin{align}
Z^{2,4;2,(4)}_{(1,3,0)|(1,3,0)}(q)&=\chi^{2,4;2,(4)}_{1,1,0}(q)+\chi^{2,4;2,(4)}_{1,3,0}(q)+\chi^{2,4;2,(4)}_{1,5,0}(q)
=(1\!+\!q^2\!+\!q^4)+2q^{\frac{1}{2}}(1\!+\!q\!+\!2q^2\!+\!q^3\!+\!q^4)\nonumber\\
&=1+2q^{\frac{1}{2}}+2q^{\frac{3}{2}}+q^2+4q^{\frac{5}{2}}+2q^{\frac{7}{2}}+q^4+2q^{\frac{9}{2}}
\end{align}
corresponding to
\bea
L_0^{(4)}=0\oplus\begin{pmatrix}\frac{1}{2}&1\\ 0&\frac{1}{2}\end{pmatrix}\oplus\begin{pmatrix}\frac{3}{2}&1\\ 0&\frac{3}{2}\end{pmatrix}\oplus 2\oplus\begin{pmatrix}\frac{5}{2}&1\\ 0&\frac{5}{2}\end{pmatrix}\oplus\begin{pmatrix}\frac{5}{2}&1\\ 0&\frac{5}{2}\end{pmatrix}\oplus\begin{pmatrix}\frac{7}{2}&1\\ 0&\frac{7}{2}\end{pmatrix}\oplus 4\oplus\begin{pmatrix}\frac{9}{2}&1\\ 0&\frac{9}{2}\end{pmatrix}
\eea
Matching the ordering of the energies, confirms that rank-2 Jordan cells form between all of the six pairs of degenerate energy eigenvalues in the $s=3$ and $s=5$ blocks of the Hamiltonian and that this exactly coincides with the six pairs of degenerate eigenvalues between the $s=3$ and $s=5$ finitized characters. The exact degeneracy of 4 for the eigenvalue of the Hamiltonian corresponding to the conformal energy $E=\frac{5}{2}$ will only emerge in the limit $N\to\infty$. Moreover, as $N\to\infty$, the early terms in the direct sum expansion of $L_0^{(N)}$ will stabilize as $L_0^{(N)}\to L_0$.

\subsection{Some representation theory}
\label{Sec:rep}

Although our boundary conditions explicitly break supersymmetry, the representation theory of the $N=1$ superconformal 
algebra~\cite{IK03}
can nevertheless be used to explain our findings in the continuum scaling limit.
The $N=1$ superconformal algebra is defined by the (anti-)commutator relations
\be
\begin{array}{rcl}
 [L_n,L_m]&=&(n-m)L_{n+m}+\frac{c}{12}n(n^2-1)\delta_{n+m,0}
 \\[.3cm]
 [L_n,G_\sigma]&=&(\frac{n}{2}-\sigma)G_{n+\sigma}
 \\[.3cm]
 \{G_\sigma,G_{\sigma'}\}&=&2L_{\sigma+\sigma'}+\frac{c}{3}(\sigma^2-\frac{1}{4})\delta_{\sigma+\sigma',0}
\label{SCA}
\end{array}
\ee
where $n,m\in\mathbb{Z}$, while $\sigma,\sigma'\in\mathbb{Z}+\frac{1}{2}$ in the Neveu-Schwarz sector whereas 
$\sigma,\sigma'\in\mathbb{Z}$ in
the Ramond sector. The central charge is denoted by $c$ and is treated as a constant.

Here we focus on the NS sector of ${\cal LM}(P,P';2)$ in which case a typical highest-weight module of highest weight $\Delta$
is of the form
\be
\begin{array}{rc}
 \mbox{level\ $0$:}\quad& \ket{\Delta} \\[.15cm]
 \mbox{level\ $\frac{1}{2}$:}\quad& G_{-\frac{1}{2}}\ket{\Delta} \\[.15cm]
 \mbox{level\ $1$:}\quad& L_{-1}\ket{\Delta} \\[.15cm]
 \mbox{level\ $\frac{3}{2}$:}\quad& G_{-\frac{3}{2}}\ket{\Delta},\quad L_{-1}G_{-\frac{1}{2}}\ket{\Delta} \\[.15cm]
 \mbox{level\ $2$:}\quad& L_{-2}\ket{\Delta},\quad G_{-\frac{3}{2}}G_{-\frac{1}{2}}\ket{\Delta},\quad L_{-1}^2\ket{\Delta} \\[.15cm]
 \mbox{level\ $\frac{5}{2}$:}\quad& G_{-\frac{5}{2}}\ket{\Delta},\quad L_{-2}G_{-\frac{1}{2}}\ket{\Delta},\quad
  G_{-\frac{3}{2}}L_{-1}\ket{\Delta},\quad L_{-1}^2G_{-\frac{1}{2}}\ket{\Delta}\\[.15cm]
  $\vdots$\qquad &
\end{array}
\label{superVerma}
\ee
Examples of NS highest-weight modules are provided by the superconformal Kac modules $(r,s)$ for $r=1$ and $s$ odd.
Such a module $(1,s)$ is constructed as the quotient module obtained by quotienting out the submodule generated from 
the singular vector at level $\frac{s}{2}$ in the highest-weight Verma module of conformal weight $\Delta_{1,s}$ where
\be
 \Delta_{r,s}=\Delta_{r,s}^{P,P';2}=\frac{(rP'-sP)^2-(P'-P)^2}{8PP'},\qquad r-s\in2\mathbb{Z}
\label{sDelta}
\ee
The associated superconformal Kac modules arising in the continuum scaling limit of the lattice 
model ${\cal LM}(P,P';2)$ in the NS sector have characters
\be
 \chit_{r,s}^{P,P';2}(q)=q^{\Delta_{r,s}-\frac{c}{24}}(1-q^{\frac{rs}{2}})\,\phi_{NS}(q),\qquad 
 \phi_{NS}(q)=\prod_{j=1}^\infty\frac{1+q^{j-\frac{1}{2}}}{1-q^j}
\label{phiNS}
\ee
We believe these modules are given by Feigin-Fuchs type modules as in the logarithmic minimal models~\cite{Ras1012}.

We can separate the set of states in (\ref{superVerma}) into integer and half-integer level parts. 
This corresponds to decomposing the module with respect
to its Virasoro subalgebra and gives rise to the characters $\chi_{1,s,\ell}^{P,P';2}(q)$, $\ell=0,2$, and more
generally to $\chi_{r,s,\ell}^{P,P';2}(q)$. As illustration, let us consider
the superconformal Kac module $(1,3)$ constructed by setting the singular vector at level $\frac{3}{2}$ to zero
in the ambient Verma module of conformal weight $\Delta_{1,3}$. 
Since $\ell=2$ corresponds to the integer-level part in this case, it readily follows 
that, in the notation of (\ref{chihat}) and in accordance with (\ref{qexp4}) and (\ref{qexp5}),
\be
 \chih_{1,3,0}(q)=1+q+3q^2+\ldots,\qquad \chih_{1,3,2}(q)=1+q+2q^2+\ldots
\ee
Correspondingly, the full decomposition is given by
\be
 \chit_{1,3}^{P,P';2}(q)=q^{\Delta_{1,3}-\frac{c}{24}}\big[\chih_{1,3,2}(q)+q^{\frac{1}{2}}\chih_{1,3,0}(q)\big]
\ee
More generally, the superconformal Kac characters (\ref{phiNS}) are obtained as the plus expressions in (\ref{superCharspm}).

\subsubsection{Superconformal dense polymers ${\cal LSM}(2,3)$}

In the case of superconformal dense polymers ${\cal LSM}(2,3)={\cal LM}(1,3;2)$ where $c=-\frac{5}{2}$, 
the embedding or structure diagram of the Verma module of highest weight $\Delta_{1,1}=0$ is given by
\be
\begin{pspicture}(8,0)(13,1)
\psset{unit=.7cm}
\setlength{\unitlength}{.7cm}
\rput(0,0){$0$}
\rput(1,0){$\to$}
\rput(2,0){$\frac{1}{2}$}
\rput(3,0){$\to$}
\rput(4,0){$\frac{5}{2}$}
\rput(5,0){$\to$}
\rput(6,0){$4$}
\rput(7,0){$\to$}
\rput(8,0){$8$}
\rput(9,0){$\to$}
\rput(10,0){$\frac{21}{2}$}
\rput(11,0){$\to$}
\rput(12,0){$\ldots$}
\end{pspicture} 
\\[.4cm]
\label{V0a}
\ee
where the arrows indicate actions of the algebra generators (\ref{SCA}) on irreducible subquotients represented by their 
conformal weights.
The structure diagrams of the superconformal Kac modules $(1,s)$ for $s$ odd but $s\not\in3\mathbb{N}$ appear as sub-diagrams
of this ambient structure diagram.
The first few of these superconformal Kac modules are thus given by
\be
 (1,1):\ 0,\qquad (1,5):\ 0\to\tfrac{1}{2},\qquad (1,7):\ \tfrac{1}{2}\to\tfrac{5}{2},\qquad (1,11):\ \tfrac{5}{2}\to4
\label{15}
\ee
indicating that the module $(1,1)$ is irreducible.

Reducible yet indecomposable modules on which $L_0$ acts non-diagonalizably also arise in the NS sector of the $N=1$ 
superconformal algebra, and as above, their integer and half-integer parts match our lattice observations in the previous subsection.
Here we focus on the structure of a particular reducible yet indecomposable module of rank 2 which we conjecture appears in 
${\cal LM}(1,3;2)$, namely
\be
\psset{unit=.25cm}
\setlength{\unitlength}{.25cm}
\Rc_{1,3}^{0,2}:\qquad 
 \mbox{
\begin{picture}(13,3)(-1,6)
    \unitlength=1cm
  \thinlines
\put(1.1,2){$\tfrac{1}{2}$}
\put(0.1,1){$0$}
\put(2,1){$0$}
\put(0.85,1){$\longleftarrow$}
\put(1.5,1.5){$\nwarrow$}
\put(0.5,1.5){$\swarrow$}
\end{picture}
}
\\[0.4cm]
\label{rank2a}
\ee
Up to Virasoro level $\frac{3}{2}$, the states in this module are given by
\be
\begin{array}{rcl}
 \mbox{level\ $\frac{3}{2}$:}\qquad&G_{-\frac{3}{2}}\ket{0}'&\qquad G_{-\frac{3}{2}}\ket{0}\qquad L_{-1}\ket{\tfrac{1}{2}}
 \\[.25cm]
 \mbox{level\ $1$:}\qquad&&\qquad \qquad\qquad\quad G_{-\frac{1}{2}}\ket{\tfrac{1}{2}}
 \\[.25cm]
 \mbox{level\ $\frac{1}{2}$:}\qquad&&\qquad \qquad\qquad\qquad  \ket{\tfrac{1}{2}}
 \\[.25cm]
 \mbox{level\ $0$:}\qquad&\ket{0}'&\qquad\quad \ket{0}
 \\[-.4cm]
  &\underbrace{\phantom{G_{-\frac{3}{2}}\ket{0}'}}_{(1,1):\ 0}&\qquad 
    \underbrace{\phantom{G_{-\frac{3}{2}}\ket{0}\qquad L_{-1}\ket{\tfrac{1}{2}}}}_{(1,5):\ 0\,\to\,\frac{1}{2}}
\end{array}
\ee
where
\be
 \ket{\tfrac{1}{2}}=G_{-\frac{1}{2}}\ket{0},\qquad G_{-\frac{1}{2}}\ket{\tfrac{1}{2}}=L_{-1}\ket{0},\qquad
  L_{-1}\ket{\tfrac{1}{2}}=L_{-1}G_{-\frac{1}{2}}\ket{0}
\ee
The states in the leftmost column form the irreducible superconformal Kac module $(1,1)$ up to level $\frac{3}{2}$, 
while the states in the other two columns form the two irreducible subquotients $0$ and $\frac{1}{2}$ of the reducible yet indecomposable
superconformal Kac module $(1,5)$.
Jordan cells of rank 2 are formed between the states in the two copies of the irreducible subquotient of conformal weight 
$\Delta_{1,1}=\Delta_{1,5}=0$. Concretely, actions by the superconformal generators contributing to the horizontal arrow in 
(\ref{rank2a}) are
\be
 L_0\ket{0}=\ket{0}',\qquad L_0\big(G_{-\frac{3}{2}}\ket{0}\big)=\tfrac{3}{2}G_{-\frac{3}{2}}\ket{0}+G_{-\frac{3}{2}}\ket{0}',\qquad
  G_{\frac{3}{2}}\big(G_{-\frac{3}{2}}\ket{0}\big)=-\tfrac{5}{3}\,\ket{0}+2\,\ket{0}'
\ee
from which it readily follows that $\ket{0}$ and $\ket{0}'$ form an $L_0$ Jordan cell of rank 2 of conformal weight $0$, while
$G_{-\frac{3}{2}}\ket{0}$ and $G_{-\frac{3}{2}}\ket{0}'$ form a similar cell of conformal weight $\frac{3}{2}$.
The actions contributing to the southwest arrow in (\ref{rank2a}) are
\be
 G_{\frac{1}{2}}\ket{\tfrac{1}{2}}=2\,\ket{0}',\qquad L_1\big(G_{-\frac{1}{2}}\ket{\tfrac{1}{2}}\big)=2\,\ket{0}',\qquad
  G_{\frac{3}{2}}\big(L_{-1}\ket{\tfrac{1}{2}}\big)=4\,\ket{0}'
\ee
It is noted that, since the state $L_{-1}\ket{\frac{1}{2}}$ is an element of the irreducible subquotient of conformal weight $\frac{1}{2}$, 
it does not participate in a nontrivial Jordan cell. Instead, we simply have
\be
 L_0\big(L_{-1}\ket{\tfrac{1}{2}}\big)=\tfrac{3}{2}L_{-1}\ket{\tfrac{1}{2}}
\ee

Modules over the Virasoro algebra similar in structure to the rank-2 module in (\ref{rank2a})
were first described in~\cite{Roh96,GK9604} with extensions to certain $W$-algebras discussed in~\cite{GK9606}, and they are often
referred to as staggered modules.
Although these types of modules now are ubiquitous in logarithmic CFT~\cite{FGST0606,Ras2009,KR0905,DJS1001,GJRSV},
rank-2 modules over the $N=1$ superconformal algebra have not been described explicitly before in the literature. The module
in (\ref{rank2a}) and the ones in (\ref{rank2}), (\ref{rank2conj}) and (\ref{rank2conj2}) below are therefore the first of their kind.

The notation $\Rc_{1,s}^{0,b}$ (with certain restrictions on the labels $s$ and $b$) mimics the one used in~\cite{RP0706,RP0707}
to denote indecomposable higher-rank modules in the logarithmic minimal models
${\cal LM}(p,p')$ and indicates that the module can be thought of as an indecomposable combination of the 
two superconformal Kac modules $(1,s-b)$ and $(1,s+b)$.
The particular rank-2 module $\Rc_{1,3}^{0,2}$ is thus an indecomposable combination of $(1,1)$ and $(1,5)$, 
implying that the character of the module is given by
\be
 \chi[\Rc_{1,3}^{0,2}](q)=\chit_{1,1}^{1,3;2}(q)+\chit_{1,5}^{1,3;2}(q)=
  \big[\chi_{1,1,0}^{1,3;2}(q)+\chi_{1,5,0}^{1,3;2}(q)\big]+\big[\chi_{1,1,2}^{1,3;2}(q)+\chi_{1,5,2}^{1,3;2}(q)\big]
\label{1302}
\ee
The two characters given by square brackets are the ones we observe from the lattice.

We can also decompose the character (\ref{1302}) in terms of the irreducible superconformal characters
\be
 \ch_{\Delta_{1,s}}(q)=q^{\frac{5}{48}}\phi_{NS}(q)\begin{cases} q^{\Delta_{1,6k-1}}-q^{\Delta_{1,6k+1}}
    =q^{\Delta_{1,s}}(1-q^{k-\frac{1}{2}}),\ &\ s=6k-1\\[.15cm] 
  q^{\Delta_{1,6k+1}}-q^{\Delta_{1,6k+5}}=q^{\Delta_{1,s}}(1-q^{2k})\ &\ s=6k+1\end{cases}\qquad k\in\mathbb{N}
\ee
Noting that $\Delta_{1,1}=\Delta_{1,5}$, the decomposition thus reads
\be
  \chi[\Rc_{1,3}^{0,2}](q)=2\,\ch_{\Delta_{1,5}}(q)+\ch_{\Delta_{1,7}}(q)=2\,\ch_{0}(q)+\ch_{\frac{1}{2}}(q)
\ee

\subsubsection{Superconformal percolation ${\cal LSM}(3,4)$}

In the case of superconformal percolation ${\cal LSM}(3,4)={\cal LM}(2,4;2)$ where $c=0$, the structure diagrams of the superconformal Kac 
modules $(1,s)$ for $s$ odd appear as sub-diagrams of the ambient structure diagram
\be
\begin{pspicture}(5,1.4)(20,4)
\psset{unit=.9cm}
\setlength{\unitlength}{.9cm}
 \rput(0.1,.73){$\Delta_{1,s}$}
 \rput(0.9,1.28){$\nearrow$}
 \rput(0.9,0.2){$\searrow$}
 \rput(2,1.5){$\Delta_{3,4-s}$}
 \rput(3,1.5){$\to$}
 \rput(4,1.5){$\Delta_{5,s}$}
 \rput(5,1.5){$\to$}
 \rput(6,1.5){$\Delta_{7,4-s}$}
 \rput(7,1.5){$\to$}
 \rput(8,1.5){$\ldots$}
 \rput(2,0){$\Delta_{3,s}$}
 \rput(3,0){$\to$}
 \rput(4,0){$\Delta_{5,4-s}$}
 \rput(5,0){$\to$}
 \rput(6,0){$\Delta_{7,s}$}
 \rput(7,0){$\to$}
 \rput(8,0){$\ldots$}
 \rput(3,0.73){$\searrow$}
 \rput(3,0.73){$\nearrow$}
 \rput(5,0.73){$\searrow$}
 \rput(5,0.73){$\nearrow$}
 \rput(7,0.73){$\searrow$}
 \rput(7,0.73){$\nearrow$}
 \rput(10.5,.73){$s=1,3$}
\end{pspicture}
\\[.6cm]
\label{V1s}
\ee
where the conformal weights are given by (\ref{sDelta}) with $P=2$ and $P'=4$.
The ambient Verma module of highest weight $\Delta_{1,1}=\Delta_{1,3}=0$ is therefore described by
\be
\begin{pspicture}(0,1)(16,3)
\psset{unit=.7cm}
\setlength{\unitlength}{.7cm}
 \rput(0.1,.73){$0$}
 \rput(0.9,1.28){$\nearrow$}
 \rput(0.9,0.2){$\searrow$}
 \rput(2,1.5){$\frac{1}{2}$}
 \rput(3,1.5){$\to$}
 \rput(4,1.5){$5$}
 \rput(5,1.5){$\to$}
 \rput(6,1.5){$\frac{15}{2}$}
 \rput(7,1.5){$\to$}
 \rput(8,1.5){$\ldots$}
 \rput(2,0){$\frac{3}{2}$}
 \rput(3,0){$\to$}
 \rput(4,0){$3$}
 \rput(5,0){$\to$}
 \rput(6,0){$\frac{21}{2}$}
 \rput(7,0){$\to$}
 \rput(8,0){$\ldots$}
 \rput(3,0.73){$\searrow$}
 \rput(3,0.73){$\nearrow$}
 \rput(5,0.73){$\searrow$}
 \rput(5,0.73){$\nearrow$}
 \rput(7,0.73){$\searrow$}
 \rput(7,0.73){$\nearrow$}
\end{pspicture} 
\\[.4cm]
\label{V0}
\ee
from which it follows that the first few superconformal Kac modules $(1,s)$ are given by
\be
 (1,1):\ 0\to\tfrac{3}{2},\qquad (1,3):\ 0\to\tfrac{1}{2},\qquad (1,5):\ \tfrac{1}{2}\to5,\qquad (1,7):\ \tfrac{3}{2}\to3
\ee

We conjecture that the reducible yet indecomposable rank-2 module
\be
\psset{unit=.25cm}
\setlength{\unitlength}{.25cm}
\Rc_{1,4}^{0,1}:\qquad 
 \mbox{
\begin{picture}(13,5)(-1,4)
    \unitlength=1cm
  \thinlines
\put(1.1,2){$5$}
\put(0.1,1){$\frac{1}{2}$}
\put(2,1){$\frac{1}{2}$}
\put(1.1,0){$0$}
\put(0.85,1){$\longleftarrow$}
\put(1.5,1.5){$\nwarrow$}
\put(0.5,1.5){$\swarrow$}
\put(1.5,0.5){$\swarrow$}
\put(0.5,0.5){$\nwarrow$}
\end{picture}
}
\\[0.8cm]
\label{rank2}
\ee
appears in ${\cal LM}(2,4;2)$.
Jordan cells of rank 2 are formed between the states in the two copies of the irreducible subquotient of conformal weight 
$\Delta_{1,5}=\frac{1}{2}$.
Since the rank-2 module $\Rc_{1,4}^{0,1}$ can be thought of as an indecomposable combination of the two superconformal Kac modules 
$(1,3)$ and $(1,5)$, its character is given by
\be
 \chi[\Rc_{1,4}^{0,1}](q)=\chit_{1,3}^{2,4;2}(q)+\chit_{1,5}^{2,4;2}(q)
  =\big[\chi_{1,3,0}^{2,4;2}(q)+\chi_{1,5,0}^{2,4;2}(q)\big]+\big[\chi_{1,3,2}^{2,4;2}(q)+\chi_{1,5,2}^{2,4;2}(q)\big]
\ee
As before, the two characters given by square brackets are the ones we observe from the lattice.
In terms of the irreducible superconformal characters
\be
 \ch_{\Delta_{1,s}}(q)=\phi_{NS}(q)\Big(q^{\Delta_{1,s}}+\sum_{j=k+1}^\infty(-1)^{j-k}[q^{\Delta_{2j+1,1}}+q^{\Delta_{2j+1,3}}]\Big)
\label{chD1s}
\ee
where
\be
 s=s_0+4k,\qquad s_0=1,3,\qquad k\in\mathbb{N}_0
\ee 
we have
\be
 \chit_{1,s}^{2,4;2}(q)=\ch_{\Delta_{1,s_0+4k}}(q)+\ch_{\Delta_{1,8-s_0+4k}}(q)
\ee
and
\be
  \chi[\Rc_{1,4}^{0,1}](q)=\ch_{\Delta_{1,1}}(q)+2\,\ch_{\Delta_{1,5}}(q)+\ch_{\Delta_{1,11}}(q)=\ch_0(q)+2\,\ch_{\frac{1}{2}}(q)+\ch_5(q)
\ee
Among the irreducible characters (\ref{chD1s}), it is noted that
\be
 \ch_0(q)=1
\ee
as in the case of critical percolation ${\cal LM}(2,3)$.

\subsubsection{Neveu-Schwarz sector conjectures}

As we intend to discuss elsewhere, the $N=1$ representation theory outlined above for the NS sector for $r=1$
can be extended to $r>1$ and the Ramond sector and generalises to all ${\cal LM}(P,P';2)$.
Here we conjecture that, in the NS sector of ${\cal LM}(P,P';2)$, there exists an infinite family of indecomposable rank-2 modules 
$\Rc_{1,kP'}^{0,b}$ with $1\leq b<P'$, $kP'-b\in2\mathbb{N}-1$ and $k\in\mathbb{N}$, where $\Rc_{1,kP'}^{0,b}$ can be thought
of as an indecomposable combination of the two superconformal Kac modules $(1,kP'-b)$ and $(1,kP'+b)$, and where these modules
are described by
\be
 (1,kP'-b):\ \Delta_{1,kP'-b}\to\Delta_{1,kP'+b},\qquad
 (1,kP'+b):\ \Delta_{1,kP'+b}\to\Delta_{(k+2)P'-b}
\ee
The corresponding structure diagram is given by
\be
\psset{unit=.25cm}
\setlength{\unitlength}{.25cm}
\Rc_{1,kP'}^{0,b}:\qquad \quad
 \mbox{
\begin{picture}(13,5)(-1,4)
    \unitlength=1cm
  \thinlines
\put(0.6,2){$\Delta_{(k+2)P'-b}$}
\put(-0.8,1){$\Delta_{1,kP'+b}$}
\put(2,1){$\Delta_{1,kP'+b}$}
\put(0.6,0){$\Delta_{1,kP'-b}$}
\put(0.85,1){$\longleftarrow$}
\put(1.5,1.5){$\nwarrow$}
\put(0.5,1.5){$\swarrow$}
\put(1.5,0.5){$\swarrow$}
\put(0.5,0.5){$\nwarrow$}
\end{picture}
}
\\[0.9cm]
\label{rank2conj}
\ee
and the rank-2 Jordan cells are formed between the states in the two copies of 
irreducible subquotients of conformal weight $\Delta_{1,kP'+b}$.
In the special case $\Delta_{1,kP'-b}=\Delta_{1,kP'+b}$, the superconfomal Kac module $(1,kP'-b)$ is irreducible.
This happens if and only if $k=1$ and $P=1$, and the corresponding structure diagram is then given by
\be
\psset{unit=.25cm}
\setlength{\unitlength}{.25cm}
\Rc_{1,P'}^{0,b}:\qquad\quad 
 \mbox{
\begin{picture}(13,3)(-1,6)
    \unitlength=1cm
  \thinlines
\put(0.6,2){$\Delta_{(k+2)P'-b}$}
\put(-0.8,1){$\Delta_{1,kP'+b}$}
\put(2,1){$\Delta_{1,kP'+b}$}
\put(0.85,1){$\longleftarrow$}
\put(1.5,1.5){$\nwarrow$}
\put(0.5,1.5){$\swarrow$}
\end{picture}
}
\\[0.4cm]
\label{rank2conj2}
\ee

Following the analysis initiated in (\ref{su2}), we also conjecture the $N=1$ superconformal fusion rule
\be
 (1,3)\otimes(1,3)=\begin{cases} (1,3)\oplus\Rc_{1,3}^{0,2},\quad &\mbox{in}\ {\cal LM}(1,3;2) \\[.15cm]
   (1,1)\oplus\Rc_{1,4}^{0,1},\quad &\mbox{in}\ {\cal LM}(2,4;2)\\[.15cm] 
   (1,1)\oplus(1,3)\oplus(1,5),\quad &\mbox{otherwise}
  \end{cases}
\label{fus}
\ee
encapsulating the Jordan-cell structures observed in Section~\ref{Sec:ExRank2} and discussed in Section~\ref{Sec:rep}.
In the case of ${\cal LM}(2,4;2)$, the fusion rule (\ref{fus}) and the conjectured structure diagram (\ref{rank2}) of the indecomposable
rank-2 module $\Rc_{1,4}^{0,1}$ seem 
to be confirmed~\cite{CRtalk} by initial applications of the Nahm-Gaberdiel-Kausch algorithm~\cite{Nahm9402,GK9604} 
to the fusion product $(1,3)\otimes(1,3)$.

\section{Conclusion}
\label{Sec:Conclusion}

In this paper, we have argued that the higher level logarithmic minimal models ${\cal LM}(P,P';n)$\cite{PRcoset13} with $n\ge2$, as defined by the GKO cosets (\ref{Clog}), are realized as the continuum scaling limit of the $n\times n$ fused logarithmic minimal lattice models ${\cal LM}(p,p')_{n\times n}$. The identification 
\bea
{\cal LM}(p,p')_{n\times n}\equiv {\cal LM}(P,P';n)\label{Conclude}
\eea
assumes that $P,P'$ are properly related to $p,p'$. After developing an algebraic framework for general $n$ within the fused TL algebra, we have focussed on the $n=2$ logarithmic superconformal minimal models ${\cal LSM}(p,p')$ (\ref{super}) which include superconformal dense polymers ${\cal LSM}(2,3)$ and superconformal percolation ${\cal LSM}(3,4)$  as its first members. For this series, we establish numerically that the identification (\ref{Conclude}) is indeed correct with $P=|2p-p'|$ and $P'=p'$. We explicitly construct commuting double row transfer matrices and their associated quantum Hamiltonians on the strip for the simplest boundary conditions conjugate to operators labelled by the quantum numbers $(r,s,\ell)$ with $r=1$, $s=1,2,3,\ldots$ and $\ell=0,1,2$. These matrices are diagonalizable with real eigenvalues. The spectra separates into Neveu-Schwarz ($r+s$ even, $\ell=0,2$) and Ramond ($r+s$ odd, $\ell=1$) sectors. The transfer matrices and Hamiltonians act on suitable vector spaces of link states which are explicitly constructed for the boundary conditions considered. By using combinatorial arguments based on the counting of the link states, $q$-trinomial coefficients as well as Motzkin and Riordan polynomials, we conjecture general fermionic finitized Kac characters for the $r=1$ boundary conditions which reproduce the expected superconformal Kac characters as the system size $N\to\infty$. The fact that such finitized characters exist in each $r=1$ sector, satisfying all of the required properties, gives remarkable confirmation of the consistency of the lattice approach.

The bulk free energies of the logarithmic superconformal minimal models are obtained analytically. Similarly, the boundary free energies are obtained analytically in the Neveu-Schwarz sectors with $r=1$. In the Neveu-Schwarz sectors $(r,s,\ell)=(1,1,\ell)$, $(1,3,\ell)$ with $\ell=0,2$ and the Ramond sector $(1,2,1)$, we have carried out extensive finite-size numerical studies of the Hamiltonian spectra to extract the central charges $c$ and conformal dimensions $\Delta_{1,1,2}=\frac{3}{2}$, $\Delta_{1,2,1}$, $\Delta_{1,3,0}$ and $\Delta_{1,3,2}$. In all cases, to within numerical error, we find complete agreement with the theoretical predictions. In addition, the numerics correctly reproduce the first few excited levels for the $n=2$ Kac characters. Lastly, examination of the Jordan canonical forms of the Hamiltonians for small system sizes confirms the expected patterns for the appearance of Jordan cells in the Virasoro dilatation operator $L_0$, thus confirming that the logarithmic superconformal minimal models are indeed logarithmic. Further work is required to study the logarithmic superconformal minimal models with boundary conditions labelled by $(r,s,\ell)$ with $r>1$ building on the results of \cite{PRV1210}. It would also be of interest to construct the lattice boundary conditions associated with the superconformal and ${\cal W}$-extended chiral algebras and to study the fusion rules and representation theory in full generality.

\section*{Acknowledgments}
This work is supported by the Australian Research Council.
JR is supported by the Australian Research Council
under the Future Fellowship scheme, project number FT100100774. 
The authors thank Alexi Morin-Duchesne and Adam Ong for discussions.

\end{document}